\documentclass[11pt]{article}
\renewcommand{\textwidth}{6.35 in}

\usepackage{pdflscape}

\usepackage{caption}
\usepackage{graphicx}
\usepackage{amsmath,amsthm,amsfonts,bm}

\usepackage{mathrsfs}
\usepackage{mathtools}
\usepackage{float}
\usepackage[pdftex]{color}

\usepackage{rotating}  
\usepackage{lipsum}    

\usepackage{textcomp}

\makeatletter
\def\normaljustify{%
  \let\\\@centercr\rightskip\z@skip \leftskip\z@skip%
  \parfillskip=0pt plus 1fil}
\makeatother

\usepackage[T1]{fontenc}
\usepackage[utf8]{inputenc}
\usepackage{tabularx,ragged2e,booktabs,caption}
\newcolumntype{C}[1]{>{\Centering}m{#1}}

\restylefloat{table}
\restylefloat{figure}

\begin{document}

\pagestyle{empty}

\long\def\symbolfootnote[#1]#2{\begingroup%
\def\thefootnote{\fnsymbol{footnote}}\footnote[#1]{#2}\endgroup}

\hfuzz 50pt

\rightline{First draft:  March 2, 2026}
\rightline{This draft: March 6, 2026}

\begin{center}

\vskip 1.4cm

{\bf The Gibbs Posterior and Parametric Portfolio Choice}

\vskip .25in

Christopher G. Lamoureux\symbolfootnote[1]{Department of Finance, The University
of Arizona, Eller College of Management, Tucson, 85721,
520--621--7488, lamoureu@arizona.edu.
\noindent
When ready for circulation, the current version of this paper will be available for download from lamfin.arizona.edu/rsch.html .}
\end{center}

\small

\vskip .25in


\baselineskip 15pt

\parindent 18pt

\vskip .7in

\normalsize

\begin{center}
{\bf Abstract}
\end{center}

\noindent

Parametric portfolio policies may experience estimation risk. I develop a generalized Bayesian framework that updates priors, delivering a posterior
distribution over characteristic tilts and out-of-sample returns that is the unique belief-updating rule consistent with the investor's utility function,
requiring no model for the return generating process.  The Gibbs posterior is the closest distribution to the prior in Kullback-Leibler divergence subject
to utility maximization.  The posterior's scaling 
parameter $\lambda$ controls the weight placed on data relative to the  prior. I develop a KNEEDLE algorithm to select optimal $\lambda^*$ in-sample by trading off posterior precision 
against numerical fragility, eliminating the need for out-of-sample validation.
I apply this to U.S. equities (1955–2024), and confirm characteristic-based gains concentrate pre-2000. I find that $\lambda^*$ varies 
meaningfully with risk aversion and depends on higher-order moments.

\vskip 1.25in


\newpage

\setcounter{footnote}{0}

\baselineskip 19pt

\parindent 18pt

\pagestyle{plain}

\setcounter{page}{1}

\leftline{{\bf 1. Introduction}}

\vskip .2 in

Brandt, Santa-Clara, and Valkanov (2009) develop an algorithm for building
utility-optimizing portfolios using measurable stock characteristics. Their parametric portfolio
policy (PPP) maps each asset's characteristics into portfolio weights through a low-dimensional
parameter vector $\theta \in \mathbb{R}^K$, thereby avoiding an explicit return-generating model,
likelihood function, or even moment estimation. Because $K$ is small, PPP is often viewed as a
practical way to reduce estimation risk relative to unconstrained portfolio optimization
(e.g., A\"it-Sahalia and Brandt, 2001). Yet estimation risk does not disappear simply because the
policy is parsimonious: it can persist in the portfolio variance and can be amplified when the
investor is more risk tolerant. Lamoureux and Zhang (2024) show that empirical optimization of
PPP parameters can overfit in-sample and generate large out-of-sample losses.

This paper's contributions are primarily \emph{methodological}: I provide a coherent Bayesian decision
framework for parametric portfolio choice that (i) remains model-free in the sense of PPP, (ii)
delivers a posterior distribution over portfolio policies and out-of-sample portfolio outcomes, and
(iii) regularizes in-sample without requiring costly held-out validation. The core idea is to replace
ad hoc resampling- or tuning-based fixes with a generalized Bayesian update that treats the
investor's utility as the loss function. I begin from a prior on the PPP parameter space that
encodes prior belief (including, potentially, strong belief) in market efficiency---for example, a
prior centered on the market portfolio---and I use the data to update that prior in a way that is
coherent for decision-making even in the absence of a likelihood function.

Specifically, I use generalized Bayesian inference (Bissiri, Holmes, and Walker, 2016), also known
as the Gibbs posterior (Zhang, 2006a, 2006b). The Gibbs posterior updates prior beliefs using an
exponentiated loss:
\begin{eqnarray}
p(\theta \mid \text{data}) \propto \exp\{\lambda \mathcal{L}(\theta,\text{data})\}\,\pi(\theta),
\end{eqnarray}
where $\mathcal{L}$ is the loss function and $\pi(\cdot)$ is the prior.
A key implication of loss-based updating is that the scale parameter $\lambda$ is not pinned down
by Bayesian conditioning under a likelihood. In standard Bayesian analysis, the likelihood fixes the unit of fit and
therefore fixes the update’s scale. Since here, I replace the likelihood with a utility function, I have to choose a learning-rate
(or “temperature”) parameter, $\lambda$, and its calibration "moves from a modeling detail to a
structural component of principled analysis" (McAllin and Takanashi 2026, p. 2).  Rather than treating
$\lambda$ as a nuisance parameter, I treat it as the central \emph{regularization} device, and solve for the optimal value of
$\lambda$, $\lambda^\ast$, in sample.

In this paper I develop a coherent, utility-based posterior distribution over PPP
parameters, portfolio weights, and any posterior function of interest, including the \emph{posterior
predictive distribution of out-of-sample portfolio returns}. This delivers uncertainty quantification
for economically meaningful objects such as certainty equivalents, Sharpe ratios, and factor
exposures, without positing a data generating process, likelihood function, or moment restrictions.  I also
construct a decision-theoretic portfolio using the posterior mean coefficients.  These portfolios' out-of-sample
certainty equivalent returns are higher than the mean out-of-sample certainty equivalents.  I specify a prior
that the market is efficient and I update this prior by finding the distribution that is closest to it in the 
Kullback-Leibler sense, subject to maximizing my utility.

Unlike existing approaches that regularize using out-of-sample data, I develop analytically a link between the
posterior geometry and optimal regularization,
$\lambda^\ast$.   I solve for $\lambda^\ast$ analytically in the quadratic-utility case, and 
show how it depends on risk aversion through the Hessian of the utility function.
For general power utility, I
develop a data-driven {\em identification frontier} that trades off gains in posterior precision against
increases in numerical fragility (overfitting). I measure precision by the log determinant of the
posterior covariance matrix and fragility by its condition number, and I tailor a KNEEDLE-style
elbow algorithm to identify the inflection point $\lambda^\ast$ where marginal fragility begins to
dominate marginal precision.  By identifying $\lambda^\ast$ in sample, I integrate learning, regularization, and
portfolio selection into a single decision-theoretic framework--informed by the utility function.

I obtain posteriors from US equities over the period 1955-2024.  I confirm Lamoureux and Zhang's finding that 
characteristic tilts afford large utility gains prior to 2001 for investors with varying tolerance for risk.
These gains vanish in the 21st Century.  I find that $\lambda^*$ varies meaningfully with risk aversion and departs 
from the quadratic benchmark in ways that are informative about higher-order moments.
The pattern of departure is stable prior to 2001, and changes in ways consistent with structural changes in the
data generating process in the 21st Century.

A practical advantage of selecting $\lambda^\ast$ from posterior geometry is that it avoids reliance
on additional out-of-sample data and bootstrapping or simulating synthetic data that was not observed. 
This is especially material in financial settings where the
data generating process is unstable and where out-of-sample validation may be both expensive
and sensitive to structural change. It also provides an alternative to bootstrap-based fixes that
implicitly impose a resampling model for the data. I do not claim that validation or resampling
are ``wrong''; rather, relative to PPP's model-agnostic motivation, a single-shot coherent update
provides a clean baseline in which regularization is driven by the posterior itself. In this spirit,
the approach is aligned with likelihood-principle reasoning (e.g., Birnbaum, 1962; Berger and
Wolpert, 1988), while remaining applicable precisely when a likelihood is unavailable or
undesirable.

A large literature in portfolio choice emphasizes that estimation error is a first-order
determinant of out-of-sample performance: small errors in inputs can translate into large,
unstable changes in optimized weights. A classic example is Jagannathan and Ma (2003),
who show why seemingly ad hoc constraints (e.g., no-short-sale restrictions) can reduce
risk in estimated optimal portfolios by effectively shrinking noisy mean estimates.
More generally, DeMiguel, Garlappi, Nogales, and Uppal (2009) formalize this idea by
showing that norm constraints act as statistical regularizers that improve out-of-sample
performance in the presence of estimation error. 

Since $\lambda$ regularizes the empirical optimization, it has antecedents in robust control (Hansen and
Sargent, 2001, 2008) and ambiguity aversion (Maenhout 2004; and Garlappi, Uppal, and Wang 2007).   Gilboa
and Schmeidler (1989) argue that a choice problem is ambiguous when the agent does not have a prior.
In my setting, a prior is quite natural: I believe the market is efficient until the data convince me otherwise.
As with the Gibbs sampler, this ambiguity literature uses a Kullback-Leibler divergence from a reference model to regularize.
In that literature the penalty captures ambiguity aversion.  My $\lambda^*$ follows from my own utility function and the
geometry of empirical utility optimization.

The number of papers that use PPP is growing and expanding into a wide array of asset classes
and applications. 
In this setting, DeMiguel, Martín-Utrera, Nogales, and Uppal (2020) show that 
transaction costs can play a related economic regularization role.
They incorporate transaction costs directly into mean–variance
parametric portfolio policies and show that trading frictions materially affect the
portfolio-relevant role of characteristics. In this paper I work with power utility
and regularize using $\lambda^\ast$--obtained from posterior geometry.

The paper also contributes to the burgeoning literature that represents returns and risks in
characteristic space. Like Kelly, Pruitt, and Su's (2019) Instrumented Principal Components Analysis (IPCA),
I shift the investment universe from a set of CUSIPs to a cloud of observations in characteristic space.
In contrast to IPCA, which uses this transformation to identify latent statistical factors,
I use the geometry of characteristic-managed returns to construct a Gibbs posterior over
\emph{utility-maximizing} portfolios. Kozak, Nagel, and Santosh (2018, 2020) argue that
characteristics span the investment opportunity set and estimate the associated low-dimensional
structure via shrinkage and low-rank methods. In a complementary way, their work tells us what
returns look like in characteristic space and IPCA tells us how risk moves in characteristic space;
this paper tells us how to \emph{optimally invest} in characteristic space \emph{under uncertainty}.

Nigmatullin (2003) provides a Bayesian
interpretation of first-order conditions like those in PPP by constructing an empirical-likelihood
function from moment restrictions and then applying Bayes' rule. This approach fits within the
``M-open'' tradition of likelihood-free Bayesian procedures that approximate a likelihood-based
posterior when the likelihood is unavailable or unattractive. In Bayesian empirical likelihood
(Lazar, 2003) and exponentially tilted variants (Schennach, 2005), inference proceeds
by specifying identifying moment conditions $E[m(Z,\theta)]=0$ and using maximum-entropy
arguments to construct a surrogate likelihood that enforces the sample analogs of these moments.

The Gibbs posterior used here is conceptually different. It is not a likelihood surrogate and does
not require moment restrictions. Instead, it is the exact posterior implied by the decision-theoretic
update of Bissiri, Holmes, and Walker (2016), which selects the distribution that maximizes the investor's
expected utility penalized by the 
Kullback--Leibler divergence from the prior. This links the update 
directly to the portfolio choice problem, with $\lambda$ acting as an explicit
learning-rate/regularization parameter rather than a device to approximate a likelihood-based
posterior.  Furthermore, unlike applications of an empirical likelihood, this entropy maximization is
axiomatic: it is the only coherent updating rule under my utility function.

My empirical application's (U.S. equities, 1955--2024) posterior predictive analysis yields
three broad findings. First, the utility gains from conditioning on characteristics are concentrated
before the onset of the 21st century. Second, the $\lambda^\ast$ selected from the identification
frontier varies meaningfully with risk aversion and deviates from the quadratic benchmark,
consistent with an important role for higher-order moments. Third, conventional summaries such
as the Sharpe ratio can be misleading for non-quadratic investors when characteristic tilts alter
tail risk, underscoring the value of a posterior over utility-relevant outcomes rather than point
estimates.

While my interest is in constructing a coherent posterior on out-of-sample portfolio returns and functions of these, I also explore a decision-theoretic approach, in which
I use the posterior mean of the $\theta$ vector to build a portfolio.  I place that portfolio in the context of the posterior.  Because weights are linear in the
coefficients, the mean returns on this decision-theoretic portfolio match the posterior out-of-sample portfolio return means.  This decision-theoretic
portfolio's out-of-sample certainty equivalent returns always lie above the posterior mean of this function, suggesting that the posterior mean provides a viable
tool for decision-making.

The remainder of the paper proceeds as follows. Section~2 describes the data and characteristics.
Section~3 presents the Gibbs posterior framework for parametric portfolio choice and develops
the $\lambda^\ast$ selection method. Section~4 reports the posterior predictive results and the
interactions between risk aversion and regularization. Section~5 concludes.  I provide details on the numerical properties of my Metropolis chains in an appendix.

\vskip .2 in

\leftline{{\bf 2. Data}}

\vskip .2 in

My sample starts with January 1955 and ends with December 2024.  To be eligible for inclusion in the sample
in month $t$, a stock must have no missing returns in the CRSP database for the previous 60 months,
and it must have a positive book value in the Compustat database for a fiscal year-end between $t-6$ and $t-18$, in order to
construct the characteristics in month $t$.  I obtain Gibbs posteriors for each 20-year period starting with 1960-1979, ending with 2005-2024.
There are 46 overlapping 20-year periods used to construct Gibbs posteriors on the parameters,
and 45 years with out-of-sample data used to construct Gibbs posteriors on the out-of-sample returns.  The Gibbs posterior
constructed with the 240 months ending in month $t$ is used to construct the Gibbs posterior out-of-sample portfolio returns in months $t+1$ through $t+12$.
As in Lamoureux and Zhang (2024), I use the following six characteristics: momentum, the book-to-market ratio, log size, beta, market model residual standard deviation,
and the average same-month return over the preceding five years, ($\overline{r}$).
Momentum is the stock's compounded return from month $t-13$ though $t-2$.  Equity market capitalization is the market value
of the company's outstanding shares (aggregated across all share classes)
at time $t-2$.   Book value is obtained from the Compustat database for the most recent fiscal year-end between $t-6$ and $t-18$.
Letting $B$ be one plus the ratio of book value to equity market capitalization,
the book-to-market ratio is the natural log of $B$.
Size is equity market capitalization in month $t-2$.
I estimate beta and the residual standard deviation by regressing monthly returns from months
$t-60$ through $t-1$ on the CRSP value-weighted index.  Thus the stocks in the January 1960 sample have no missing data from January 1955 through
December 1959.

All characteristic and return data are from the merged CRSP--Compustat
file on WRDS.
I use the US Consumer Price Index, CPIAUCSL: all urban consumers: all items,  from the Federal Reserve (FRED),
to construct a minimum size
criterion of \$50 million in January 1990 dollars.  I exclude stocks whose market capitalization is less than this inflation-adjusted
size criterion from the sample.  This excludes stocks with market capitalization less than \$11.5 million in January 1960,
and \$124.6 million in December 2024.
I next exclude the smallest 10\% of stocks that meet all inclusion criteria prior to January 1978, when the first Nasdaq
stocks enter the sample, and the smallest 20\% afterwards.
If the stock return is missing in month $t$, I look to the CRSP delisting return.  If that is missing, I substitute $-30\%$ for NYSE--
and AMEX--listed stocks and $-50\%$ for Nasdaq stocks.

There are 412 (exclusively New York Stock Exchange) stocks in the final sample in
January 1960.  In August, 1967, my sample jumps from 678 to 882, with the addition of American Stock Exchange-listed stocks.
Nasdaq stocks enter my
sample in January 1978, which increases the sample size from 1,001 to 1,420 stocks.  The maximum number of stocks is 2,290 in April, 2006.  There are 1,956 stocks in my sample in March 2008
and 1,616 stocks in the last sample month, December 2024.

I normalize and standardize the characteristics so that they have zero means and unit standard deviations.
Weights are tilted linearly in these standardized characteristics away from the value-weighted portfolio of all sample assets.

\vskip .2 in

\leftline{\bf 3. Bayesian updating}

\vskip .2 in

I consider a fully rational Bayesian investor with a prior that the stock market is efficient.  Under her prior beliefs, the optimal portfolio of risky assets is the
value-weighted portfolio of eligible stocks.  The investor has historical data on the cross-section of characteristics, measurable at $t$, and returns in month $t+1$
to update beliefs that she might be able to improve expected utility by allowing portfolio weights to depend on characteristics.  Brandt, Santa-Clara, and Valkanov
(2009) model expected utility as a direct function of portfolio weights.  By standardizing characteristics to have a zero-mean, the investor's problem is
\begin{eqnarray}
\max_\theta \sum\limits_{t=0}^{T-1}U\left(1 + r_{p,t+1}\right) \left( \frac{1}{T} \right)
\end{eqnarray}
by allowing portfolio weights to depend on observable stock characteristics:
\begin{eqnarray}
r_{p,t+1} = \sum\limits_{i=1}^{N_t} \left(\overline{\omega}_{i,t} + \frac{1}{N_t} \theta^{'} x_{i,t} \right) \cdot r_{i,t+1} \, ,
\end{eqnarray}
where: $x_{i,t}$ is the $K$-vector of zero-mean, unit standard deviation characteristics on firm $i$, measurable at time $t$; $\overline{\omega}_{i,t}$ is
the weight of stock $i$ in the (value-weighted) portfolio of all sample stocks at time $t$; and $N_t$ is the number of stocks in
the sample at time $t$.  $U(1+r)$ is the investor's utility function.  I consider four power utility functions: $U(1+r) = ln(1+r)$ and $U(1+r) = \frac{(1+r)^{1-\gamma}}{1-\gamma},$
for $\gamma = \left\{2, \, 3, \, 6\right\}$.

Brandt, Santa-Clara, and Valkanov (2009) stress that an investor can use this algorithm to maximize in-sample utility without taking a stand on the data generating process.  By reducing the 
portfolio problem to that of estimating the $K-$vector $\theta$ of characteristic tilts, thereby avoiding the estimation of moments, or a complex likelihood function, they 
hope to mitigate the problem of overfitting
that plagues traditional portfolio optimization.  As described above, as a Bayesian decision maker, I also want to be completely agnostic about the data generating process---not simply
to make the estimation more parsimonious, but also to avoid traps associated with a misspecified model.
Lamoureux and Zhang (2024) show that despite its parsimony this algorithm can suffer from overfitting.
They show that estimation risk lives in the portfolio variance, so more risk tolerant investors' optimal portfolios produce large losses out of sample.

\vskip .2 in

\leftline{\em 3.1 Metropolis within Gibbs draws}

The investor updates beliefs about $\theta$ after seeing 240 months of data using the Gibbs posterior:
\begin{eqnarray}
p_\lambda(\theta | \mathrm{data}) \propto e^{\left\{\lambda U(\mathrm{data} , \theta) \right\}} \pi(\theta)  = \operatorname*{arg\,min}_{q} \left\{-\int \lambda U(\mathrm{data} , \theta) q(\theta) d\theta + \mathrm{KL}(q(\theta)|| \pi(\theta)) \right\}
\end{eqnarray}
For a given $\lambda$, the posterior is the distribution that minimizes the Kullback-Leibler divergence from the prior subject to maximizing utility.
The parameter $\lambda$ serves two purposes.  First, it scales the loss function so that it is conformable to the prior.  This is unnecessary in traditional Bayesian
analysis, as the loss function is a density---that is not the case here.  I also use it to mitigate estimation risk.  At very high values of $\lambda$, the Gibbs posterior
will mirror empirical utility maximization (analogous to Brandt, Santa-Clara, and Valkanov 2009).  At very low values of $\lambda$ the posterior will simply be the
prior.  A standard approach to choose parameters like $\lambda$ is out-of-sample validation.  That is costly in portfolio optimization as predictive relationships 
for stock returns are not temporally stable.  Instead, I obtain Gibbs posteriors for an array of $\lambda$ values and choose the optimal by trading off precision
and instability (overfitting).  In the sense of the variational characterization of (4), $\lambda$ is a constraint placed on expected utility.

I start with a ${\cal N}\left(\theta_0,\Sigma_0\right)$ prior on $\theta$.  I specify $\Sigma_0 = I_K$ and $\theta_0 = 0$.
For a given power utility investor, I obtain a Gibbs posterior---conditional on 240 months of data ending in month $t$ and $\lambda$.  I use Metropolis within
Gibbs to draw component-wise from the parameter space.  I specify a symmetric stable Paretian proposal density for each element of $\theta$, with characteristic exponent
($\alpha$) of 1.75, calibrated scale ($\gamma_{s,j}$), and mean equal to the latest draw.
I calibrate $\gamma_{s,j}$ based on experience to target an acceptance rate 
between 0.35 and 0.6 for each parameter.  
For the Metropolis step, I draw a $\theta_j^p$ from the proposal distribution.  I evaluate in-sample expected utility replacing $\theta_j$ with $\theta_j^p$, $U^p$.
Let $U$ be the in-sample expected utility at the current $\theta_j$ vector.
The Metropolis--Hastings acceptance ratio (symmetric proposal) is:
\begin{eqnarray}
\rho = \exp\!\left( \lambda (U^p - U) -\frac{1}{2}\Big[ (\theta_j^{p})^{'} \Sigma_0^{-1} \theta_j^{p} - \theta_j^{'} \Sigma_0^{-1} \theta_j \Big] \right).
\end{eqnarray}
Accept $\theta_j^p$ with probability $\alpha(\theta_j, \theta_j^p) = \min\{1, \rho\}.$

For all Gibbs posterior distributions I discard the first 200,000 draws to burn-in the Markov chain.  
For base analysis I keep the next 100,000 draws, comprising the posterior.
For the optimal cases, the Gibbs posterior comprises at least 300,000 Metropolis draws.  I report properties of the Metropolis draws in the  appendix.  Following Roberts
and Rosenthal (2001), I select the scales of the proposal densities to target an acceptance rate of 44\%.  In practice, the posterior distributions are quite robust
to acceptance rates that range from 19 to 80\%.  I do not use an adaptive algorithm to ensure that my chain is Markovian and standard Markov Chain Monte Carlo convergence
properties apply.

This Metropolis within Gibbs approach is quite robust to recalcitrant functions.  As Chernozhukov and Hong (2003) note, MCMC methods are well-suited to optimizing 
highly non-concave functions.  In earlier work (Lamoureux and Zhang 2024), we did not include log utility because empirical optimization was unstable.  Since optimization 
was embedded within a bootstrap for millions of optimizations, and achieving convergence required extensive manual intervention, systematic implementation was
impractical.  Unlike optimization, MCMC (including this Gibbs posterior approach) concentrates draws into the maximal region(s) of the function.\footnote{This aspect of
generalized Bayesian inference is also stressed by Knoblauch, Jewson, and Damoulas (2022), Martin and Syring (2022) and Syring and Martin (2023).}

\vskip .2 in

\leftline{\em 3.2 $\lambda$ selection}

I develop an algorithm to select the optimal value of $\lambda$, $\lambda^*$ by analyzing the properties of the posterior
covariance matrix of $\theta$, $\Sigma$.  As noted above, at low levels of $\lambda$ ($\lambda = 500$), the Gibbs posterior fully reflects the prior---the data
has little or no effect.  At very high levels of $\lambda$ ($\lambda = 100,000$), the Gibbs posterior is tightly distributed about the empirical optimal with no 
regularization or discipline from the prior.  Selecting an optimal value for $\lambda$ entails finding an optimal tradeoff between precision (learning from the data)
and overfitting (fragility).  I use the condition number of $\Sigma$ (the ratio of the largest to the smallest eigenvalue) to characterize overfitting and 
$\log \det \Sigma$ to measure precision.   I characterize the tradeoff with an identification frontier to select the value of $\lambda$, $\lambda^*$, where the marginal increase
in precision equals the marginal loss from overfitting under the investor's utility function.

I have
a grid of $J = 11$ $\lambda$ values for the log utility function shown in Figure 1, and $J = 13$  $\lambda$ values for the power utility function with $\gamma = 6$ shown in Figure 2. 
Let $y(\Sigma) = $~-log~det~$\Sigma$ be the negative log determinant and $\kappa$ the condition number of the posterior covariance matrix $\Sigma$.  Define $x = $~log~$\kappa$.
I use this change of variable and  project $y$ on $x$, which yields the slope $m$.  Then $\frac{dy}{d\kappa} = \frac{m}{\kappa}$.  Differentiating again with respect to $\kappa$:
$\frac{d^2 y}{d\kappa^2} = \frac{1}{\kappa^2}(m' - m).$  Assuming the curvature of the projection in log space is negligible compared to $m$: Information Deceleration $\approx
\frac{-m}{\kappa^2}$.  Alternatively, this is the exact second derivative if $y$ and $\kappa$ follow a power law relationship: $y \sim \kappa^m$.

Mapping this information frontier in $(\kappa, \frac{d^2(-log det \Sigma)}{d \kappa^2}$ space isolates the phase transition in the specific investor's Bayesian updating.  KNEEDLE identifies
the inflection point, which shows where the marginal gain in learning from the data (signal quality) starts to decline relative to the marginal increase in fragility.  In this way the
resulting $\lambda^{*}$ is scaled to this specific investor's utility function.  

Figures 1 and 2 plot information deceleration on the vertical axis and the condition number on the horizontal axis to construct the identification frontier.
I use a KNEEDLE algorithm (minimum chord) to isolate the "knee" (or "elbow") on this frontier (Satop\"{a}\"{a}, Albrecht, Irwin, and Raghavan 2011).
For $j = 1, \ldots, J$, let $\kappa_j$ be the condition number of $\Sigma_j$, the
Gibbs posterior covariance matrix conditional on $\lambda_j$;
and $\Delta_j = \frac{d^2 y_j}{d\kappa_j^2}$, the information deceleration conditional on $\lambda_j$.  
Define $d_{1,j} = \frac{\kappa_j - \kappa^{-}}{\kappa^{+} - \kappa^{-}}$ and $d_{2,j} = \frac{\Delta_j - \Delta^{-}}{\Delta^{+} - \Delta^{-}}$.  Where $x^{-}$ and $x^{+}$ are the minimum
and maximum values of the set of $J$ variables.  My perpendicular distance KNEEDLE rule: 
\begin{eqnarray}
\lambda^* = \operatorname*{arg\,max}_{j} \frac{| d_{1,j} - d_{2,j} |}{\sqrt{2}}.
\end{eqnarray}
Figure 1 shows the identification frontier on the $\lambda$ grid
for the case of log utility conditional on data from the 240 months from January 1977 through December 1996, and $\lambda^*$ is 7,500.
Figure 2 shows the information from the same data for the
power utility function with coefficient of relative risk aversion, $\gamma =6$, and $\lambda^*$ is 2,500.  
The two information frontiers are explicit functions of each utility function.
In Figure 1, for the log utility investor,
at $\lambda = 500$, $\Sigma$'s condition number is 1.32  and the approximated second derivative is -1.65 (-0.40),  and at $\lambda = 10,000$, these values are 6.99 (23.28) and 3.20.
In Figure 2 for the power utility investor with $\gamma = 6$, the condition number is 2.96 when $\lambda = 500$ and the approximate second derivative is -0.40; and at $\lambda = 10,000$,
the condition number is 7.73 and the approximate second derivative is 23.28.
Both figures show that at values of $\lambda$ below $\lambda^*$, increasing $\lambda$ has a large effect on the log determinant without 
increasing the condition number dramatically.  By contrast, gains in precision from increasing $\lambda$ above $\lambda^*$ are overwhelmed by large increases in the condition number.

\vskip .2 in

\leftline{\em 3.3 Analytics of $\lambda^* | \gamma$}

Table 1 shows the effects of $\gamma$ and $\lambda$ on the Gibbs posterior mean and standard deviation of $\theta$, conditional on the 240 months of data, January 1977 - 
December 1996, corresponding to out-of-sample year 1997 in Figures 3-6.  
I provide the Gibbs posterior mean and standard deviation of $\theta$ for each of the four utility functions, and for four values of $\lambda$: 2,500;
3,500; 4,500; and 7,500.  The subscripts on $\theta$ refer to the characteristics: 1: momentum; 2: book-to-market ratio; 3: log size; 4: $\beta$; 5: average same-month return;
and 6: residual volatility.
We saw that for log utility $\lambda^* = 7,500$ in this period in Figure 1, and that similarly for power utility with $\gamma = 6$ $\lambda^* = 2,500$ in
Figure 2.  The posterior values for $\lambda^*$ for each utility function are emboldened in the table.
For fixed $\lambda$ increasing risk aversion results in smaller absolute values of five of the six coefficients: the $\theta$ weight tilts on momentum, log size, the book-to-market
ratio, average same-month return, and residual volatility.  The $\theta$ coefficient on $\beta$ is close to 0 in all cases and there is no regular effect of changing $\gamma$
on its $\theta_4$ coefficient.  In a related manner, with $\lambda$ fixed, increasing $\gamma$ increases the Gibbs posterior precision.  For fixed $\gamma$, increasing $\lambda$
reduces the magnitude of $\theta$ and posterior precision. I retain $\beta$ in the sample characteristic set for several reasons.  First, and
from a substantive perspective, Liu, Stambaugh, and Yuan (2018) note that $\beta$ and residual volatility are correlated measures of past variance so using only one may be confounding by omission.
Second, from a statistical perspective, the covariances between the $\theta$ coefficient on $\beta$ and other elements of $\theta$ are relevant to the posterior.
In the Appendix, I show that the correlation dynamic between these two volatility measures affects posterior inference on the $\theta$ vector.

To demonstrate the role of $\lambda^*$, and its relationship to risk aversion and the Hessian, 
I use the Laplace approximation to approximate the Gibbs posterior near the mode, $\hat{\theta}$, by expanding the utility function as a quadratic
function at $\hat{\theta}$.  Since first-order conditions are satisfied,
\begin{eqnarray}
U_\gamma(\theta) \approx U_\gamma(\hat{\theta}) + \tfrac{1}{2} \left(\theta - \hat{\theta}\right)' H^U_\gamma \left(\theta - \hat{\theta}\right)
\end{eqnarray}
where $H^U_\gamma$ is the Hessian of the utility function: $H^U_\gamma = \frac{\partial^2 U_\gamma}{\partial \theta \partial \theta^{'}}$.  Since utility is concave in $\theta$,
it is convenient to define the positive definite curvature matrix $H_\gamma = -H^U_\gamma$.
Under this approximation, $\theta \sim {\cal N}(\overline{\theta}, \Sigma)$, where 
\begin{eqnarray}
\Sigma(\lambda,\gamma) = \left(\Sigma_0^{-1} + \lambda H_\gamma \right)^{-1} \, \, \, \, \mathrm{and} \, \, \, \, \overline{\theta}(\lambda,\gamma) = \Sigma(\lambda,\gamma) \left(\Sigma_0^{-1} \theta_0 + \lambda H_\gamma \hat{\theta}\right)
\end{eqnarray}
Let $\Sigma_{r,t}$ be the $N_t \times N_t$ covariance matrix of returns and $\mu_t$ be the $N_t-$vector of mean returns at time $t$.  
Define $X_t$ to be the $N_t \times K$ matrix of characteristics at $t$.\footnote{The columns of $X$ have zero mean and unit variance.}
Projecting returns onto characteristics,
define: $g = X_t' \mu$ and $Q = X_t' \Sigma_{r,t} X_t.$
Locally, a mean-variance expansion of the utility function gives:
\begin{eqnarray}
U_\gamma(\theta) \approx g' \theta - \frac{\gamma}{2} \theta^{'} Q \theta + \text{(skewness and kurtosis terms)}.
\end{eqnarray}
And: 
\begin{eqnarray}
\frac{\partial^2 U_\gamma(\theta)}{\partial \theta \partial \theta^{'}} \approx -\gamma Q.
\end{eqnarray}
So that $H \approx \gamma Q$.  This simplifies the posterior:
\begin{eqnarray}
\Sigma(\lambda , \gamma) = \left(\Sigma_0^{-1} + \lambda \gamma Q\right)^{-1}, \, \, \mathrm{and}  \, \, \overline{\theta}(\lambda , \gamma) = \Sigma(\lambda , \gamma) \lambda \gamma Q \hat{\theta}.
\end{eqnarray}
Both $\Sigma$ and $\overline{\theta}$ depend on $\lambda$ and $\gamma$ only through $\tau = \gamma \lambda$.
Define the posterior certainty equivalent $CE(\lambda,\gamma):$
\begin{eqnarray}
U_\gamma \left(1 + CE(\lambda,\gamma)\right) = EU(\lambda , \gamma)
\end{eqnarray}
Continuing with these simplifying assumptions, $\overline{\theta}(\lambda , \gamma) = \overline{\theta}(\tau)$ and $\Sigma(\lambda , \gamma) = \Sigma(\tau)$.  So:
\begin{eqnarray}
CE(\lambda , \gamma) = g' \overline{\theta}(\tau) - \frac{\gamma}{2} \left(\overline{\theta}(\tau)^{'} Q \overline{\theta}(\tau) + tr(Q \Sigma(\tau)) \right).
\end{eqnarray}
For expositional purposes, expand around the empirical mean-variance optimizer $\hat{\theta}$, which satisfies $Q \hat{\theta} \propto g$.
Now,
\begin{eqnarray}
CE(\lambda , \gamma) = \Psi(\tau) - \frac{\gamma}{2} \Phi(\tau).
\end{eqnarray}
The first order condition for $\lambda^*$ is: $\Psi'(\tau^*) = \frac{\gamma}{2}\Phi'(\tau^*)$, (since $\gamma \ge 0$).  Risk aversion, $\gamma$, and $\lambda$ enter
$\Phi'(\tau)$ and $\Psi'(\tau)$ only through $\tau$.  The data, through this first order condition, defines a single optimal value $\tau^*$ for all mean-variance investors.
So as risk aversion increases, the optimal value of $\lambda^*$ decreases proportionally:
\begin{eqnarray}
\lambda^*(\gamma) = \frac{\tau^*}{\gamma} \propto \frac{1}{\gamma}.
\end{eqnarray}
In this special (mean-variance) case, as the utility function becomes more curved, the optimal weight 
on the prior relative to the data increases.  This seems counterintuitive as increasing $\gamma$ already penalizes variance risk more heavily.  But $\lambda^*$ has to shrink
estimation risk proportionately.

This quadratic setting also allows me to link my KNEEDLE algorithm to select $\lambda^*$ in sample to the axiomatic origins of the Gibbs posterior.
In Equation (4) I showed that the Gibbs posterior is the distribution closest to the prior in KL divergence, subject to the utility-based updating constraint.  Referring to
(8) above and specifying the prior $\Sigma_0 = I_K$, the differential entropy is:
\begin{eqnarray}
h(\mathcal N(\overline{\theta},\Sigma)) = \tfrac{1}{2}\log\! \big((2 \pi e)^K\ \det \Sigma\big)
\end{eqnarray}
Thus, relative to the prior the reduction in entropy is:
\begin{eqnarray}
h(\pi)-h(p_\lambda)=\tfrac{1}{2}\log\det\Sigma_0-\tfrac{1}{2}\log\det \Sigma(\lambda,\gamma)
= -\tfrac{1}{2}\log\det \Sigma(\lambda,\gamma)
\end{eqnarray}
Furthermore, the KL divergence from the posterior to the prior is:
\begin{eqnarray}
KL\!\left(\mathcal N(m,\Sigma)\,\|\,\mathcal N(0,I_K)\right)
=\tfrac{1}{2}\left(\mathrm{tr}(\Sigma)+\overline{\theta}'\, \overline{\theta}-K-\log\det \Sigma\right).
\end{eqnarray}
Relying on (8):
\begin{eqnarray}
-\log\det \Sigma(\lambda,\gamma)=\log\det(I_K+\lambda\gamma Q).
\end{eqnarray}

This shows the sense in which my precision metric is a proxy for information gain (i.e., entropy reduction).
The Hessian determines the curvature of the utility surface around the mode and Equation (8) links the Hessian to the posterior covariance matrix.
So the posterior covariance's condition number summarizes the heterogeneity of that curvature along the directions of the six characteristics.
It is in this sense that trading off fragility against the information gain avoids anisotropic entropy.
In the mean-variance case the Hessian is strictly proportional to $Q$, the covariance matrix of characteristic-managed portfolio returns, and so
$\lambda^{*}$ scales proportionately to $\gamma$.  However for power utility functions higher order moments affect the Hessian, which motivates the
KNEEDLE algorithm to select $\lambda^{*}$ for each investor and each conditioning data set.

\newpage


\leftline{\bf 4. Results}

\vskip .2in

\leftline{\em 4.1 Empirical $\lambda^*$ -- higher order moments}

Table 1 shows that for the 1977-1996 estimation period $\lambda^*$ does monotonically decrease in $\gamma$, but the rate of decrease is not proportional.
Table 2 shows the optimal $\lambda$ for each 20-year estimation period, for each of the four utility functions.  Year 18 (emboldened) corresponds to the results in Table 1
and Figures 1 and 2.
As in Table 1, generally $\lambda^*$ decreases in $\gamma$, but the decrease is less than proportional and there are some violations of monotonicity.  For example, for the
twenty-eighth estimation period (1987-2006), $\lambda^*$ is 6,250 for log utility, and 2,000; 3,000; and 2,250 for power utility functions with $\gamma = 2, 3, 6$, respectively.

In practice, these deviations provide a sense of the importance of higher-order moments for the portfolio problem.  The local Hessian is: $H_\gamma = \gamma Q + 
H_{\mathrm skew}(\gamma) + H_{\mathrm kurt}(\gamma).$ And along tail directions, $H_{\mathrm kurt} \sim \gamma(\gamma + 1) (\gamma + 2)$, which would generate a larger value
of $\lambda^*$ at $\gamma = 3$ than at $\gamma = 2$, or flattening of $\lambda^*$ in this region, as seen in several of the 20-year periods ending in 2001 through 2019.
The flattening of $\lambda^*(\gamma)$ in all years is the result of skewness and kurtosis.  This behavior of $\lambda^*$ as a function of $\gamma$ implies that skewness and kurtosis
make information about the distribution's shape more valuable as risk aversion increases relative to the mean-variance case.  This is intuitive as the leptokurtosis in returns puts a
higher penalty on not knowing the distribution's shape as risk aversion increases.

This approach and algorithm depend solely on the investor's utility function and in-sample data.  
I construct a coherent posterior for portfolio decision without introducing a separate
layer for estimation risk.  Approaches such as Lamoureux and Zhang (2024) regularize out-of-sample performance via bootstrap-based minimax tuning.  While a valid 
robustness device, it is not a single-step belief-updating procedure as uncertainty and optimization are handled separately rather than within a unified posterior.


\vskip .2in

\leftline{\em 4.2 Characteristic tilts}

I report the marginal Gibbs posterior distributions on $\theta$ conditional on $\lambda^*$ for the four utility functions in Figures
3-6, respectively, from each of the 46 overlapping 20-year estimation periods.  
Out-of-sample Year 1997 in these figures corresponds to the data in Table 1.  The figures reveal several patterns.  First, as in Lamoureux and Zhang (2024),
there appears to be a structural break around the turn of the century.  This pattern is most evident for the case of log utility in Figures 3a and 3b, and especially
concerning the $\theta$ coefficients on momentum, average same-month return, and residual volatility, all of which tend to zero in the 21st Century.  As risk aversion
increases (through Figures 4, 5 and 6), the trends are present but not as distinct since the characteristic tilts are less aggressive, and the posteriors are
comparatively wide (relative to the medians) throughout both subperiods.

Second, when the model worked well (for the 21 out-of-sample years ending 1979-2000) the coefficients on
momentum, the book-to-market ratio, and average same-month returns are significantly greater than 0 and decline monotonically in risk aversion.  Similarly, in this period
the coefficients on log-size and residual volatility are significantly less than 0 and increase monotonically in risk aversion.
We glean insights into the historical connections between characteristics and future returns by comparing optimal portfolio properties across the four utility functions.
Focusing on the 20th Century, when this link was efficacious, Figures 3-6 show that while characteristic-based portfolios afford higher Sharpe ratios than the market,
the relationship does not afford arbitrage.  The log utility function puts a weight tilt in the order of magnitude of 10 on momentum.  Since the average sample size in the
1960-1999 period is 1,605 stocks, a $\theta$ of 10 means that the posterior weight on a stock whose momentum is one standard deviation above the mean tilts 1.2\% higher than
a stock whose momentum is one standard deviation below the mean.  So for example, if both stocks have a market weights 0.07\%, the former stock's weight would be 1.32\% and
the latter's -1.18\%, ceteris paribus.  By contrast, for the more risk averse investor with $\gamma = 6$, the average weights in the 20th Century on these two
stocks would be 0.23\% and -0.09\%, respectively.

Figures 3-6 also provide insight into the structural changes that occur around the turn of the century.  Rolling through this period shows that the appeal of characteristic
shifts has largely vanished.  Most of the 95\%ile posterior bands on $\theta$ include zero when utility is maximized using years within the 21st Century.  Table 2 shows that
$\lambda^*$ provided an early warning signal that relationships between characteristics and future returns are changing around the year 2000--although primarily for the
more risk-tolerant utility functions.  For example, $\lambda^*$ for the log utility function is 7,500 in samples 1-21, that is the 20-year periods starting in 1960-1979 and
ending in 1980-1999.  In this case, $\lambda^*$ drops to 6,250 in the 20-year period 1981-2000, and drops further to 5,000 in the 1985-2004 period, and 4,000 in the
1996-2015 period.

These figures also show the role that the time variation in $\lambda^*$, as shown in Table 2, has on the posteriors.  Specifically, in the second subperiod $\lambda^*$ for the log
utility investor drops from 7,500 to as low as 4,000.  This maintains the scale of the posterior on $\theta$ to be fairly constant over time.  This is evident in the
height of the boxes and whiskers each year in all figures.  The boxes show the interquartile ranges and the whiskers the 95\% posterior bands.

\vskip .2in

\leftline{\em 4.3 Optimal portfolio return properties}

As a Bayesian econometrician I use $\lambda^*$, and the marginal Gibbs posterior conditional on $\lambda^*$
on $\theta$ from months $t-239$ through $t$ to construct the marginal Gibbs  posterior on the out-of-sample returns from the optimal portfolio policy in months $t+1, \ldots,
t+12$, conditional on the data from these months.  As a Bayesian decision maker, I use the mean of the posterior distribution of $\theta$ from the most recent year-end 
to construct my optimal portfolio each month.  Figures 7-14 and Table 3 provide characterizations of the posterior distribution of out-of-sample returns, and Table 3 
includes the decision-theoretic portfolio's properties.   Because weights are linear in $\theta$ and returns are linear in weights, the mean return from the decision-theoretic
portfolio equals the mean of the return posterior from the full marginal posterior on out-of-sample returns (seen in all panels of Table 3).
In Table 3, skewness is measured as difference between the posterior mean and the posterior median return scaled by the posterior standard deviation.  I measure posterior 
out-of-sample kurtosis using the Hogg coefficient:
\begin{eqnarray}
\text{Kurt}  = \left\{\frac{\overline{r}^{+}_{.95} - \overline{r}^{-}_{.05}}{\overline{r}^{+}_{.5} - \overline{r}^{-}_{.5}} - 2.63\right\} \cdot 100
\end{eqnarray}
where:
$\overline{r}^{+}_{.95}$ is the mean of the highest 5\% of returns, $\overline{r}^{-}_{.05}$ is the mean of the smallest 5\% of returns,
$\overline{r}^{+}_{.5}$ is the mean of the top half of returns, and $\overline{r}^{-}_{.5}$ is the mean of the bottom half of returns.

Table 3 Panel A shows that the 95\% marginal Gibbs posterior band on the certainty equivalent return for the log utility investor is $(5.8 \, , \, 6.3\%)$ per month, compared
to the two benchmarks (weighted by market capitalization or equally) constructed from the sample stocks of 1.3\% per month.  The decision-theoretic approach (which is
just a single path) has a monthly certainty equivalent of 6.1\% per month.  This gain in out-of-sample expected utility is achieved by using a lot of leverage to increase
the portfolio's expected return from a benchmark mean of 1.4\% to 7.4\% per month, and a weak shift in skewness (from negatively skewed to symmetric).  The variance and
kurtosis increase versus the benchmarks, but at a lower rate.  The annualized Sharpe ratio increases from 0.65 to 1.46 $(1.40 \, , \, 1.53)$ relative to the value-weighted
benchmark.  This is evident in the top panel of Figure 7, where the right flank and the right tail dominate the benchmark distribution, and the central tendency is 
noticeably to the right of the benchmark.  In this panel the dominance of the optimal portfolio's left tail is trivial relative to the right.  The lower panel in Figure 7 compares the
Gibbs posterior out-of-sample return distribution with that of the value-weighted benchmark of all sample securities in the second subperiod, 2001-2024.  Comparing this to the upper panel 
makes the PPP's deterioration in this period clear.

The upper panel in Figure 8 shows the relative likelihood of returns of the PPP and the equally-weighted benchmark in the first suboperiod.  In this subperiod, monthly returns
over the range -18\% - 10\% are more likely under the benchmark than under the PPP.  Monthly returns of at least 25\% are at least 10 times more likely under PPP than the benchmark.
Both of the Gibbs posterior PPP tails are thicker than the benchmark's in both subperiods.  In the first subperiod the upper tail dominates, whereas the tails are more symmetric in the
second subperiod.  The lower panel shows that in the second subperiod PPP does reduce the probability of small negative returns relative to the benchmark, but the tails in this subperiod
are still large and symmetric.

As in Lamoureux and Zhang (2024), the data show that all investors are better served by holding the value weighted portfolio in the 21st Century.
As noted above, we also see a change in the conditional distributions' tail behavior with the lack of monotonicity of $\lambda^*$ in $\gamma$ when data 
that includes the first decade of the 21st Century are used in updating the priors.
The kurtosis of the parametric portfolio posterior out-of-sample returns jumps dramatically from the first to the second subperiod.  Both the equally and value weighted
benchmarks' sample kurtosis are unchanged from subperiod 1 to subperiod 2.  However,  posterior kurtosis of the log utility increases from 0.35 $[0.26 \, , \, 0.44]$ to 1.20
$[0.98 \, , \, 1.43]$.  For the power utility function with $\gamma = 2$, the posterior out-of-sample kurtosis in the first and second subperiods are 
0.42 $[0.28 \, , \, 0.57]$ and 0.83 $[0.51 \, , \, 1.18]$.  These values for the power utility function with $\gamma = 3$ are 0.43 $[0.25 \, , \, 0.61]$ and 0.69 
$[0.38 \, , \, 1.03]$.  And the power utility function with $\gamma = 6$, posterior out-of-sample kurtosis is virtually constant across the two periods:
0.50 $[0.28 \, , \, 0.74]$ in the first subperiod and 0.50 $[0.21 \, , \, 0.83]$ in the second.  These results are
manifest by the symmetry in tail dominance relative to the benchmark in the lower panel of Figures 7-10.

The second subperiod is marked by large negative skewness of the value-weighted index.
The value weighted portfolio's median return over these 288 months is 1.37\%, 56 basis points higher than the mean.  More aggressive characteristic weighting reduces this
skewness.  In the second subperiod, the posterior mean out-of-sample return skewness of the four portfolios in increasing risk tolerance are: -.112, -.088, -.074, and -.04,
for log utility.  Whereas none of the optimal portfolios had significant skewness in the first subperiod, the 97.5\% posteriors of all of four of these skewness measures are 
negative in the second subperiod.  Characteristic-based investing symmetrized the optimal portfolios prior to 2000, but is unable to do so in the 21st Century..
As noted in the introduction, higher order moments also play an important role in the failure of PPP in the second subperiod.

Figures 9 and 10 show the distributions and density ratios on a log scale for the power utility investor with $\gamma = 2$.  The gains afforded by the PPP in the first subperiod are clear
in Figure 9's upper panel, being the dominant right flank, and the benchmark's dominant left flank.  Figure 10 shows positive monthly returns between 7\% and 30\% are more likely under the
benchmark than the PPP in the second subperiod, while the opposite was true in the first subperiod.  The lower panel also shows that large losses of 30\% or more are
10 times more likely under PPP than the benchmark for this investor.  Table 3 shows that the PPP distribution is significantly more leptokurtic than the benchmark in both subperiods. 
Further, it is significantly higher in the second subperiod than in the first subperiod, unlike the benchmark.  Coupling the significantly higher kurtosis with a significant left skew
explains dominance of the benchmark in the second subperiod.

We see similar patterns relative to the benchmark for the PPP of the power utility investor with $\gamma = 3$ in Figures 11 and 12 and Panel C of Table 3.  The significant drop in the
PPP's monthly certainty equivalent return from 4.0 $[3.7 \, , \, 4.2]$ in the first subperiod to 0.3 $[-0.1 \, , \, 0.7]$ in the second subperiod generates the drop in certainty
equivalent monthly return from 3.0 $[2.8 \, , \, 3.3]$ to 0.3 $[-0.1 \, , \, 0.7]$.  The lower panels in Figures 11 and 12 reveal the extreme leptokurtosis of the PPP relative to
the benchmark in this case.  And as with the more risk tolerant cases, PPP shifts from being symmetric in the first subperiod (effectively removing the benchmark's negative skew), to 
being significantly negatively skewed in the second subperiod, insignificantly different from the benchmark's skewness.

Table 3, Panel D and Figures 13 and 14 provide the Gibbs posterior out-of-sample return distributions in the two subperiods for the most risk-averse investor I consider.  In these two
figures the benchmark is the value-weighted portfolio of sample stocks in {\em both} subperiods.
In the first
subperiod the PPP's Sharpe ratio, 1.5 $[1.3 \, , \, 1.7]$ is significantly higher than the benchmark's 0.6, and tightly distributed around its mean.  
While the PPP's Gibbs posterior certainty equivalent, 1.8 $[0.8 \, , \, 2.2]$ is also significantly higher than the benchmark's 0.7,
it is negatively skewed with a large variance.  This
reflects the extreme left tail in the top panel of Figure 14 combined with this investor's strong aversion to large losses.    The lower panel in Figure 14 shows that left tail is 
no longer present in the second subperiod.  However, in this second subperiod positive returns of 18\% or more are 10 times more likely under the benchmark than the out-of-sample PPP.
Unlike the more risk tolerant investors, in this second subperiod, the decision-theoretic PPP has a higher certainty equivalent than the benchmark (0.23 vs. 0.20 basis points per month), 
however the Gibbs posterior mean certainty equivalent is lower (0.11).

\vskip .2 in

\leftline{\em 4.4 Optimal portfolios' factor exposures}

The four panels of Table 4 show the marginal posteriors on the optimal portfolios' Fama-French 6-factor exposures, for both subperiods.\footnote{I obtained the factor
data as well as the monthly riskfree rates from  Professor Kenneth French's web-based database, for which I am grateful.}
Linearity means that the regression estimates  from the decision-theoretic portfolio match the Gibbs posterior mean estimates, so the table provides the GMM $t-$statistics of
the regression coefficients for this portfolio constructed using the Gibbs posterior mean $\theta$ to define out-of-sample returns.
Consistent with the certainty
equivalent results (Table 3), in the 20th Century all four portfolios have large significant $\alpha$ values that decline monotonically in $\gamma$.   In the 21st Century all
Gibbs posterior mean $\alpha$ values are negative, but only for the case of log utility does the 95\%ile posterior band not include zero.

The large statistically significant $\alpha$ estimates have two sources.  First, two of the characteristics, month-of-the-year return and residual volatility, are not accounted for by the
six-factor model.  More importantly, the additive, affine nature of standard factor models may be inadequate to analyze a non-linear utility-optimized strategy.  The PPP framework 
exploits complementarities between characteristics, for example the joint distribution of momentum and size, etc.  The large $\alpha$ provides evidence that these cross-characteristic
synergies were large in the 20th Century.

Table 4 also reports the percentage of portfolio variances attributed directly to each factor---ignoring covariance effects.  More than half of the portfolio variance is outside the span
of the Fama-French factors for the log utility investor in both subperiods.  In the first subperiod this portfolio has large of exposures to the value and momentum factor, and
almost no exposure to the other four factors including the market factor.  I omit the covariance terms, which are material, explaining why the percentage explained sums to greater than one.

Market $\beta$ is flat in $\gamma$ in both subperiods, even as the market portfolio's role in portfolio variance increases monotonically in $\gamma$.
As $\gamma$ increases and leverage decreases the exposure to the market factor increases, so that in
the first subperiod the market factor accounts for 35\% of the variance of the optimal portfolio for the power utility investor with $\gamma = 6$.  Exposure to the value factor
declines significantly in the second subperiod for all four portfolios, as expected in light of the pattern of $\theta$ coefficients on the book-to-market ratio (Figures 3a, 4a,
5a, and 6a).

All four utility functions' posterior out-of-sample portfolio returns load significantly on RMW--the operating profitability factor--in the second subperiod, which is 
interesting as operating profitability is not obviously linked to any of the characteristics in $X$.  Although perhaps this reflects the changing industrial structure of firms as
large stocks tend to be information technology companies with low book-to-market values in the 21st Century.  By contrast, only the most risk averse utility function's
portfolio loads significantly on this factor in the first subperiod, with a mean posterior RMW beta of 0.5.

\vskip .2 in


\leftline{{\bf 5. Conclusions}}

\vskip .2 in

In this paper I show how to apply the tools of Bayesian estimation and decision making to Brandt, Santa-Clara, and Valkanov's (2009) parametric portfolio policy---with no
restrictions on the data generating process.
The idea of optimizing a policy function directly on the data, without specifying a likelihood function, has tremendous appeal.  
As Brandt, Santa-Clara, and Valkanov (2009) and A\"it-Sahalia and Brandt (2001) highlight,
doing so can streamline estimation and reduce the risks from model misspecification.  My Gibbs posterior  has several advantages over approaches that rely on traditional machine learning
tools such as the bootstrap.  From a statistical perspective I avoid assumptions about the data generating process that a bootstrap requires.  From a decision theoretic
perspective I have a coherent approach that updates my beliefs in light of new data.  I conform to the likelihood principle in that only sample data characterize the
decision problem.  My priors are placed on the parameters that link characteristics to portfolio weights--not the parameters of a model of the return generating process.
From a practical perspective, I regularize (mitigate the effects of overfitting)
in-sample, using the geometry of the Gibbs posterior covariance matrix.  This facilitates decision making, as less data is needed to estimate and regularize.  It also enables 
more powerful inference as more data is available at the testing stage.

I show that this in-sample regularization works and I link the Gibbs posterior weighting parameter to the curvature of the loss function both theoretically and empirically.
In the mean-variance case, the product of the optimal weight on the data  and the risk aversion coefficient is constant across investors.  Higher order moment structure of the Hessian
breaks this relationship in the power utility case.
I obtain the exact Gibbs posterior for optimal portfolio weights under the utility function.  I use these to construct the predictive distribution of {\em out-of-sample}
portfolio returns.  Thus I have a fully coherent uncertainty distribution without a likelihood.
The paper's substantive findings rely solely on the observed data, and confirm and update Lamoureux and Zhang (2024) that the traditional characteristics lose predictive
efficacy around the turn of this century.
A likely explanation for this is that the strategies were largely
infeasible given the technological and regulatory constraints of that era.  Since the year 2000, not only has information about the usefulness of characteristic-based investing
been widely disseminated (McLean and Pontiff 2016) and investors have learned about the data generating process (Martin and Nagel 2022; Nagel 2021), 
transactions costs have dropped dramatically and information processing speeds have increased.
A final note on the effectiveness of the in-sample regularization introduced in this paper, 
Sharpe ratios and certainty equivalents are significantly lower than the benchmarks in the second subperiod only for the log utility function.  
The heightened regularization helped to mitigate the drop in efficacy for the more risk averse utility functions relative to Lamoureux and Zhang's (2024) machine learning approach.

\newpage

\leftline{{\bf Appendix: Chain diagnostics}}
        
\vskip .2 in

I build all Gibbs posteriors using a Metropolis within Gibbs sampling.  I use a burn-in sample of 200,000 draws for all cases.  I draw the $\theta$ parameters individually
using a symmetric stable Paretian density as the proposal density.  This stable densities
have characteristic exponent $\alpha$ of 1.75, so that their means exist.  The reason for this leptokurtic distribution is to ensure tail dominance of the proposal.
I draw $\theta_j$ conditional
on the (in-sample) data and the other five $\theta$ values at this step.  The proposal mean for $\theta_j$  is the current draw.
I have a Markovian design so the scale of the proposal is fixed through the chain.  Higher values of this scale lead to a lower acceptance rate.  Optimal acceptance rates
for this one parameter at a time design are in the 35 - 55\% range.  All chains have either 100,000, 200,000, 300,000, 400,000, or 600,000  post burn-in draws.
All cases under $\lambda^*$ use at least 300,000 post burn-in draws.
       
In this appendix
I explore the convergence properties of the algorithm using the multivariate extension of the Gelman and Rubin (1992) statistic developed in Brooks
and Gelman (1998).  This Multivariate Potential Scale Reduction Factor (MPSRF) statistic evaluates the extent of homogeneity across multiple independent
chains to characterize the hypothetical reduction in uncertainty due to nonconvergence.
I have m independent chains each with n Metropolis draws on the hyperparameter $K-$vector $\theta$ ($K = 6$).  
For chains of unequal length, I truncate the longer chains and use the
first $n$ draws.  For each chain $i$, $i = 1, \ldots , m$, the Gibbs posterior mean $\theta$ is $\overline{\theta_i}$ and its covariance matrix
is $S_i$.  Then $W = \frac{1}{m} \sum_{i=1}^{m} S_i$ and $\overline{\theta} = \frac{1}{m} \sum_{i=1}^{m} \overline{\theta}_i$.  Define:
\begin{eqnarray*}
B = \frac{n}{m-1} \sum_{i=1}^m \left(\overline{\theta}_i - \overline{\theta} \right) \left(\overline{\theta}_i - \overline{\theta} \right)^{'}
\end{eqnarray*}
The MPSRF is
\begin{eqnarray*}
\widehat{R}_{\mathrm{multi}}
=
\sqrt{
\frac{n-1}{n}
+
\frac{m+1}{mn}\,
\lambda_{\max}\!\left(W^{-1}B\right)
}
\end{eqnarray*}
where $\lambda_{\max}(\cdot)$ is the largest eigenvalue of the generalized eigenproblem.

Table A-1 reports the MPSRF for several representative cases.  The largest value in Panel A, for log utility is 1.00047, indicating that any residual lack of convergence could
inflate posterior uncertainty by less than  .05\% in the worst linear combination of parameters.  The largest value across all periods and the four utility functions is 1.00058
from the case where the in-sample period comprises Months 1 - 240, and the utility function is power utility with coefficient of relative risk aversion equal to 3 (Panel C).
This table provides strong evidence that the chains have converged optimally.

I present additional properties of the posterior chains in Table A-2.  For the seven representative periods the table shows the Metropolis acceptance rate for each parameter
as well as the effective sample size.  I measure the effective sample size, $N_{\mathrm{eff}} = \frac{N}{1 + \sum\limits_{i=1}^P \rho_i}$ where $N$ is the number of Metropolis draws,
$\rho_i$ is the $i^{\mathrm{th}}-$order autocorrelation of the Metropolis draws, and $P+1$ is the order of the first non-positive autocorrelation.  We approximate the numerical
standard error of the Metropolis chain as $\frac{1}{\sqrt{N_{\mathrm{eff}}}}$.  The only cases where the effective sample sizes are less than 10,000 are in Periods 22
(in-sample period: 1981-2000) and 24 (in-sample period: 1983-2002) in the two most risk-averse cases ($\gamma = 3$ and $\gamma = 6$), for
$\theta_4$ (weight coefficient on $\beta$) and $\theta_6$ (weight coefficient on residual volatility).
This occurs as the posterior correlation between these two parameters is high in absolute value in these cases.  More importantly, the smallest effective sample size for
the posterior draws on the in-sample certainty equivalent return is 17,042, in the case of the most risk-averse investor in Period 22.  This reflects a numerical standard
error of 0.77\%.

Most of the 15 correlations in the $\theta$ posterior distribution are less than 0.2 in absolute value.  The strongest correlations are between the $\theta$
coefficients on $\beta$, $\theta_4$, and residual volatility, $\theta_6$.  I report these correlations in Table A-3 Panel A for the 7 representative periods, for all four utility functions.
In all cases the correlations are negative.  They are highest in absolute value in the second regime, in which the model fares poorly out-of-sample.  They also tend to
increase in absolute value in risk-aversion.  This helps explain the lowered Metropolis efficiency shown in Table A-2 for these two parameters in periods 22, 24, and 34.
These correlations fall to the lowest (absolute value) levels in the first regime in the last few 20-year periods in the sample.  If the two characteristics act in similar ways,
then the sum of these coefficients is potentially better identified and more important for utility maximization than the individual values.  This is manifest by the fact that
the effective sample size of the in-sample certainty equivalent return is always much larger than the effective sample sizes of $\theta_4$ and $\theta_6$.

Two other $\theta$ pairs have some posterior correlations that exceed 20\% in absolute value.  Panel B reports the posterior correlations between the $\theta$ coefficients on
momentum and the book-to-market ratio.  Panel C provides the posterior correlations between the $\theta$ coefficients on log size and $\beta$.  The largest of these are
in the 240 month period 1993-2012 for the more risk averse investors.  It is 61\% in this period for the most risk-averse utility function considered.  Following the momentum
crash of 2009, the investor will only put money into momentum stocks if they are also value stocks.  Alternatively the only value stocks that are attractive in this period
are also momentum stocks.  The complementarity between log size and $\beta$ is only material in the early periods, when the algorithm is very successful.  It is also flat
across the four levels of risk aversion.  Big stocks are less undesirable if they have high betas.

\newpage

\hoffset -.4in

\begin{center}
{\bf References}
\end{center}

\parindent -22pt
\parskip .1in


A\"{i}t-Sahalia, Yacine and Michael W. Brandt, 2001, Variable selection for portfolio choice, {\em Journal of
    Finance} 56, 1297--1351.

  




Berger, James O. and Robert L. Wolpert, 1988, {\em The Likelihood Principle}, 2nd edition, Hayward, Cal;
       Institute of Mathematical Statistics, Lecture notes monograph.





Birnbaum, Allan, 1962, On the foundations of statistical inference, {\em Journal of the American Statistical Association}
      57, 269--306.

Bissiri, P.G., C.C. Holmes, and S.G. Walker, 2016, A general framework for updating belief distributions,
      {\em Journal of the Royal Statistical Society, Series B} 78, 1103--1130.


Brooks, Stephen and Andrew Gelman, 1998, General methods for monitoring convergence of iterative simulations, {\em Journal of
      Computational and Graphical Statistics} 7, 434--455.


Chernozhukov, Victor and Han Hong, 2003, An MCMC approach to classical estimation, {\em Journal of Econometrics} 115, 293--346.


DeMiguel, Victor, Lorenzo Garlappi, Francisco J. Nogales, and Raman Uppal, 2009, A generalized approach to portfolio optimization:
      Improving performance by constraining portfolio norms, {\em Management Science} 55, 798--812.

DeMiguel, Victor, Alberto Martín-Utrera, Francisco J. Nogales, and Raman Uppal, 2020, A transaction-cost perspective on the
       multitude of firm characteristics, {\em Review of Financial Studies} 33, 2180--2222.

Garlappi, Lorenzo, Raman Uppal, and Tan Wang, 2007, Portfolio selection with parameter and model uncertainty: A multi-prior
       approach, {\em Review of Financial Studies} 20, 41--81.

Gelman, Andrew and Donald Rubin, 1992, Inference from iterative simulation using multiple sequences, {\em Statistical Science}
        7, 457--511.

Gilboa, Itzhak and David Schmeidler, 1989, Maxmin expected utility with non-unique prior, {\em Journal of Mathematical Economics},
        18, 141--153.


Hansen, Lars P. and Thomas J. Sargent, 2001, Robust control and model uncertainty, {\em American Economic Review} 91, 60--66.

Hansen, Lars P. and Thomas J. Sargent, 2008, {\em Robustness}, Princeton University Press.

\newpage

Jagannathan, Ravi and Tongshu Ma, 2003, Risk reduction in large portfolios: Why imposing the wrong constraints helps, 
        {\em Journal of Finance} 63, 1651--1683.

Kelly, Bryan, Seth Pruitt, and Yinan Su, 2019, Characteristics are covariances: A unified model of risk and return, 
       {\em Journal of Financial Economics} 134, 501--524.




Knoblauch, Jeremias, Jack Jewson, and Theodoras Damoulas, 2022, An optimization-centric view on Bayes' rule: Reviewing and
    generalizing variational inference, {\em Journal of Machine Learning Research} 23, 1--109.

Kozak, Serhiy, Stefan Nagel, and Shrihari Santosh, 2018, Interpreting factor models, {\em The Journal of Finance} 73,
       1183--1223.


Kozak, Serhiy, Stefan Nagel, and Shrihari Santosh, 2020, Shrinking the cross-section,, {\em Journal of Financial Economics} 135,
       271--292.

Lamoureux, Christopher and Huacheng Zhang, 2024, An empirical assessment of characteristics and optimal portfolios,
     {\em Review of Asset Pricing Studies} 14, 450-480.

Lazar, Nicole A., 2003, Bayesian empirical likelihood, {\em Biometrika} 90, 319--326.




Liu, Jianan, Robert F. Stambaugh, and Yu Yuan, 2018, Absolving beta of volatility's effects, {\em Journal
     of Financial Economics} 128, 1--15.





Maenhout, Pascal J., 2004, Robust portfolio rules and asset pricing, {\em Review of Financial Studies} 17, 951--983.

Martin, Ian and Stefan Nagel, 2022, Market efficiency in the age of big data, {\em Journal of Financial Economics} 
       145, 154--177.

Martin, Ryan and Nicholas Syring, 2022, Direct Gibbs posterior inference on risk minimizers: Construction, concentration and
       calibration, {\em Handbook of Statistics v. 47}, Cambridge, MA, Elsevier, 1--41.

McAlinn, Kenichiro and K\=osaku Takanashi, 2026, When is generalized Bayes Bayesian?  A decision-theoretic characterization of
       loss-based updating, Working Paper, ArXiv:2602.01573v1.

McLean, R. David and Jeffrey Pontiff, 2016, Does academic research destroy stock return
    predictability?  {\em Journal of Finance} 71, 5--31.

Nagel, Stefan, 2021, {\em Machine Learning in Asset Pricing}, Princeton University Press.


\newpage

Nigmatullin, Eldar, 2003, Estimation of Markov decision processes in the presence of model uncertainty,
      Doctoral thesis, University of Wisconsin-Madison.












Roberts, Gareth O. and Jeffrey S. Rosenthal, 2001, Optimal scaling for various Metropolis-Hastings algorithms,
    {\em Statistical Science} 16, 351--367.


Satop\"{a}\"{a}, Ville, Jeannie Albrecht, David Irwin, and Barath Raghavan, 2011, Finding a "Kneedle" in a
      haystack: Detecting knee points in sytem behavior, IEEE Computer Scoiety: Proceedings of the 31st
      International Conference on Distributed Computing Systems Workshops, 166-171.


Schennach, Susanne, 2005, Bayesian exponentially tilted empirical likelihood, {\em Biometrika} 92, 31--46.


Syring, Nicholas and Ryan Martin, 2023, Gibbs posterior concentration rates under sub-exponential type losses,
        {\em Bernoulli} 29,  1080--1108.






Zhang, Tong, 2006a, From $\epsilon$-entropy to KL-entropy: Analysis of minimum information complexity density estimation,
       {\em The Annals of Statistics} 34, 2180--2210.

Zhang, Tong, 2006b, Information-theoretic upper and lower bounds for statistical estimation, {\em IEEE Transactions on Information
       Theory} 52, 1307--1321.

\newpage

\textheight 9.75in

\pagestyle{empty}

\begin{landscape}

\setlength{\intextsep}{0pt plus 2pt}
\setlength{\abovecaptionskip}{8pt}

\flushbottom

\captionsetup{width=\textwidth}

\restylefloat{figure}

\voffset=-1.35in
\hoffset=-0.75in

\long\def\symbolfootnote[#1]#2{\begingroup%
\def\thefootnote{\fnsymbol{footnote}}\footnote[#1]{#2}\endgroup}

\begin{figure}[H]
\includegraphics*[scale=0.533]{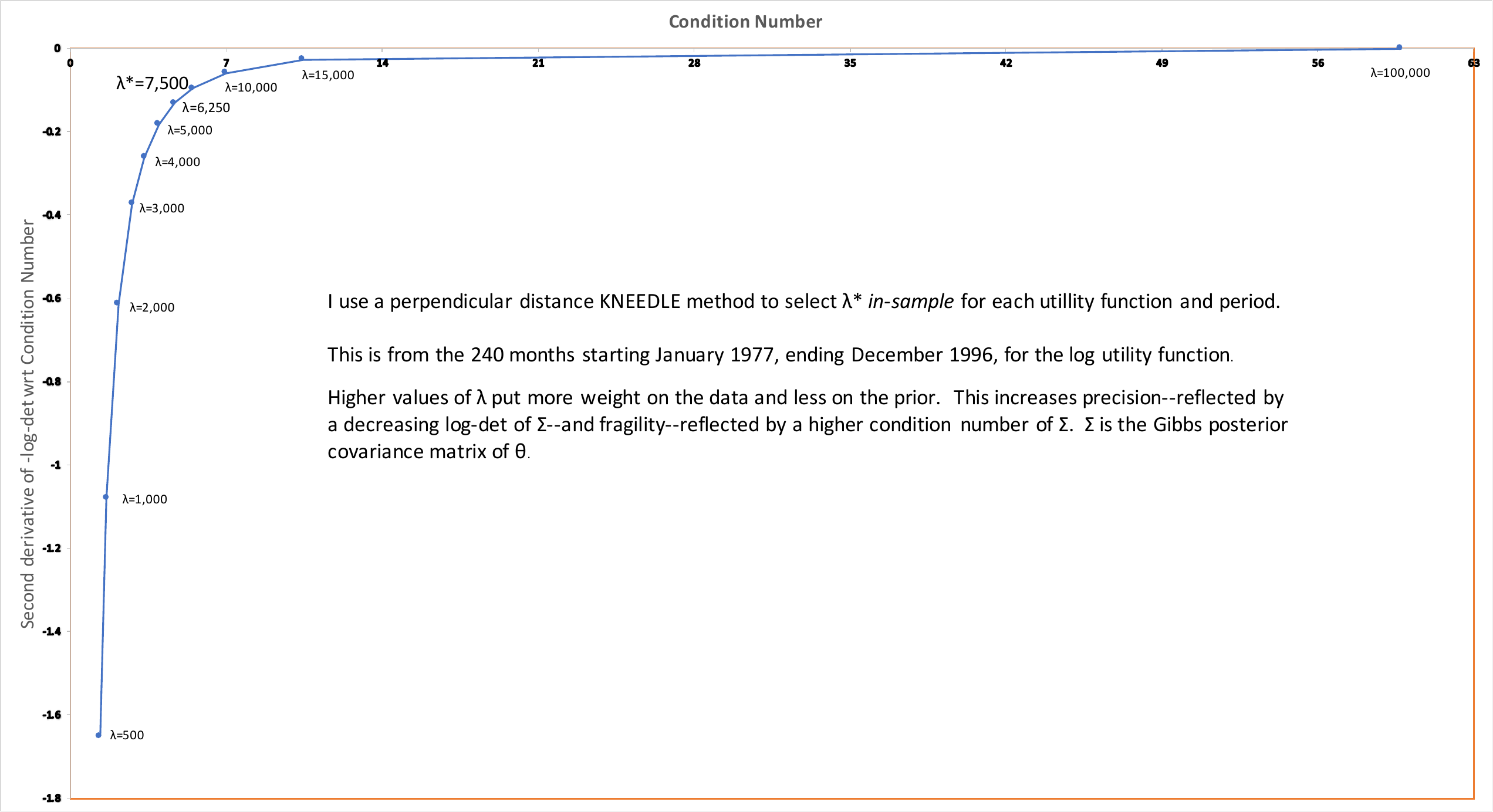}

\vspace{1.8\baselineskip}
\caption{{\bf Figure 1.  Identification Frontier}
$\Sigma$ is the posterior covariance matrix.  I consider the -log~det~$\Sigma$ a function of $\Sigma$'s condition number, $\kappa$.  I project
-log~det~$\Sigma$ onto log~$\kappa$ to obtain the slope $m$.  I construct the second derivative with respect to $\kappa$ as $\tfrac{-m}{\kappa^2}$.  This
information deceleration is the vertical axis in this figure. I use a KNEEDLE method to identify $\lambda^*$.
This is for the case of the 240 months ending in December 1996, and log utility. }

\end{figure}

\newpage

\hoffset -1in

\begin{figure}[H]
\includegraphics*[scale=0.85]{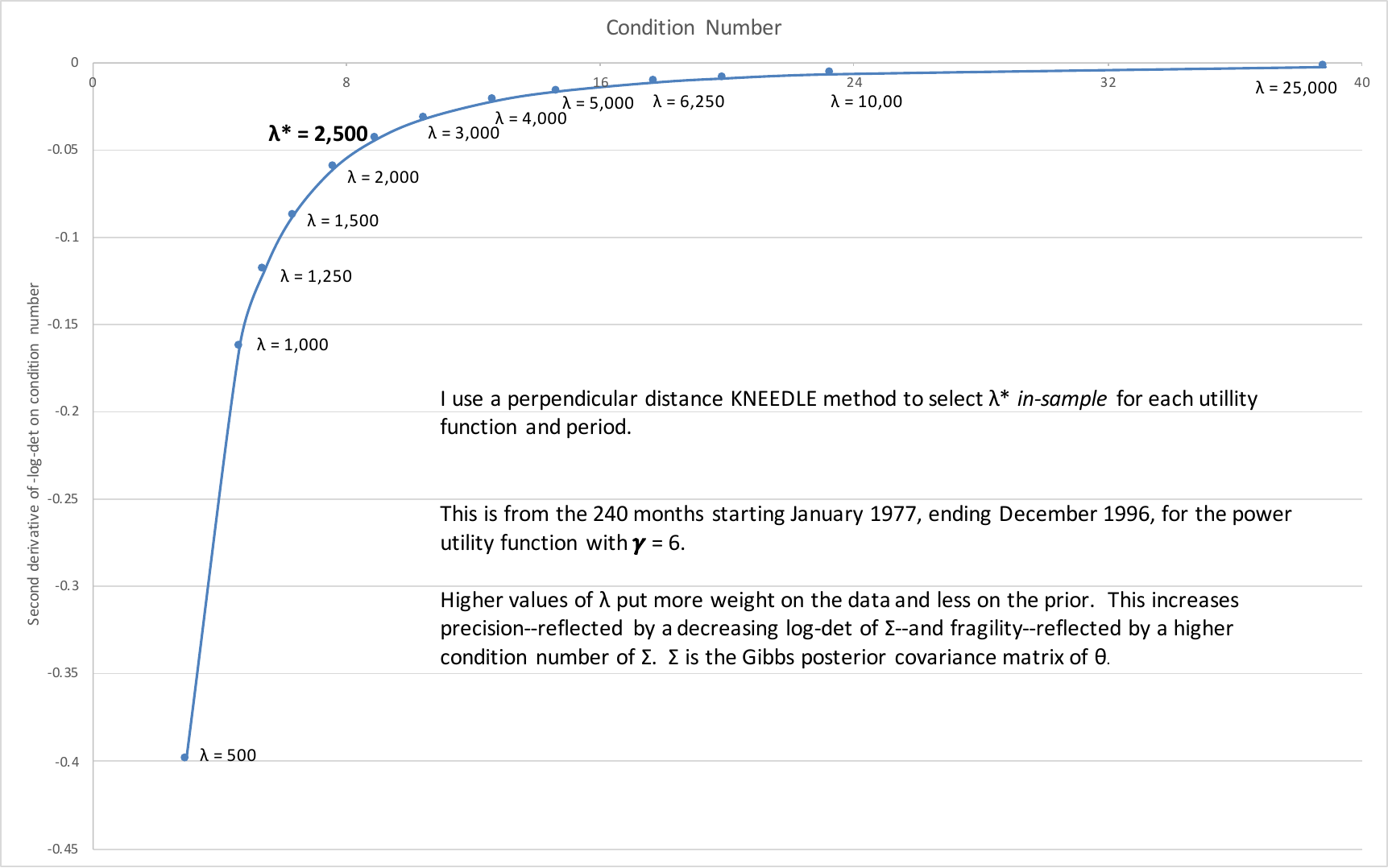}

\vspace{.8\baselineskip}
\caption{{\bf Figure 2.   Identification Frontier}
$\Sigma$ is the posterior covariance matrix.  I consider the -log~det~$\Sigma$ a function of $\Sigma$'s condition number, $\kappa$.  I project
-log~det~$\Sigma$ onto log $\kappa$ to obtain the slope $m$.  I construct the second derivative with respect to $\kappa$ as $\tfrac{-m}{\kappa^2}$.  This
information deceleration is the vertical axis in this figure.  I use a KNEEDLE method to identify $\lambda^*$.
This is for the case of the 240 months ending in December 1996, and a power utility function with coefficient of relative risk aversion, $\gamma$, = 6.}

\end{figure}

\end{landscape}

\newpage

\hoffset -1.25in

\begin{figure}[H]
\includegraphics*[scale=1.07]{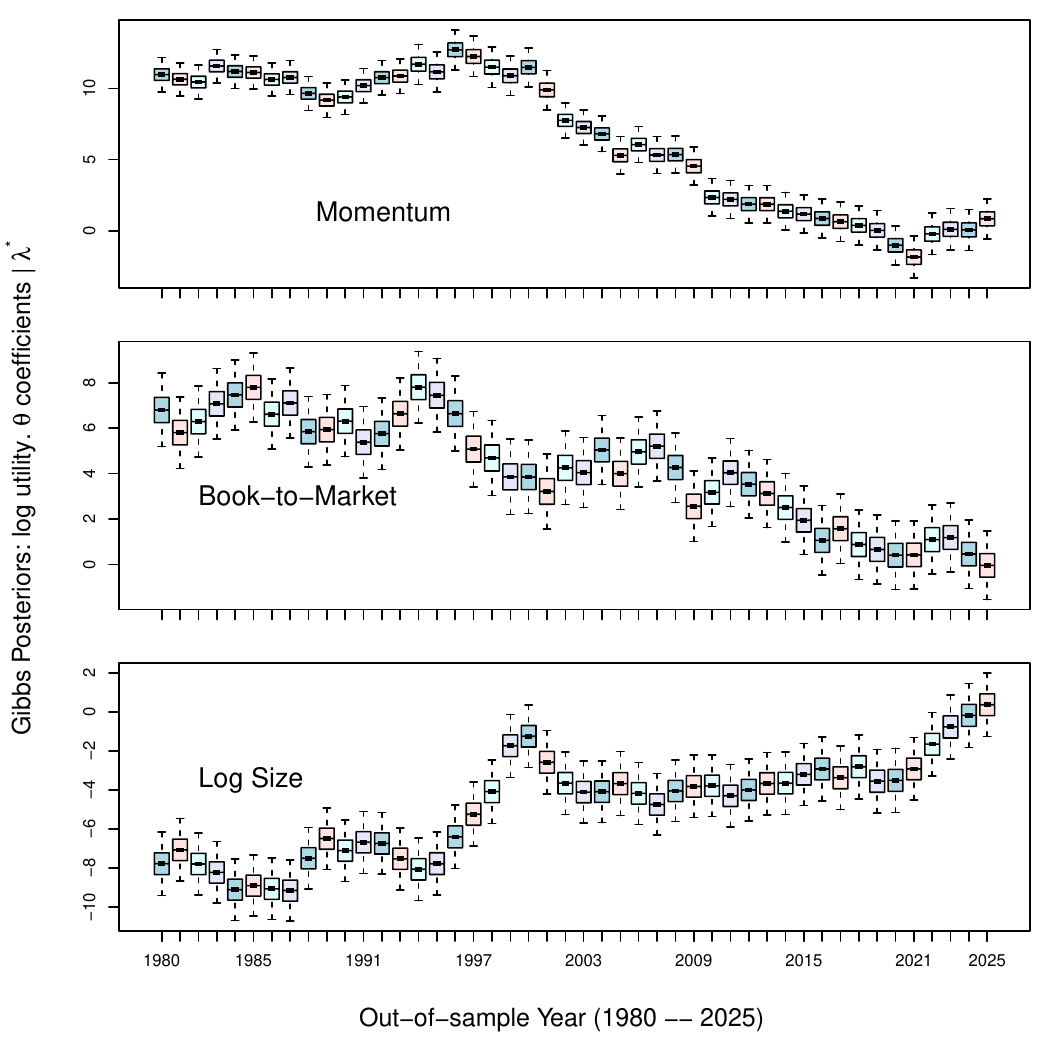}
\vspace{2.8\baselineskip}
\caption{{\bf Figure 3a.  Gibbs posterior on $\theta$, by year.}
Gibbs posterior of the $\theta$ coefficients for the log utility function.  Posteriors are constructed conditional on the preceding 240 months of data, and $\lambda^*$.
The "whiskers" show the 2.5\%ile - 97.5\%ile Gibbs posterior bands.  The "box" shows the Gibbs posterior interquartile range, and the bar inside the box is the Gibbs posterior
median.}

\end{figure}

\newpage

\begin{figure}[H]
\includegraphics*[scale=1.07]{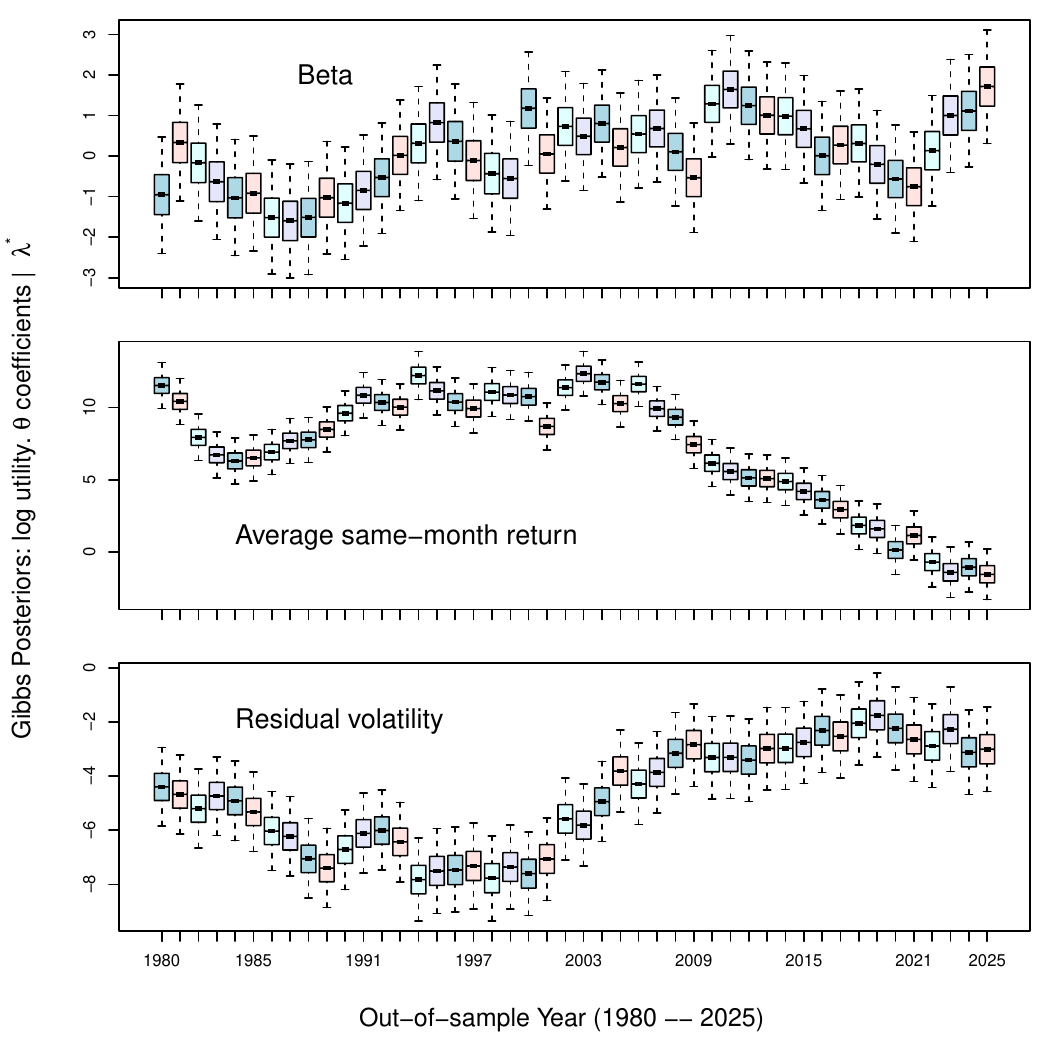}
\vspace{1.8\baselineskip}
\caption{{\bf Figure 3b.  Gibbs posterior on $\theta$, by year.}
Gibbs posterior of the $\theta$ coefficients for the log utility function.  Posteriors are constructed conditional on the preceding 240 months of data, and $\lambda^*$.
The "whiskers" show the 2.5\%ile - 97.5\%ile Gibbs posterior bands.  The "box" shows the Gibbs posterior interquartile range, and the bar inside the box is the Gibbs posterior
median.}

\end{figure}

\newpage

\begin{figure}[H]
\includegraphics*[scale=1.07]{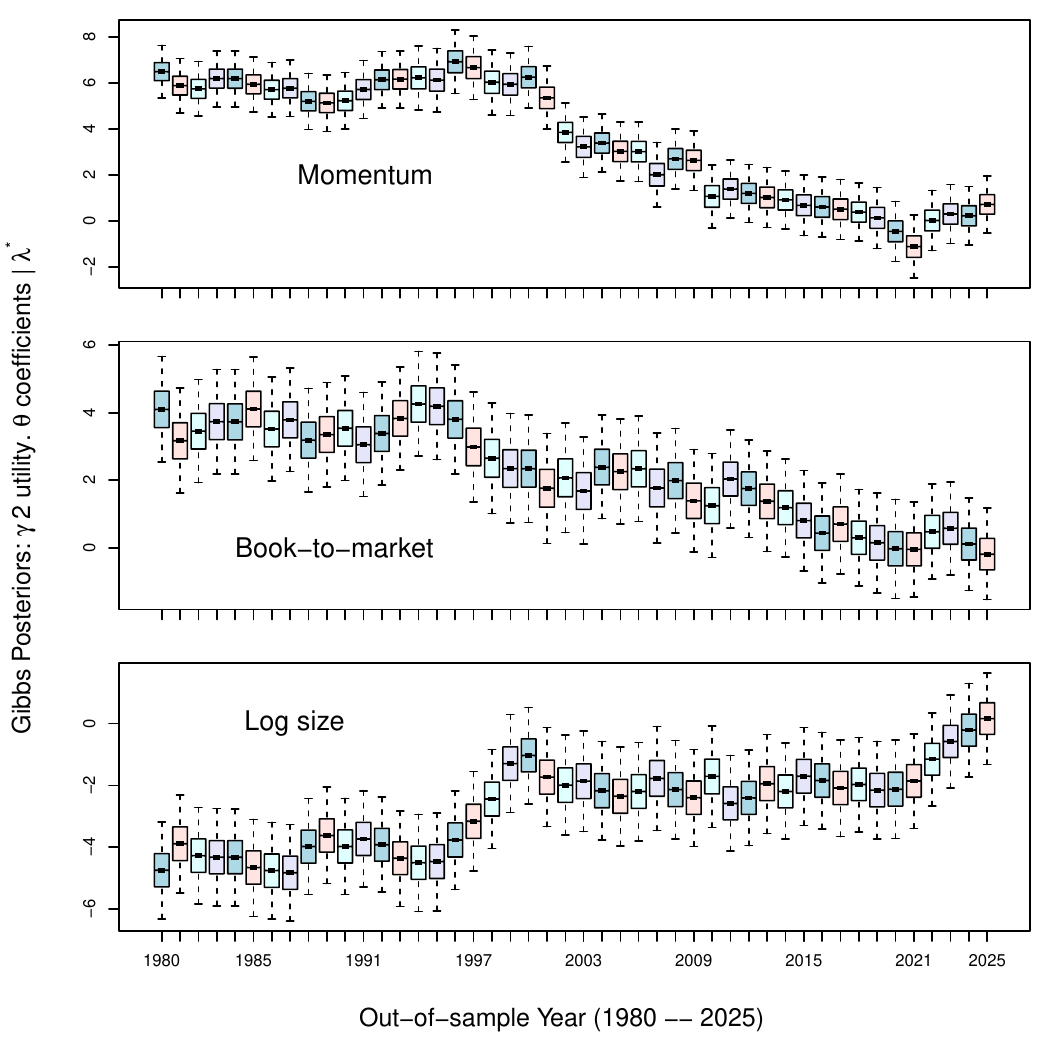}
\vspace{1.8\baselineskip}
\caption{{\bf Figure 4a.  Gibbs posterior on $\theta$, by year.}
Gibbs posterior of the $\theta$ coefficients for the power utility function with $\gamma = 2$.  Posteriors are constructed conditional on the preceding 240 months of data, and $\lambda^*$.
The "whiskers" show the 2.5\%ile - 97.5\%ile Gibbs posterior bands.  The "box" shows the Gibbs posterior interquartile range, and the bar inside the box is the Gibbs posterior
median.}

\vspace{1.8\baselineskip}
\end{figure}

\newpage

\begin{figure}[H]
\includegraphics*[scale=1.07]{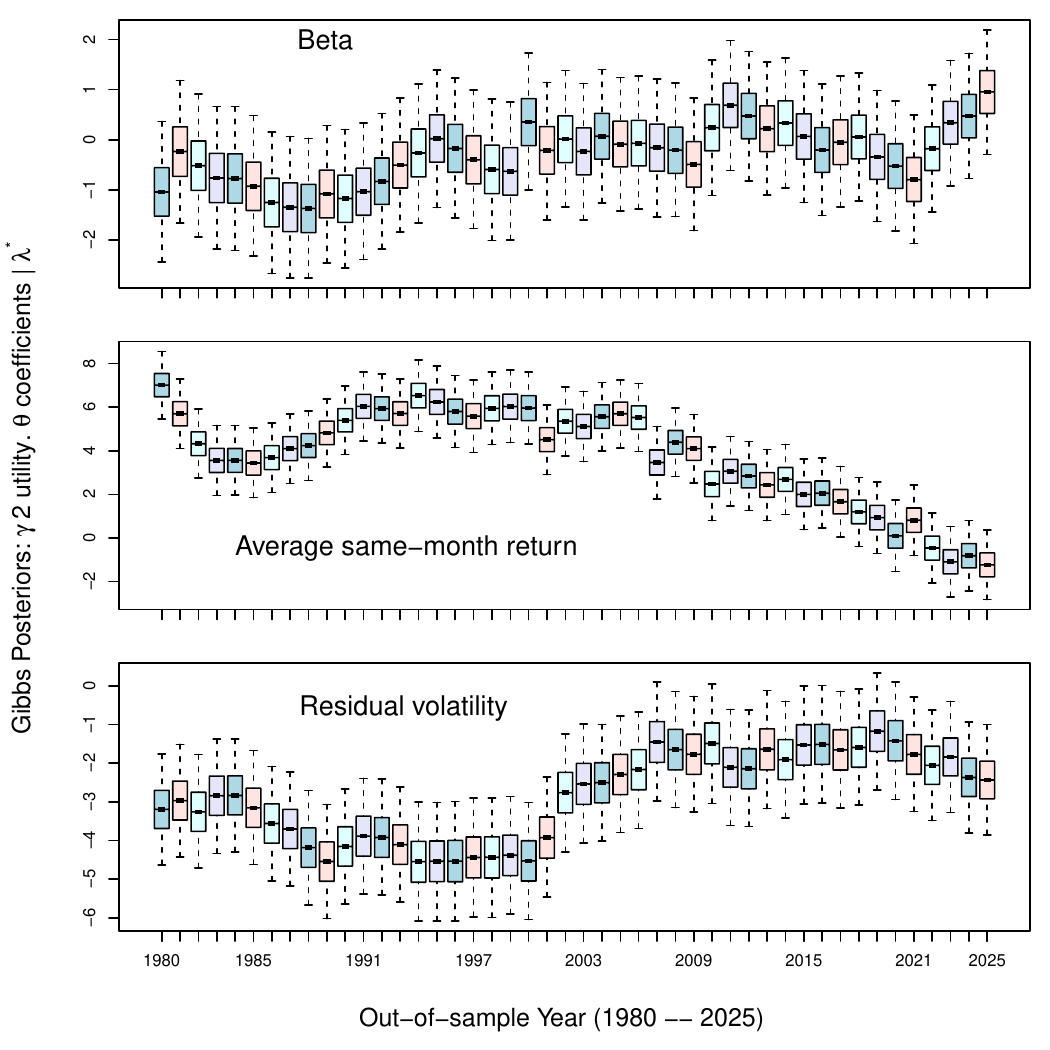}
\vspace{1.8\baselineskip}
\caption{{\bf Figure 4b.  Gibbs posterior on $\theta$, by year.}
Gibbs posterior of the $\theta$ coefficients for the power utility function with $\gamma = 2$.  Posteriors are constructed conditional on the preceding 240 months of data, and $\lambda^*$.
The "whiskers" show the 2.5\%ile - 97.5\%ile Gibbs posterior bands.  The "box" shows the Gibbs posterior interquartile range, and the bar inside the box is the Gibbs posterior
median.}

\end{figure}

\newpage

\begin{figure}[H]
\includegraphics*[scale=1.08]{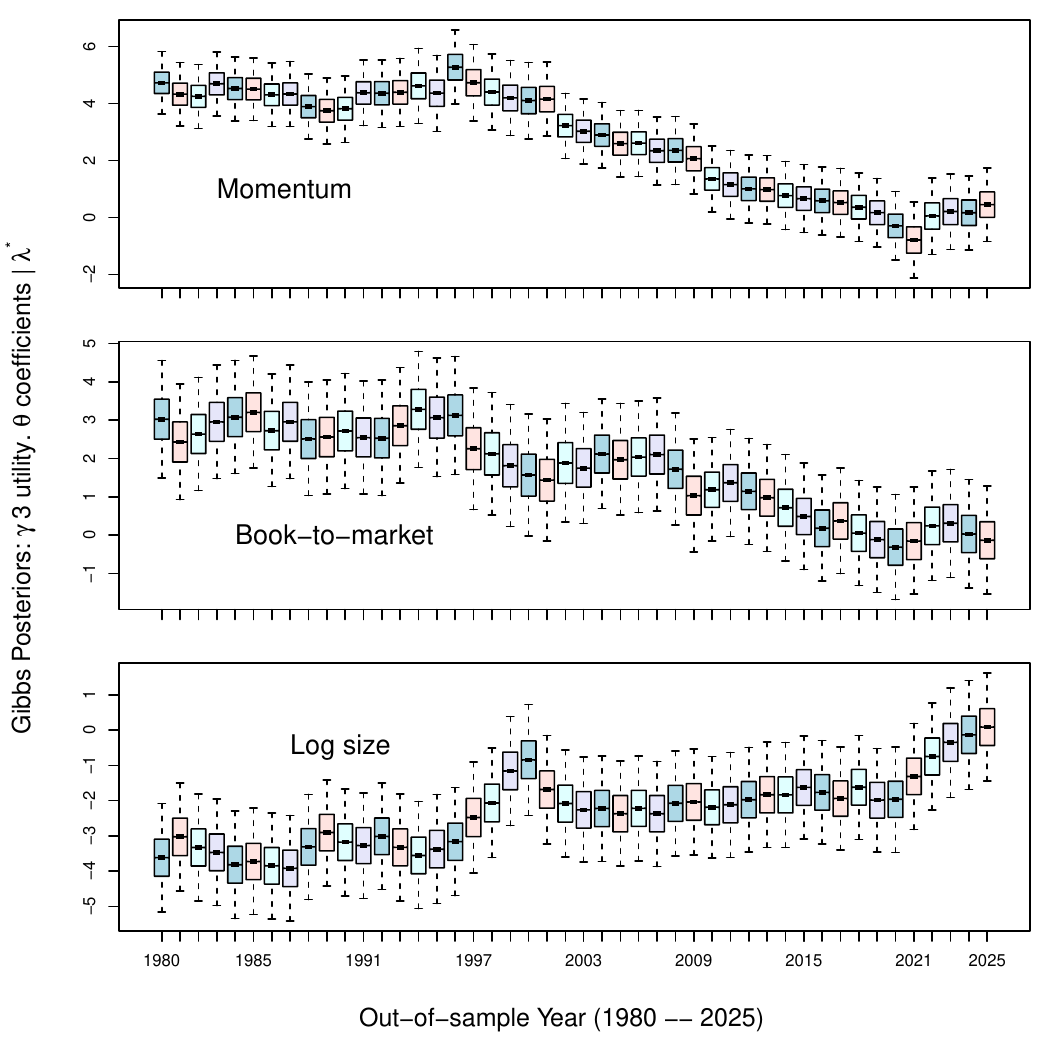}
\vspace{1.8\baselineskip}
\caption{{\bf Figure 5a.  Gibbs posterior on $\theta$, by year.}
Gibbs posterior of the $\theta$ coefficients for the power utility function with $\gamma = 3$.  Posteriors are constructed conditional on the preceding 240 months of data, and $\lambda^*$.
The "whiskers" show the 2.5\%ile - 97.5\%ile Gibbs posterior bands.  The "box" shows the Gibbs posterior interquartile range, and the bar inside the box is the Gibbs posterior
median.}

\end{figure}

\newpage

\begin{figure}[H]
\includegraphics*[scale=1.08]{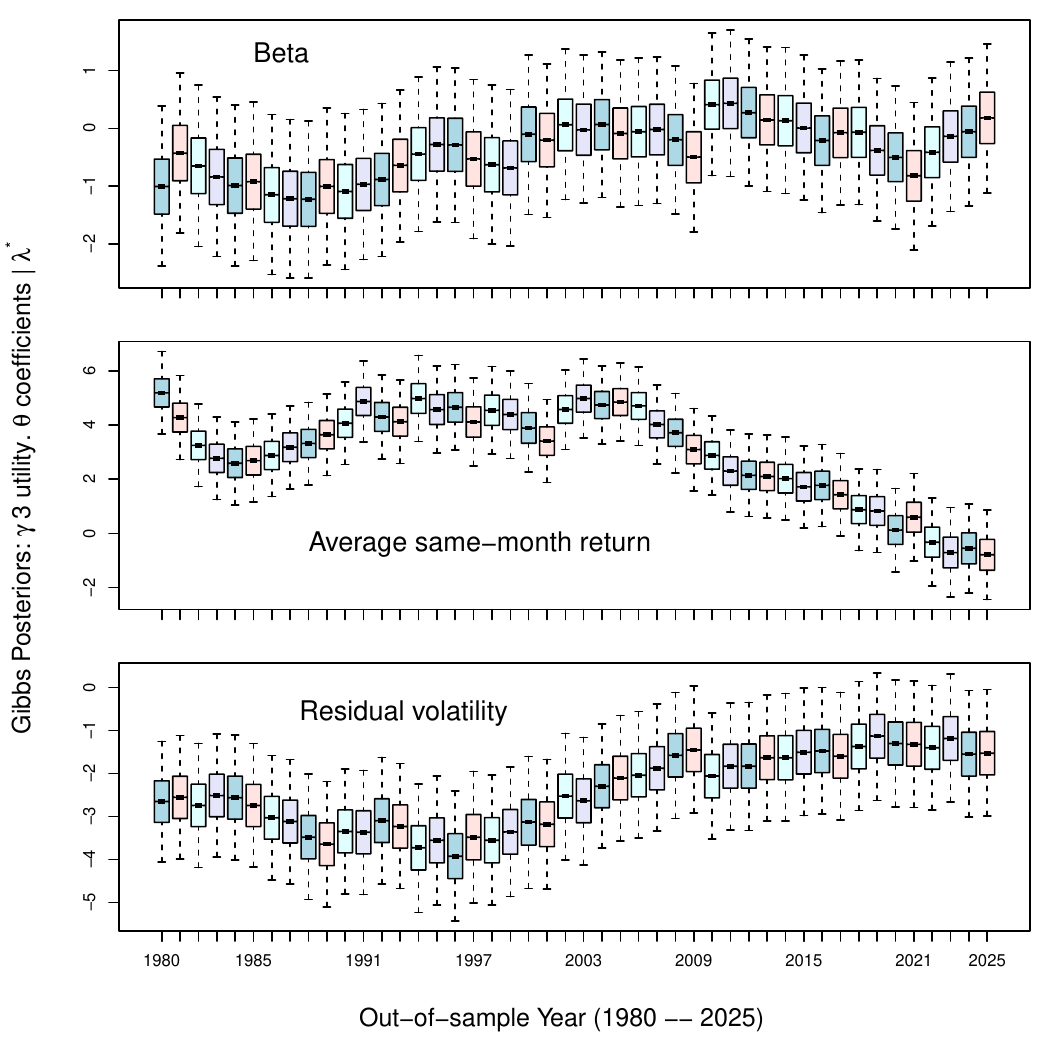}
\vspace{1.8\baselineskip}
\caption{{\bf Figure 5b.  Gibbs posterior on $\theta$, by year.}
Gibbs posterior of the $\theta$ coefficients for the power utility function with $\gamma = 3$.  Posteriors are constructed conditional on the preceding 240 months of data, and $\lambda^*$.
The "whiskers" show the 2.5\%ile - 97.5\%ile Gibbs posterior bands.  The "box" shows the Gibbs posterior interquartile range, and the bar inside the box is the Gibbs posterior
median.}

\end{figure}

\newpage

\begin{figure}[H]
\includegraphics*[scale=1.08]{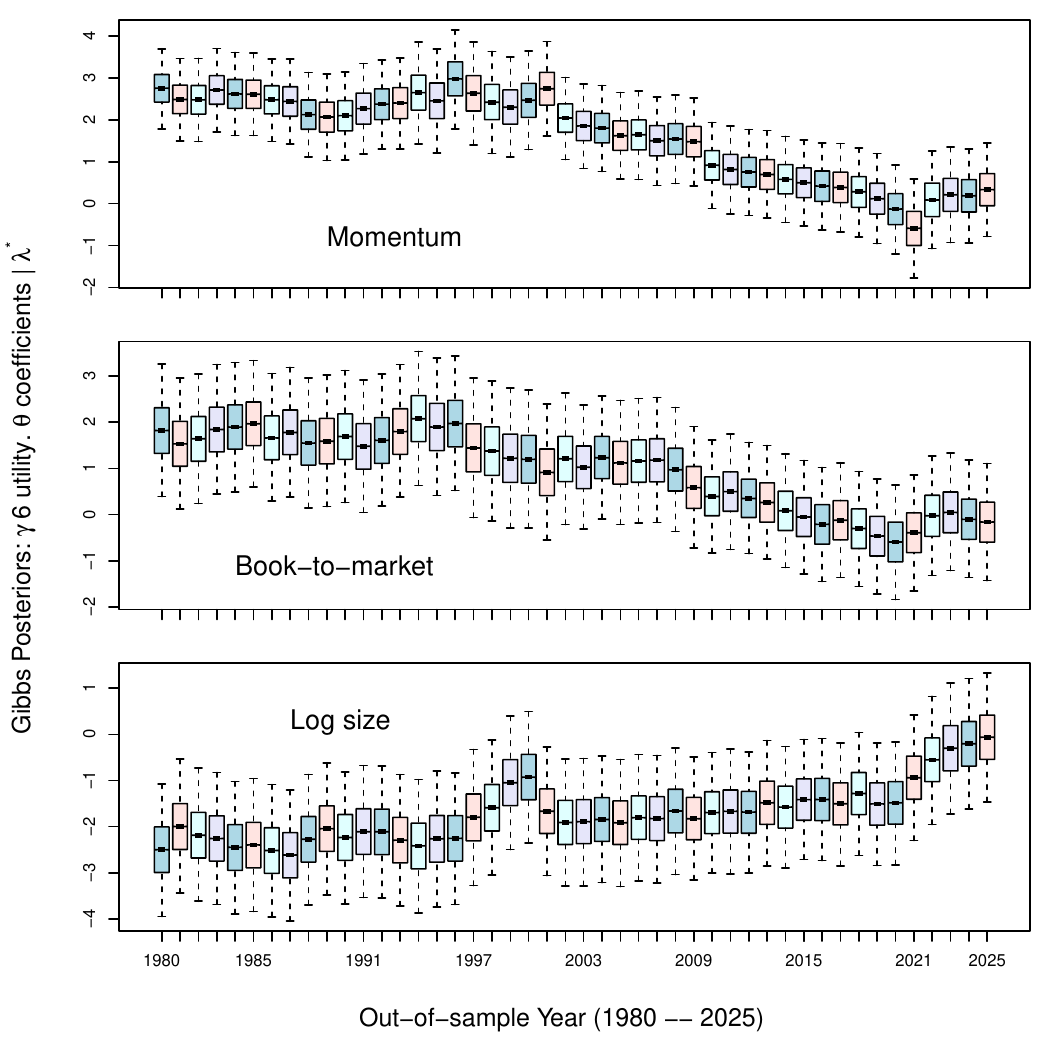}
\vspace{1.8\baselineskip}
\caption{{\bf Figure 6a.  Gibbs posterior on $\theta$, by year.}
Gibbs posterior of the $\theta$ coefficients for the power utility function with $\gamma = 6$.  Posteriors are constructed conditional on the preceding 240 months of data, and $\lambda^*$.
The "whiskers" show the 2.5\%ile - 97.5\%ile Gibbs posterior bands.  The "box" shows the Gibbs posterior interquartile range, and the bar inside the box is the Gibbs posterior
median.}

\end{figure}

\newpage

\begin{figure}[H]
\includegraphics*[scale=1.08]{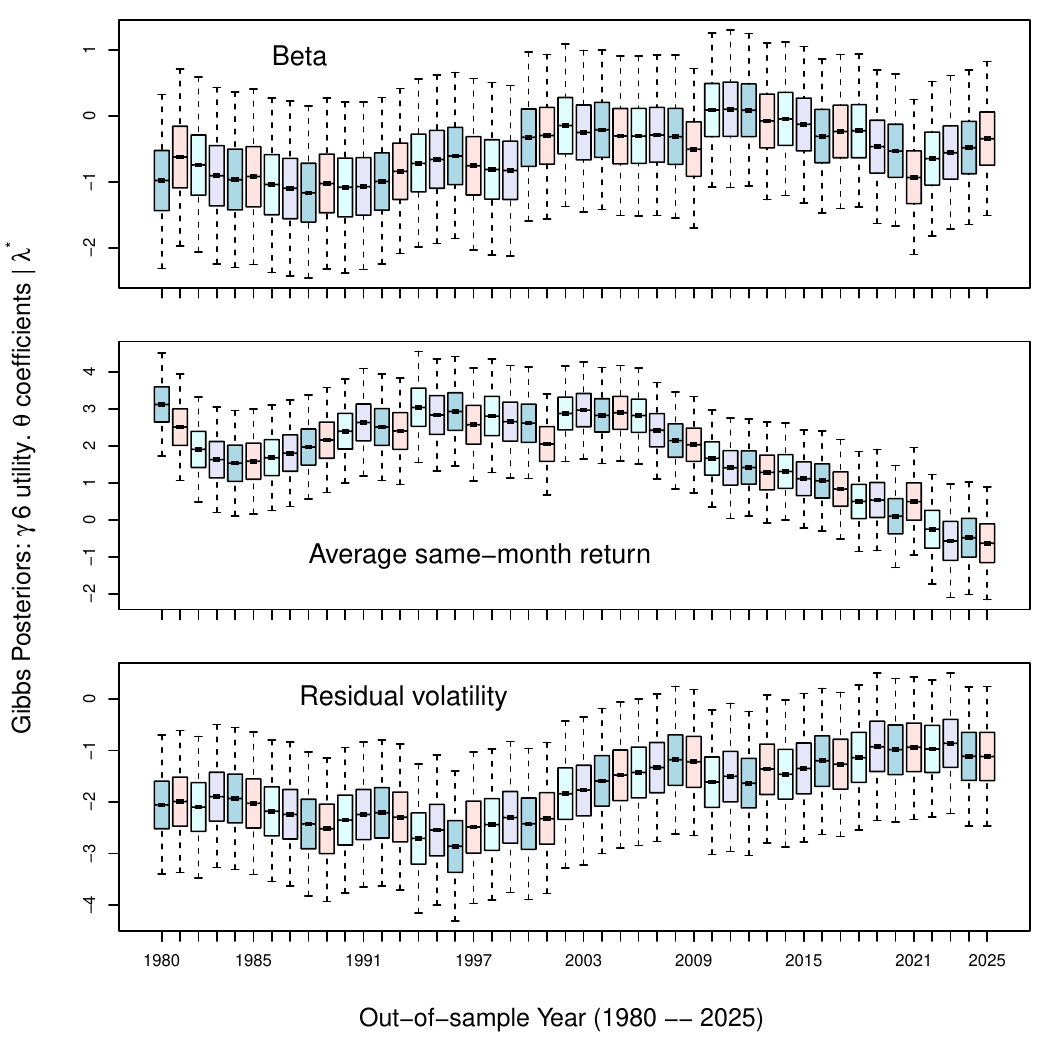}
\vspace{1.8\baselineskip}
\caption{{\bf Figure 6b.  Gibbs posterior on $\theta$, by year.}
Gibbs posterior of the $\theta$ coefficients for the power utility function with $\gamma = 6$.  Posteriors are constructed conditional on the preceding 240 months of data, and $\lambda^*$.
The "whiskers" show the 2.5\%ile - 97.5\%ile Gibbs posterior bands.  The "box" shows the Gibbs posterior interquartile range, and the bar inside the box is the Gibbs posterior
median.}

\end{figure}


\newpage

\hoffset -1.25in

\begin{figure}[H]
\includegraphics*[scale=1.0]{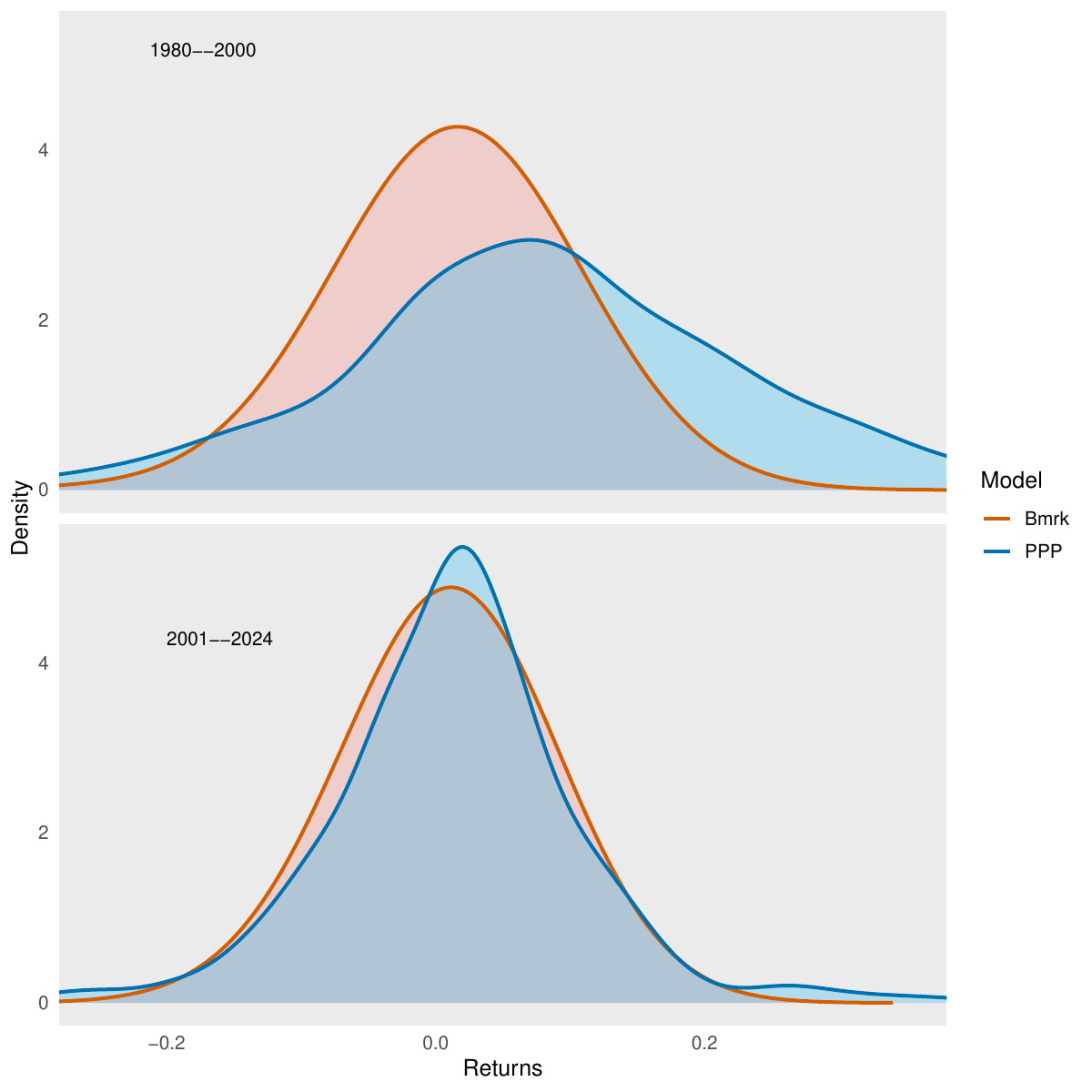}
\vspace{1.8\baselineskip}
\caption{{\bf Figure 7.  Gibbs posterior return densities}
Gibbs posterior densities of out-of-sample optimal portfolio returns $( | \lambda^*)$, and the benchmark return densities for the two subperiods.  This is for the log utility investor.  Each
out-of-sample year is constructed using the data from that year and the Gibbs posterior on $(\theta | \lambda^*).$
The Parametric Portfolio Policy portfolios are shaded blue and the benchmark in red.  In the first subperiod the benchmark is the equally-weighted portfolio of all eligible
stocks.  In the second subperiod the benchmark is the value weighted portfolio of all eligible stocks.}

\end{figure}

\newpage

\hoffset -1.25in

\begin{figure}[H] 
\includegraphics*[scale=1.0]{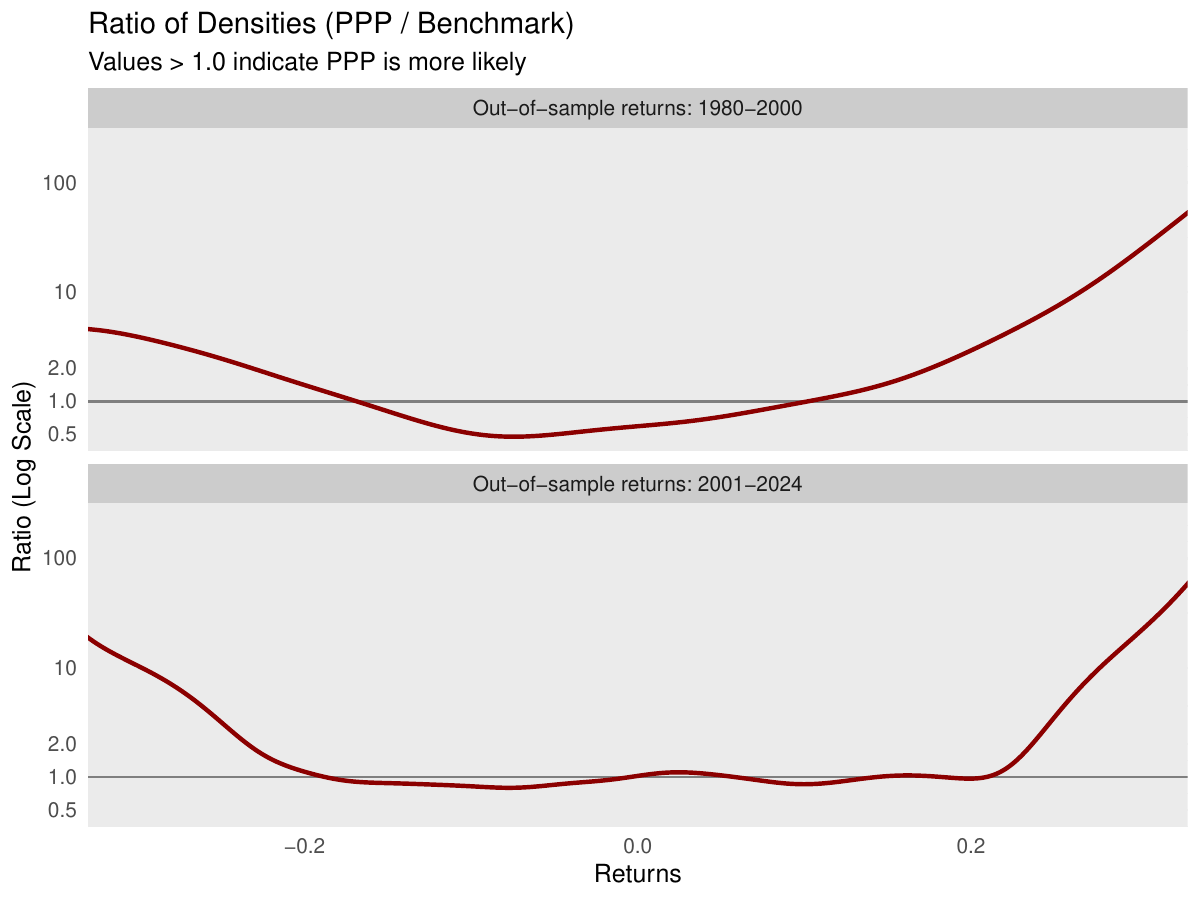}
\vspace{1.8\baselineskip}
\caption{{\bf Figure 8.  Gibbs posterior relative likelihoods}
Gibbs posterior densities of out-of-sample optimal portfolio returns $( | \lambda^*)$, and the benchmark return densities for the two subperiods.  This is for the log utility investor.  Each
out-of-sample year is constructed using the data from that year and the Gibbs posterior on $(\theta | \lambda^*).$
Plot of the ratio of densities for returns on a log scale, PPP/Bmrk.  A value of 2 means that that return is twice as likely in the out-of-sample period under the parametric
portfolio policy than the benchmark.  In the first subperiod the benchmark is the equally-weighted portfolio of all eligible
stocks.  In the second subperiod the benchmark is the value weighted portfolio of all eligible stocks.}

\end{figure}

\newpage

\begin{figure}[H]
\includegraphics*[scale=1.0]{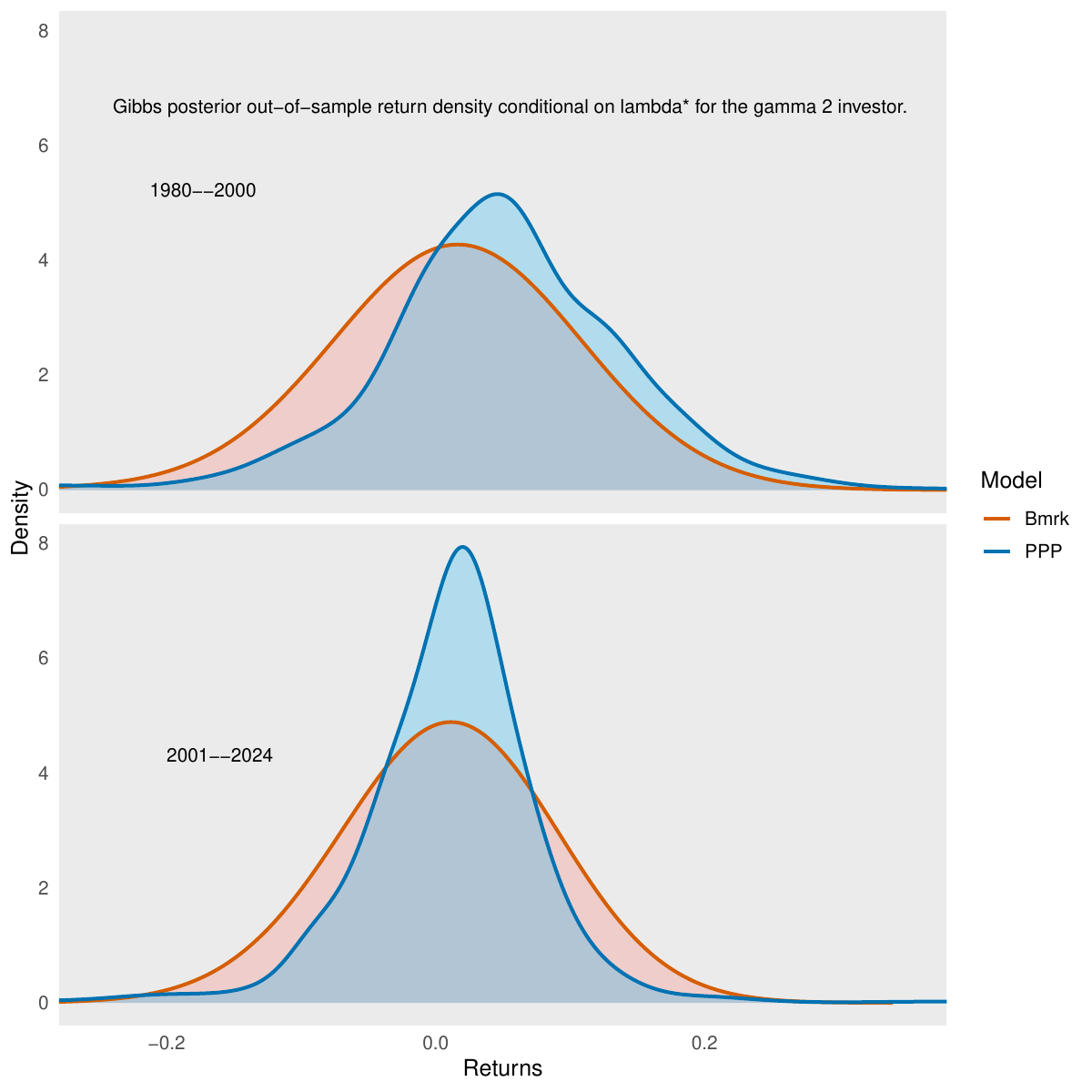}
\vspace{1.8\baselineskip}
\caption{{\bf Figure 9.  Gibbs posterior return densities}
Gibbs posterior densities of out-of-sample optimal portfolio returns $( | \lambda^*)$, and the benchmark return densities for the two subperiods.  This is for the power utility investor
with coefficient of relative risk aversion $\gamma = 2$.  Each out-of-sample year is constructed using the data from that year and the Gibbs posterior on $(\theta | \lambda^*).$  The Parametric Portfolio Policy portfolios are shaded blue and the benchmark in red.  In the first subperiod the benchmark is the equally-weighted portfolio of all eligible
stocks.  In the second subperiod the benchmark is the value weighted portfolio of all eligible stocks.}

\end{figure}

\newpage

\hoffset -1.25in

\begin{figure}[H] 
\includegraphics*[scale=1.0]{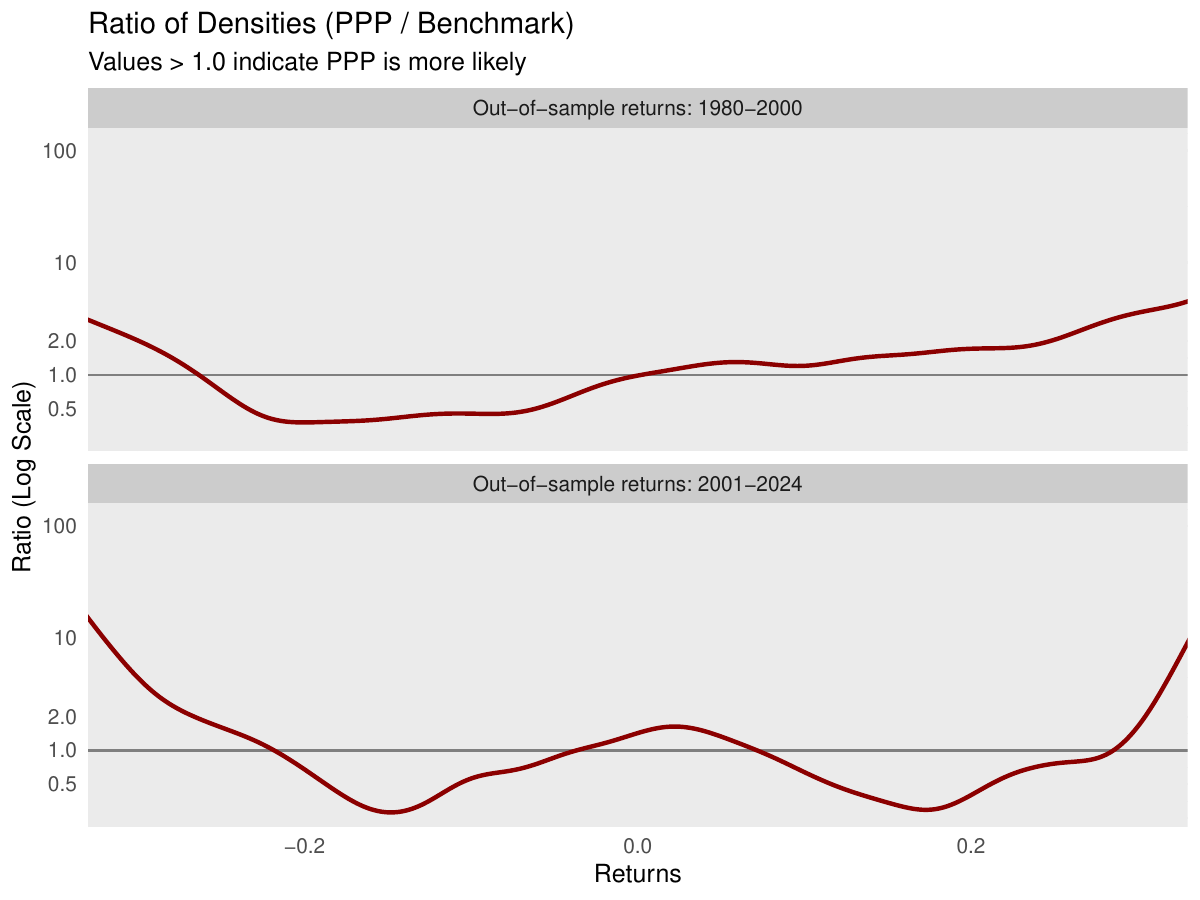}
\vspace{1.8\baselineskip}
\caption{{\bf Figure 10.  Gibbs posterior relative likelihoods}
Gibbs posterior densities of out-of-sample optimal portfolio returns $( | \lambda^*)$, and the benchmark return densities for the two subperiods.  This is for the power
utility investor with coefficient of relative risk aversion, $\gamma = 2$. Each
out-of-sample year is constructed using the data from that year and the Gibbs posterior on $(\theta | \lambda^*).$
Plot of the ratio of densities for returns on a log scale, PPP/Bmrk.  A value of 2 means that that return is twice as likely in the out-of-sample period under the parametric
portfolio policy than the benchmark.  In the first subperiod the benchmark is the equally-weighted portfolio of all eligible
stocks.  In the second subperiod the benchmark is the value weighted portfolio of all eligible stocks.}

\end{figure} 

\newpage

\hoffset -1.25in

\begin{figure}[H]
\includegraphics*[scale=1.0]{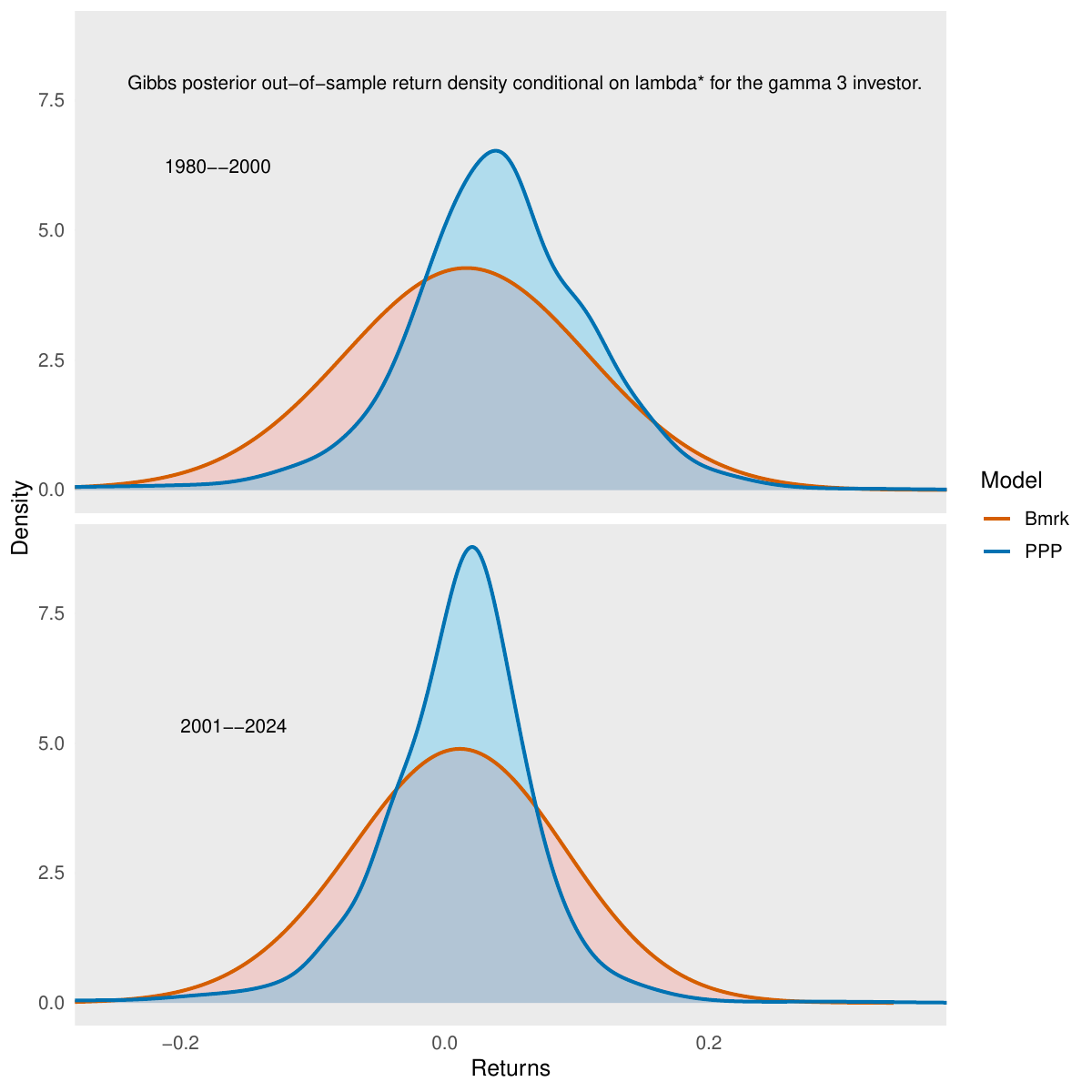}
\vspace{1.8\baselineskip}
\caption{{\bf Figure 11.  Gibbs posterior return densities}
Gibbs posterior densities of out-of-sample optimal portfolio returns $( | \lambda^*)$, and the benchmark return densities for the two subperiods.  This is for the case of power utility 
with coefficient of relative risk aversion, $\gamma = 3$.  Each out-of-sample year is constructed using the data from that year and the Gibbs posterior on $(\theta | \lambda^*).$The Parametric Portfolio Policy portfolios are shaded blue and the benchmark in red.  In the first subperiod the benchmark is the equally-weighted portfolio of all eligible
stocks.  In the second subperiod the benchmark is the value weighted portfolio of all eligible stocks.}

\end{figure}

\newpage

\hoffset -1.25in

\begin{figure}[H]
\includegraphics*[scale=1.0]{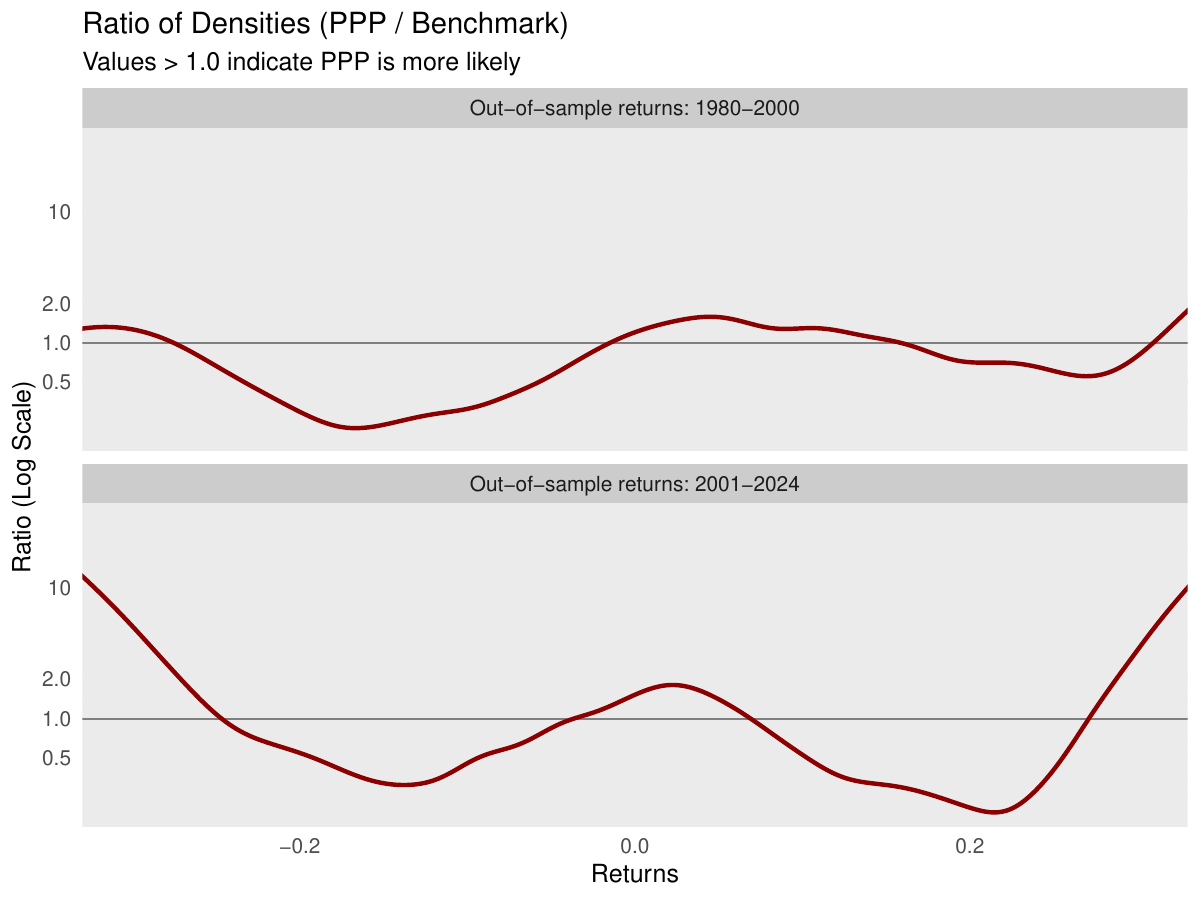}
\vspace{1.8\baselineskip}
\caption{{\bf Figure 12.  Gibbs posterior relative likelihoods}
Gibbs posterior densities of out-of-sample optimal portfolio returns $( | \lambda^*)$, and the benchmark return densities for the two subperiods.  This is for the power
utility investor with coefficient of relative risk aversion, $\gamma = 3$. Each
out-of-sample year is constructed using the data from that year and the Gibbs posterior on $(\theta | \lambda^*).$
Plot of the ratio of densities for returns on a log scale, PPP/Bmrk.  A value of 2 means that that return is twice as likely in the out-of-sample period under the parametric
portfolio policy than the benchmark.  In the first subperiod the benchmark is the equally-weighted portfolio of all eligible
stocks.  In the second subperiod the benchmark is the value weighted portfolio of all eligible stocks.}

\end{figure}

\newpage

\hoffset -1.25in

\begin{figure}[H]
\includegraphics*[scale=1.0]{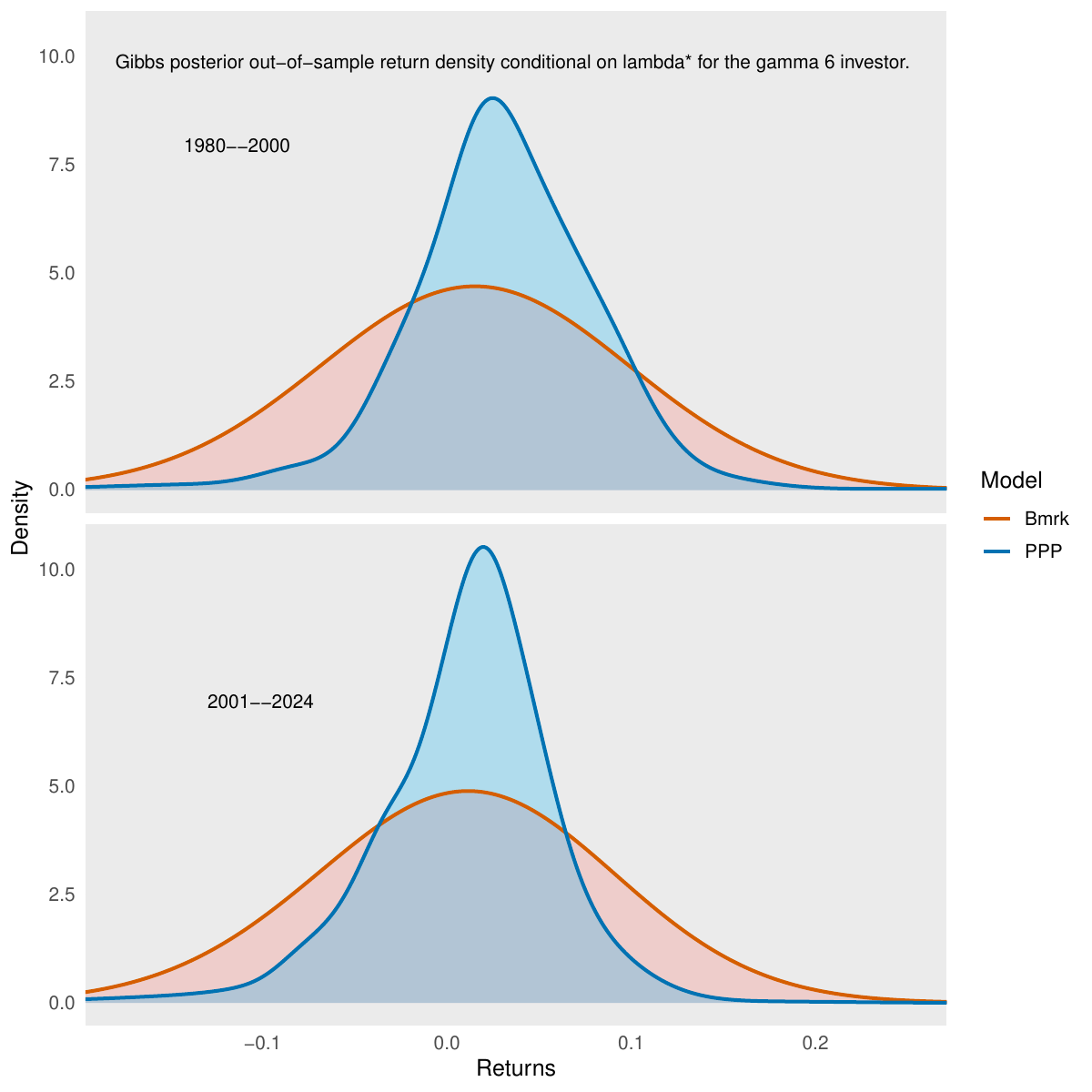}
\vspace{1.8\baselineskip}
\caption{{\bf Figure 13.  Gibbs posterior return densities}
Gibbs posterior densities of out-of-sample optimal portfolio returns $( | \lambda^*)$, and the benchmark return densities for the two subperiods.  This is for the case of power utility
with $\gamma = 6$.  Each out-of-sample year is constructed using the data from that year and the Gibbs posterior on $(\theta | \lambda^*).$
The Parametric Portfolio Policy portfolios are shaded blue and the benchmark in red.  In the first subperiod the benchmark is the equally-weighted portfolio of all eligible
stocks.  In the second subperiod the benchmark is the value weighted portfolio of all eligible stocks.}

\end{figure}

\newpage

\hoffset -1.25in

\begin{figure}[H]
\includegraphics*[scale=1.0]{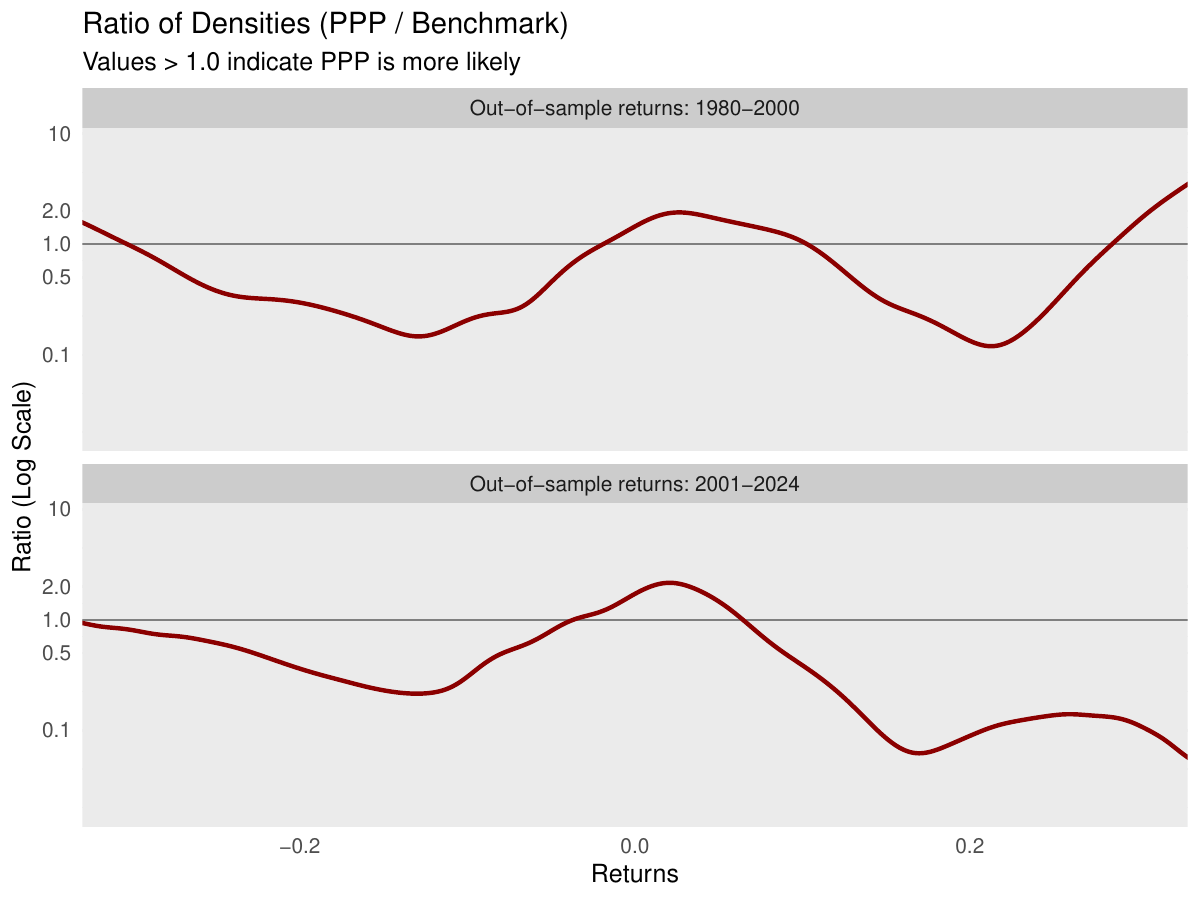}
\vspace{1.8\baselineskip}
\caption{{\bf Figure 14.  Gibbs posterior relative likelihoods}
Gibbs posterior densities of out-of-sample optimal portfolio returns $( | \lambda^*)$, and the benchmark return densities for the two subperiods.  This is for the power
utility investor with coefficient of relative risk aversion, $\gamma = 6$. Each
out-of-sample year is constructed using the data from that year and the Gibbs posterior on $(\theta | \lambda^*).$
Plot of the ratio of densities for returns on a log scale, PPP/Bmrk.  A value of 2 means that that return is twice as likely in the out-of-sample period under the parametric
portfolio policy than the benchmark.  In the first subperiod the benchmark is the equally-weighted portfolio of all eligible
stocks.  In the second subperiod the benchmark is the value weighted portfolio of all eligible stocks.}

\end{figure}

\newpage

\baselineskip 12pt

\begin{center}
{\Large {\bf Table 1}}\\
{\bf Effects of $\gamma$ and $\lambda$ on $\theta$ posterior}\\
\end{center}

\noindent
Properties of the Gibbs posteriors on $\theta$ from the 240-month period January 1977 -- December 1996.
$\gamma$ is the coefficient of relative risk aversion for power utility, with log utility when $\gamma = 1$.\\
$\lambda$ governs the relative weight of the data and the prior.  The Gibbs posterior values corresponding to the optimal $\lambda$ values, $\lambda^*$, for each
utility function are emboldened in the table.\\

\baselineskip 12pt

\vskip .1in

\hoffset=-0.3in

\begin{tabular}{crrrrrrrrrrr}

\boldmath{$\lambda$} \unboldmath & \multicolumn{2}{c}{log utility} & &  \multicolumn{2}{c}{$\gamma = 2$} &  & \multicolumn{2}{c}{$\gamma = 3$} & & \multicolumn{2}{c}{$\gamma = 6$}\\
\cline{1-1}
{\bf 7500} & Mean & Std. Dev. & & Mean & Std. Dev. & & Mean & Std. Dev. & & Mean & Std. Dev\\
 \cline{2-3} \cline{5-6} \cline{8-9} \cline{11-12}
$\theta_1$ & {\bf 12.25} & {\bf 0.722} &  & 8.63 & 0.623 &  & 6.70 & 0.555 &  & 3.98 & 0.448\\
$\theta_2$ & {\bf 5.08} & {\bf 0.845} &  & 4.51 & 0.775 &  & 3.86 & 0.717 &  & 2.51 & 0.617\\
$\theta_3$ & {\bf -5.24} & {\bf 0.838} &  & -4.49 & 0.756 &  & -3.99 & 0.688 &  & -3.25 & 0.594\\
$\theta_4$ & {\bf -0.11} & {\bf 0.731} &  & 0.30 & 0.655 &  & 0.37 & 0.619 &  & 0.10 & 0.541\\
$\theta_5$ & {\bf 9.92} & {\bf 0.862} &  & 7.85 & 0.783 &  & 6.54 & 0.714 &  & 4.44 & 0.587\\
$\theta_6$ & {\bf -7.32} & {\bf 0.803} &  & -6.53 & 0.750 &  & -5.90 & 0.725 &  & -4.67 & 0.658\\
 &  &  &  &  &  &  &  &  &  &  & \\
{\bf 4500} &  &  &  &  &  &  &  &  &  &  & \\
$\theta_1$ & 8.89 & 0.803 &  & {\bf 6.66} & {\bf 0.706} &  & 5.36 & 0.643 &  & 3.38 & 0.538\\
$\theta_2$ & 3.08 & 0.883 &  & {\bf 2.97} & {\bf 0.826} &  & 2.73 & 0.783 &  & 2.01 & 0.702\\
$\theta_3$ & -3.50 & 0.889 &  & {\bf -3.15} & {\bf 0.806} &  & -2.95 & 0.771 &  & -2.56 & 0.668\\
$\theta_4$ & -0.56 & 0.778 &  & {\bf -0.42} & {\bf 0.720} &  & -0.34 & 0.669 &  & -0.45 & 0.595\\
$\theta_5$ & 6.67 & 0.910 &  & {\bf 5.58} & {\bf 0.850} &  & 4.86 & 0.796 &  & 3.58 & 0.682\\
$\theta_6$ & -4.72 & 0.833 &  & {\bf -4.42} & {\bf 0.800} & & -4.17 & 0.756 &  & -3.51 & 0.698\\
 &  &  &  &  &  &  &  &  &  &  & \\
{\bf 3500} &  &  &  &  &  &  &  &  &  &  & \\
$\theta_1$ & 7.44 & 0.834 &  & 5.75 & 0.724 &  & {\bf 4.73} & {\bf 0.682} &  & 3.06 & 0.574\\
$\theta_2$ & 2.28 & 0.898 &  & 2.38 & 0.838 &  & {\bf 2.26} & {\bf 0.814} &  & 1.76 & 0.724\\
$\theta_3$ & -2.83 & 0.907 &  & -2.61 & 0.850 &  & {\bf -2.47} & {\bf 0.800} &  & -2.24 & 0.707\\
$\theta_4$ & -0.62 & 0.807 &  & -0.56 & 0.738 &  & {\bf -0.53} & {\bf 0.704} &  & -0.6 & 0.628\\
$\theta_5$ & 5.40 & 0.932 &  & 4.63 & 0.869 &  & {\bf 4.12} & {\bf 0.827} &  & 3.14 & 0.729\\
$\theta_6$ & -3.75 & 0.854 &  & -3.65 & 0.808 &  & {\bf -3.49} & {\bf 0.784} &  & -3.04 & 0.727\\
 &  &  &  &  &  &  &  &  &  &  & \\
{\bf 2500} &  &  &  &  &  &  &  &  &  &  & \\
$\theta_1$ & 5.76 & 0.875 &  & 4.64 & 0.792 &  & 3.91 & 0.735 &  & {\bf 2.64} & {\bf 0.627}\\
$\theta_2$ & 1.66 & 0.919 &  & 1.76 & 0.874 &  & 1.72 & 0.846 &  & {\bf 1.44} & {\bf 0.769}\\
$\theta_3$ & -2.09 & 0.933 &  & -2.00 & 0.880 &  & -1.92 & 0.834 &  & {\bf -1.80} & {\bf 0.752}\\
$\theta_4$ & -0.58 & 0.838 &  & -0.65 & 0.773 &  & -0.67 & 0.733 &  & {\bf -0.75} & {\bf 0.659}\\
$\theta_5$ & 4.06 & 0.954 &  & 3.57 & 0.899 &  & 3.23 & 0.863 &  & {\bf 2.57} & {\bf 0.776}\\
$\theta_6$ & -2.75 & 0.869 &  & -2.79 & 0.821 &  & -2.72 & 0.798 &  & {\bf -2.49} & {\bf 0.751}\\
\hline

\end{tabular}

\newpage

\begin{center}
{\Large {\bf Table 2}}\\
{\bf Optimal $\lambda$ (temperature) by year and utility function}\\
\end{center}

\small

\noindent
At the beginning of each out-of-sample year, I estimate a grid of Gibbs posteriors with $\lambda$ ranging from 500 to 100,000, as shown in Figure 1.
Each year, for each utility function I apply the KNEEDLE algorithm to this grid to select the optimal value of $\lambda$, $\lambda^*$---to trade off
precision against fragility (ill-conditioning).  This table reports $\lambda^*$ in each case.  The emboldened year (18) matches Figure 1 and Table 1.

\baselineskip 12pt

\vskip .1in

\hoffset=-1in

\begin{center}

\begin{tabular}{cccrrrr}

Year No. & Start Year & End Year & $\gamma = 1$ & $\gamma = 2$ & $\gamma = 3$ & $\gamma = 6$\\
\hline
1 & 1960 &  1979 & 7,500 & 5,000 & 4,000 & 3,000\\
2 & 1961 &  1980 & 7,500 & 4,250 & 3,500 & 2,500\\
3 & 1962 &  1981 & 7,500 & 4,250 & 3,500 & 2,500\\
4 & 1963 &  1982 & 7,500 & 4,000 & 3,500 & 2,500\\
5 & 1964 &  1983 & 7,500 & 4,000 & 3,500 & 2,500\\
6 & 1965 &  1984 & 7,500 & 4,000 & 3,500 & 2,500\\
7 & 1966 &  1985 & 7,500 & 4,000 & 3,500 & 2,500\\
8 & 1967 &  1986 & 7,500 & 4,000 & 3,500 & 2,500\\
9 & 1968 &  1987 & 7,500 & 4,000 & 3,500 & 2,500\\
10 & 1969 &  1988 & 7,500 & 4,250 & 3,500 & 2,500\\
11 & 1970 &  1989 & 7,500 & 4,250 & 3,500 & 2,500\\
12 & 1971 &  1990 & 7,500 & 4,250 & 4,000 & 2,500\\
13 & 1972 &  1991 & 7,500 & 4,500 & 3,500 & 2,500\\
14 & 1973 &  1992 & 7,500 & 4,500 & 3,500 & 2,500\\
15 & 1974 &  1993 & 7,500 & 4,250 & 3,500 & 2,500\\
16 & 1975 &  1994 & 7,500 & 4,500 & 3,500 & 2,500\\
17 & 1976 &  1995 & 7,500 & 4,500 & 4,000 & 3,000\\
{\bf 18} & {\bf 1977} &  {\bf 1996} & {\bf 7,500} & {\bf 4,500} & {\bf 3,500} & {\bf 2,500}\\
19 & 1978 &  1997 & 7,500 & 4,250 & 3,500 & 2,500\\
20 & 1979 &  1998 & 7,500 & 4,500 & 3,500 & 2,500\\
21 & 1980 &  1999 & 7,500 & 4,500 & 3,000 & 2,500\\
22 & 1981 &  2000 & 6,250 & 3,500 & 3,000 & 2,500\\
23 & 1982 &  2001 & 6,250 & 3,000 & 3,000 & 2,500\\
24 & 1983 & 2002 & 6,250 & 2,500 & 3,000 & 2,250\\
25 & 1984 &  2003 & 6,250 & 3,000 & 3,000 & 2,250\\
26 & 1985 &  2004 & 5,000 & 3,000 & 3,000 & 2,250\\
27 & 1986 &  2005 & 6,250 & 3,000 & 3,000 & 2,250\\
28 & 1987 &  2006 & 6,250 & 2,000 & 3,000 & 2,250\\
29 & 1988 &  2007 & 6,250 & 3,000 & 3,000 & 2,250\\
30 & 1989 &  2008 & 5,000 & 3,000 & 2,500 & 2,250\\
31 & 1990 &  2009 & 5,000 & 2,000 & 3,000 & 2,250\\
32 & 1991 &  2010 & 5,000 & 3,000 & 2,500 & 2,000\\
33 & 1992 &  2011 & 5,000 & 3,000 & 2,500 & 2,250\\
34 & 1993 &  2012 & 5,000 & 2,500 & 2,500 & 2,000\\
35 & 1994 &  2013 & 5,000 & 3,000 & 2,500 & 2,250\\
36 & 1995 &  2014 & 5,000 & 2,500 & 2,500 & 2,250\\
37 & 1996 &  2015 & 4,000 & 2,500 & 2,500 & 2,000\\
38 & 1997 &  2016 & 4,000 & 2,500 & 2,500 & 2,000\\
39 & 1998 &  2017 & 4,000 & 3,000 & 2,500 & 2,000\\
40 & 1999 &  2018 & 4,000 & 2,500 & 2,500 & 2,000\\
41 & 2000 &  2019 & 4,000 & 2,500 & 2,500 & 2,000\\
42 & 2001 &  2020 & 5,000 & 3,500 & 2,500 & 2,000\\
43 & 2002 &  2021 & 5,000 & 4,000 & 2,500 & 2,000\\
44 & 2003 &  2022 & 5,000 & 4,500 & 2,500 & 2,000\\
45 & 2004 & 2023 & 5,000 & 4,250 & 2,500 & 2,000\\
46 & 2005 & 2024 & 5,000 & 4,500 & 2,500 & 2,000\\
\hline

\end{tabular}

\end{center}

\begin{landscape}

\hfuzz 50pt

\begin{center}
{\Large {\bf Table 3}}\\
{\bf Gibbs posteriors on out-of-sample portfolio returns conditional on} \boldmath{$\lambda^*$} \unboldmath
\end{center}

\small

\noindent
Properties of the Gibbs posteriors on optimal out-of-sample portfolio monthly \% returns over the period 1980--2024.
The posterior comprises 300,000 post-burn-in Metropolis draws; the posterior on the parameter vector $\theta$ and $\lambda^*$ are estimated using the most recent 240 months at the
beginning of each year in the out-of-sample period.
$\gamma$ is the coefficient of relative risk aversion for power utility, with log utility when $\gamma = 1$.
$\lambda$ governs the relative weight of the data and the prior.  The Gibbs posterior values corresponding to the optimal $\lambda$ values;
$\lambda^*$ is selected at the beginning of each year using a KNEEDLE algorithm.  Also reported: $\overline{\theta}$ represents the single out-of-sample
path from using the mean $\theta$ vector (at $\lambda^*$) to form portfolios, Val. Wtd. is the value-weighted portfolio of all securities in the sample (the optimal portfolio under the prior), and
Eq. Wtd is the equally weighted portfolio of all securities in the sample. S.R. is the annualized Sharpe ratio.  CE is the monthly certainty equivalent return.

\baselineskip 12pt

\normalsize

\begin{center}

\vskip .1in

\hoffset=-1in

\begin{tabular}{crrrrrrrrcc}

\multicolumn{11}{l}{\bf{Panel A: Log utility}}\\
\multicolumn{11}{c}{}\\

Parameter & Mean & Std. dev. & 2.5\%ile & 25\%ile & median & 75\%ile & 97.5\%ile & $(\overline{\theta})$ & Val. Wtd. & Eq. Wtd.\\
\multicolumn{11}{c}{ }\\
\multicolumn{11}{l}{Out-of-sample period 1: 1980-2000}\\
$E(r)$ & 7.45  & 0.14 &  7.18 &  7.36  & 7.45 &  7.55  &  7.73  & 7.45  & 1.37 & 1.40\\
$\sigma(r)$ & 16.32  & 0.29 & 15.77 & 16.13 & 16.32 & 16.51 & 16.89 & 16.19 & 4.35 & 4.72\\
S.R. & 1.463  & 0.032  & 1.400  & 1.441 &  1.463  & 1.485 &  1.526 & 1.475 & 0.647 & 0.622\\
median & 7.63 &  0.29  & 7.06 &  7.43 &  7.63 &  7.82 &  8.19 & 7.72 & 1.65 & 1.76\\
IQR & 18.75 &   0.58  & 7.62 & 18.36  & 18.75 &  19.14 &  19.88 & 18.53 & 5.45 & 6.81\\
Skew & -0.011 &  0.017 & -0.044 & -0.022 & -0.011 &  0.001 &  0.023 & -0.017 & -0.063 & -0.075\\
Kurt & 0.347  & 0.045  & 0.261  & 0.317  & 0.347  & 0.377  & 0.437 & 0.326 & 0.251 & 0.221\\
CE &   6.04 &  0.14 &  5.76 &  5.94 &  6.04 &  6.13 &  6.30 & 6.07 & 1.27 & 1.29\\
\hline
\multicolumn{11}{c}{ }\\
\multicolumn{11}{l}{Out-of-sample period 2: 2001-2024}\\
$E(r)$ & 1.03  & 0.14 &  0.75 &  0.93  & 1.03 &  1.12  &  1.30  & 1.03  & 0.81 & 0.98\\
$\sigma(r)$ & 12.28  & 0.55 & 11.23 & 11.90 & 12.27 & 12.64 & 13.38 & 12.07 & 4.38 & 5.57\\
S.R. & 0.252  & 0.043  & 0.170  & 0.223 &  0.252  & 0.281 &  0.340 & 0.256 & 0.535 & 0.527\\
median & 1.52 &  0.22  & 1.09 &  1.38 &  1.53 &  1.67 &  1.94 & 1.51 & 1.37 & 1.34\\
IQR & 10.43 &   0.44  & 9.56 & 10.13  & 10.42 &  10.72 &  11.31 & 10.45 & 5.21 & 6.89\\
Skew & -0.040 &  0.017 & -0.074 & -0.052 & -0.041 &  -0.029 &  -0.006 & -0.040 & -0.127 & -0.064\\
Kurt & 1.202  & 0.116  & 0.978  & 1.123  & 1.200  & 1.279  & 1.433 & 1.223 & 0.263 & 0.243\\
CE &  0.16 &  0.20 & -0.26 &  0.03 &  0.17 &  0.30 &  0.53 & 0.20 & 0.72 & 0.83\\
\hline
\end{tabular}

\end{center}

\newpage

\begin{center}
{\Large {\bf Table 3} (Cont'd.)}\\
{\bf Gibbs posteriors on out-of-sample portfolio returns conditional on} \boldmath{$\lambda^*$} \unboldmath
\end{center}

\small

\noindent
Properties of the Gibbs posteriors on optimal out-of-sample portfolio monthly \% returns over the period 1980--2024.
The posterior comprises 300,000 post-burn-in Metropolis draws; the posterior on the parameter vector $\theta$ and $\lambda^*$ are estimated using the most recent 240 months at the
beginning of each year in the out-of-sample period.
$\gamma$ is the coefficient of relative risk aversion for power utility, with log utility when $\gamma = 1$.
$\lambda$ governs the relative weight of the data and the prior.  The Gibbs posterior values corresponding to the optimal $\lambda$ values;
$\lambda^*$ is selected at the beginning of each year using a KNEEDLE algorithm.  Also reported: $\overline{\theta}$ represents the single out-of-sample
path from using the mean $\theta$ vector (at $\lambda^*$) to form portfolios, Val. Wtd. is the value-weighted portfolio of all securities in the sample (the optimal portfolio under the prior), and
Eq. Wtd is the equally weighted portfolio of all securities in the sample. S.R. is the annualized Sharpe ratio.  CE is the monthly certainty equivalent return.

\baselineskip 12pt

\normalsize

\begin{center}

\vskip .1in

\hoffset=-1in

\begin{tabular}{crrrrrrrrcc}

\multicolumn{11}{l}{\bf{Panel B: Power utility} \boldmath{$\gamma = 2$} \unboldmath }\\
\multicolumn{11}{c}{}\\

Parameter & Mean & Std. dev. & 2.5\%ile & 25\%ile & median & 75\%ile & 97.5\%ile & $(\overline{\theta})$ & Val. Wtd. & Eq. Wtd.\\
\multicolumn{11}{c}{ }\\
\multicolumn{11}{l}{Out-of-sample period 1: 1980-2000}\\
$E(r)$ & 4.76  &  0.14  &  4.49  &  4.67  &  4.76  &  4.86  &  5.03 & 4.76 & 1.37 & 1.40\\
$\sigma(r)$ & 9.55  &  0.29  &  9.01  &  9.35  &  9.54  &  9.73  & 10.13 & 9.33 &  4.35 & 4.72\\
S.R. & 1.526  &  0.055  &  1.419  &  1.489  &  1.526  &  1.563  &  1.633 & 1.561 &  0.647 & 0.622\\
median & 4.86  &  0.21  &  4.44  &  4.72  &  4.86  &  5.00  &  5.27 & 4.87 & 1.65 & 1.76\\
IQR & 11.08  &  0.48  & 10.14  & 10.76  & 11.08  & 11.41  & 12.02 &  11.03 & 5.45 & 6.81\\
Skew & -0.010  &  0.021  & -0.052  & -0.025  & -0.010  &  0.004  &  0.032 & -0.011 & -0.063 & -0.075\\
Kurt & 0.424  &  0.076  &  0.275  &  0.373  &  0.424  &  0.475  &  0.571 & 0.418 & 0.251 & 0.221\\
CE & 3.80 &  0.14 &  3.53 &  3.72 &  3.81 &  3.90 &  4.06 & 3.86 & 1.18 & 1.17\\
\hline
\multicolumn{11}{c}{ }\\
\multicolumn{11}{l}{Out-of-sample period 2: 2001-2024}\\
$E(r)$ & 0.94  &  0.14  &  0.67  &  0.85  &  0.94  &  1.03  &  1.21 & 0.94  & 0.81 & 0.98\\
$\sigma(r)$ &   7.21  &  0.50  &  6.30  &  6.86  &  7.19  &  7.54  &  8.26 & 6.88 & 4.38 & 5.57\\
S.R. & 0.390  &  0.075  &  0.248  &  0.338  &  0.388  &  0.440  &  0.543 & 0.406 & 0.535 & 0.527\\
median & 1.47  &  0.19  &  1.10  &  1.35  &  1.48  &  1.60  &  1.84 & 1.63  & 1.37 & 1.34\\
IQR & 7.08  &  0.40  &  6.30  &  6.80  &  7.07  &  7.35  &  7.88 & 6.72 & 5.21 & 6.89\\
Skew & -0.074  &  0.024  & -0.122  & -0.091  & -0.074  & -0.058  & -0.026 & -0.100 & -0.127 & -0.064\\
Kurt & 0.829  &  0.171  &  0.506  &  0.711  &  0.823  &  0.941  &  1.178 & 0.785 & 0.263 & 0.243\\
CE &  0.36 &  0.19 & -0.03 &  0.24 &  0.37 &  0.49 &  0.71 & 0.43 & 0.62 & 0.67\\
\hline

\end{tabular}

\end{center}

\newpage

\begin{center}
{\Large {\bf Table 3} (Cont'd.)}\\
{\bf Gibbs posteriors on out-of-sample portfolio returns conditional on} \boldmath{$\lambda^*$} \unboldmath
\end{center}

\small

\noindent
Properties of the Gibbs posteriors on optimal out-of-sample portfolio monthly \% returns over the period 1980--2024.
The posterior comprises 300,000 post-burn-in Metropolis draws; the posterior on the parameter vector $\theta$ and $\lambda^*$ are estimated using the most recent 240 months at the
beginning of each year in the out-of-sample period.
$\gamma$ is the coefficient of relative risk aversion for power utility, with log utility when $\gamma = 1$.
$\lambda$ governs the relative weight of the data and the prior.  The Gibbs posterior values corresponding to the optimal $\lambda$ values;
$\lambda^*$ is selected at the beginning of each year using a KNEEDLE algorithm.  Also reported: $\overline{\theta}$ represents the single out-of-sample
path from using the mean $\theta$ vector (at $\lambda^*$) to form portfolios, Val. Wtd. is the value-weighted portfolio of all securities in the sample (the optimal portfolio under the prior), and
Eq. Wtd is the equally weighted portfolio of all securities in the sample. S.R. is the annualized Sharpe ratio.  CE is the monthly certainty equivalent return.

\baselineskip 12pt

\normalsize

\begin{center}

\vskip .1in

\hoffset=-1in

\begin{tabular}{crrrrrrrrcc}

\multicolumn{11}{l}{\bf{Panel C: Power utility} \boldmath{$\gamma = 3$} \unboldmath}\\
\multicolumn{11}{c}{}\\

Parameter & Mean & Std. dev. & 2.5\%ile & 25\%ile & median & 75\%ile & 97.5\%ile & $(\overline{\theta})$ & Val. Wtd. & Eq. Wtd.\\
\multicolumn{11}{c}{ }\\
\multicolumn{11}{l}{Out-of-sample period 1: 1980-2000}\\
$E(r)$ & 3.97  &  0.13  &  3.71  &  3.88  &  3.97  &  4.05  &  4.22 & 3.97 & 1.37 & 1.40\\
$\sigma(r)$ & 7.51  &  0.30  &  6.98  &  7.31  &  7.49  &  7.70  &  8.16 & 7.26 &  4.35 & 4.72\\
S.R. & 1.572  &  0.070  &  1.431  &  1.526  &  1.573  &  1.620  &  1.708 & 1.626 &  0.647 & 0.622\\
median & 4.06  &  0.20  &  3.66  &  3.92  &  4.06  &  4.19  &  4.45 & 4.06 & 1.65 & 1.76\\
IQR & 8.58  &  0.42  &  7.77  &  8.29  &  8.57  &  8.86  &  9.41 &  8.30 & 5.45 & 6.81\\
Skew & -0.012  &  0.024  & -0.060  & -0.029  & -0.012  &  0.004  &  0.036 & -0.013 & -0.063 & -0.075\\
Kurt & 0.432  &  0.091  &  0.250  &  0.371  &  0.433  &  0.493  &  0.610 & 0.418 & 0.251 & 0.221\\
CE & 3.06 &  0.15 &  2.75 &  2.97 &  3.07 &  3.16 &  3.32 &  3.14 & 1.07 & 1.04\\
\hline
\multicolumn{11}{c}{ }\\
\multicolumn{11}{l}{Out-of-sample period 2: 2001-2024}\\
$E(r)$ & 0.97  &  0.12  &  0.73  &  0.88  &  0.97  &  1.05  &  1.21 & 0.97  & 0.81 & 0.98\\
$\sigma(r)$ &  6.22  &  0.42  &  5.48  &  5.93  &  6.20  &  6.49  &  7.116 & 5.89 & 4.38 & 5.57\\
S.R. & 0.468  &  0.081  &  0.314  &  0.412  &  0.466  &  0.522  &  0.632 & 0.491 & 0.535 & 0.527\\
median & 1.52  &  0.18  &  1.17  &  1.40  &  1.52  &  1.64  &  1.86 & 1.51  & 1.37 & 1.34\\
IQR & 6.43  &  0.37  &  5.72  &  6.17  &  6.42  &  6.68  &  7.18 & 6.04 & 5.21 & 6.89\\
Skew & -0.088  &  0.026  & -0.139  & -0.106  & -0.088  & -0.071  & -0.038 & -0.092 & -0.127 & -0.064\\
Kurt & 0.692  &  0.165  &  0.382  &  0.580  &  0.687  &  0.799  &  1.032 & 0.681 & 0.263 & 0.243\\
CE &  0.33 &  0.19 & -0.07 &  0.21 &  0.34 &  0.45 &  0.66 & 0.41 & 0.52 & 0.50\\
\hline

\end{tabular}

\end{center}

\newpage

\begin{center}
{\Large {\bf Table 3} (Cont'd.)}\\
{\bf Gibbs posteriors on out-of-sample portfolio returns conditional on} \boldmath{$\lambda^*$} \unboldmath
\end{center}

\small

\noindent
Properties of the Gibbs posteriors on optimal out-of-sample portfolio monthly \% returns over the period 1980--2024.
The posterior comprises 300,000 post-burn-in Metropolis draws; the posterior on the parameter vector $\theta$ and $\lambda^*$ are estimated using the most recent 240 months at the
beginning of each year in the out-of-sample period.
$\gamma$ is the coefficient of relative risk aversion for power utility, with log utility when $\gamma = 1$.
$\lambda$ governs the relative weight of the data and the prior.  The Gibbs posterior values corresponding to the optimal $\lambda$ values;
$\lambda^*$ is selected at the beginning of each year using a KNEEDLE algorithm.  Also reported: $\overline{\theta}$ represents the single out-of-sample
path from using the mean $\theta$ vector (at $\lambda^*$) to form portfolios, Val. Wtd. is the value-weighted portfolio of all securities in the sample (the optimal portfolio under the prior), and
Eq. Wtd is the equally weighted portfolio of all securities in the sample. S.R. is the annualized Sharpe ratio.  CE is the monthly certainty equivalent return.

\baselineskip 12pt

\normalsize

\begin{center}

\vskip .1in

\hoffset=-1in

\begin{tabular}{crrrrrrrrcc}

\multicolumn{11}{l}{\bf{Panel D: Power utility} \boldmath{$\gamma = 6$} \unboldmath}\\
\multicolumn{11}{c}{}\\

Parameter & Mean & Std. dev. & 2.5\%ile & 25\%ile & median & 75\%ile & 97.5\%ile & $(\overline{\theta})$ & Val. Wtd. & Eq. Wtd.\\
\multicolumn{11}{c}{ }\\
\multicolumn{11}{l}{Out-of-sample period 1: 1980-2000}\\
$E(r)$ & 2.96  &  0.12  &  2.73  &  2.88  &  2.96  &  3.04  &  3.19 & 2.96 & 1.37 & 1.40\\
$\sigma(r)$ & 5.60  &  0.33  &  5.03  &  5.37  &  5.57  &  5.81  &  6.34 & 5.35 &  4.35 & 4.72\\
S.R. & 1.488  &  0.098  &  1.292  &  1.422  &  1.489  &  1.555  &  1.676 & 1.555 &  0.647 & 0.622\\
median & 2.98  &  0.17  &  2.64  &  2.86  &  2.97  &  3.09  &  3.32& 3.00 & 1.65 & 1.76\\
IQR & 6.11  &  0.30  &  5.52  &  5.90  &  6.10  &  6.31  &  6.71 &  6.01 & 5.45 & 6.81\\
Skew & -0.003  &  0.027  & -0.056  & -0.021  & -0.003  &  0.015  &  0.050 & -0.008 & -0.063 & -0.075\\
Kurt & 0.501  &  0.118  &  0.278  &  0.422  &  0.498  &  0.578  &  0.743 & 0.441 & 0.251 & 0.221\\
CE &   1.77 &  0.38 &  0.80 &  1.64 &  1.85 &  2.00 &  2.20 & 1.95 & 0.74 & 0.61\\
\hline
\multicolumn{11}{c}{ }\\
\multicolumn{11}{l}{Out-of-sample period 2: 2001-2024}\\
$E(r)$ & 0.95  &  0.11  &  0.75  &  0.88  &  0.95  &  1.03  &  1.16 & 0.95  & 0.81 & 0.98\\
$\sigma(r)$ &  4.94  &  0.31  &  4.39  &  4.72  &  4.92  &  5.14  &  5.61 & 4.63 & 4.38 & 5.57\\
S.R. & 0.578  &  0.088  &  0.411  &  0.518  &  0.576  &  0.637  &  0.754 & 0.614 & 0.535 & 0.527\\
median & 1.50  &  0.15  &  1.21  &  1.40  &  1.50  &  1.60  &  1.79 & 1.59  & 1.37 & 1.34\\
IQR & 5.45  &  0.31  &  4.86  &  5.24  &  5.44  &  5.65  &  6.06 & 5.25 & 5.21 & 6.89\\
Skew & -0.112  &  0.027  & -0.165  & -0.130  & -0.111  & -0.093  & -0.060 & -0.138 & -0.127 & -0.064\\
Kurt & 0.498  &  0.157  &  0.211  &  0.388  &  0.492  &  0.600  &  0.826 & 0.497 & 0.263 & 0.243\\
CE &  0.11 &  0.21 & -0.37 & -0.01 &  0.13 &  0.25 &  0.45 & 0.23 & 0.20 & -0.02\\
\hline

\end{tabular}

\end{center}

\newpage

\begin{center}
{\Large {\bf Table 4}}\\
{\bf Gibbs posteriors on Fama-French factor regression coefficients}\\
\end{center}

\small
\noindent
Properties of the Gibbs posteriors on optimal out-of-sample portfolio monthly \% returns over the period 1980--2024.
The posterior comprises 300,000 post-burn-in Metropolis draws; the posterior on the parameter vector $\theta$ and $\lambda^*$ are estimated using the most recent 240 months at the
beginning of each year in the out-of-sample period.
Monthly \% returns are projected on the 6 Fama-French factors: the market risk premium, $R_p$, SMB, HML, MOM, RMW, and CMA.
$\lambda$ governs the relative weight of the data and the prior.  The Gibbs posterior values corresponding to the optimal $\lambda$ values;
$\lambda^*$ is selected at the beginning of each year using a KNEEDLE algorithm.  Also reported: the GMM $t$-statistic for the coefficient estimates from projecting the single out-of-sample
path constructed with the mean $\theta$ vector (at $\lambda^*$) onto the Fama-French factors.  (The coefficient from that regression equals the posterior mean coefficient.)
Mean \% Var reports the posterior mean of  the percentage of the portfolio variance attributed to each factor (and for $\alpha$, independent of the factors).  These sum to 
more than 100\% as I do not report all of the covariance effects--many of which are negative.  Std dev \% Var is the posterior standard deviation of this percentage.


\baselineskip 12pt

\vskip .1in

\begin{center}

\begin{tabular}{crrrrrrrccc}

\multicolumn{11}{l}{\bf{Panel A: Log utility}}\\
\multicolumn{11}{c}{}\\

\multicolumn{11}{l}{Out-of-sample period 1: 1980-2000}\\

          & \multicolumn{7}{c}{Gibbs posterior on out-of-sample optimal portfolio returns ($| \lambda^*)$} &  &  & \\
          & \multicolumn{7}{c}{Fama-French Regressions} &  &  & \\
\cline{2-8}
Parameter & Mean & Std. dev. & 2.5\%ile & 25\%ile & median & 75\%ile & 97.5\%ile & $(\overline{\theta})$ GMM & Mean & Std dev\\
          &      &           &          &         &        &         &           & $t-$stat & \% Var  & \% Var\\
$\alpha$ & 3.09 &    0.17 &    2.76 &    2.98 &    3.10 &    3.21 &    3.43 &  2.96 & 64.41 & 1.67\\
$\beta_{R_p}$ & 0.65 &    0.06 &    0.54 &    0.61 &    0.65 &    0.69 &    0.77 &  2.17 & 3.36 & 0.60\\
$\beta_{SMB}$ &  -0.13 &    0.10 &   -0.32 &   -0.20 &   -0.13 &   -0.07 &    0.05 &  -0.19 & 0.10 & 0.10\\
$\beta_{HML}$ & 2.42 &    0.13 &    2.16 &    2.33 &    2.42 &    2.50 &    2.68 &   3.81 & 20.63 & 2.38\\
$\beta_{MOM}$ & 2.67 &    0.10 &    2.48 &    2.61 &    2.67 &    2.74 &    2.86 &   7.03 & 38.50 & 2.28\\
$\beta_{RMW}$ & 0.24 &    0.28 &   -0.32 &    0.05 &    0.24 &    0.43 &    0.80 &  0.22 & 0.34 & 0.43\\
$\beta_{CMA}$ & -0.96 &    0.12 &   -1.20 &   -1.04 &   -0.96 &   -0.88 &   -0.72 &  -1.12 & 1.52 & 0.38\\
\hline
\multicolumn{11}{c}{ }\\
\multicolumn{11}{l}{Out-of-sample period 2: 2001-2024}\\
$\alpha$ &   -0.52 &    0.16 &   -0.83 &   -0.62 &   -0.52 &   -0.41 &   -0.21 &  -0.99 & 58.58 & 2.13\\
$\beta_{R_p}$ & 0.89 &    0.05 &    0.79 &    0.85 &    0.89 &    0.92 &    0.99 & 7.24 & 10.71 & 1.34\\
$\beta_{SMB}$ &  0.39 &    0.10 &    0.19 &    0.32 &    0.39 &    0.46 &    0.60 &  1.30 & 0.84 & 0.44\\
$\beta_{HML}$ & 1.42 &    0.10 &    1.22 &    1.35 &    1.42 &    1.49 &    1.61 &  6.02 & 13.66 & 1.65\\
$\beta_{MOM}$ & 1.16 &    0.11 &    0.93 &    1.08 &    1.16 &    1.23 &    1.37 &  5.85 & 21.44 & 3.28\\
$\beta_{RMW}$ & 1.35 &    0.15 &    1.05 &    1.25 &    1.35 &    1.46 &    1.65 &  4.52 & 6.42 & 1.23\\
$\beta_{CMA}$ & -0.05 &    0.15 &   -0.35 &   -0.15 &   -0.05 &    0.05 &    0.25 &  -0.10 & 0.07 & 0.11\\
\hline
\end{tabular}

\end{center}

\newpage

\begin{center}

{\Large {\bf Table 4} (Cont'd.)}\\
{\bf Gibbs posteriors on Fama-French factor regression coefficients}\\
\end{center}

\small
\noindent
Properties of the Gibbs posteriors on optimal out-of-sample portfolio monthly \% returns over the period 1980--2024.
The posterior comprises 300,000 post-burn-in Metropolis draws; the posterior on the parameter vector $\theta$ and $\lambda^*$ are estimated using the most recent 240 months at the
beginning of each year in the out-of-sample period.
Monthly \% returns are projected on the 6 Fama-French factors: the market risk premium, $R_p$, SMB, HML, MOM, RMW, and CMA.
$\lambda$ governs the relative weight of the data and the prior.  The Gibbs posterior values corresponding to the optimal $\lambda$ values;
$\lambda^*$ is selected at the beginning of each year using a KNEEDLE algorithm.  Also reported: the GMM $t$-statistic for the coefficient estimates from projecting the single out-of-sample
path constructed with the mean $\theta$ vector (at $\lambda^*$) onto the Fama-French factors.  (The coefficient from that regression equals the posterior mean coefficient.)
Mean \% Var reports the posterior mean of  the percentage of the portfolio variance attributed to each factor (and for $\alpha$, independent of the factors).  These sum to 
more than 100\% as I do not report all of the covariance effects--many of which are negative.  Std dev \% Var is the posterior standard deviation of this percentage.


\begin{center}

\baselineskip 12pt

\vskip .1in

\begin{tabular}{crrrrrrrccc}
\multicolumn{11}{l}{\bf{Panel B: Power utility} \boldmath{$\gamma = 2$}}\\
\multicolumn{11}{c}{}\\

\multicolumn{11}{l}{Out-of-sample period 1: 1980-2000}\\
          & \multicolumn{7}{c}{Gibbs posterior on out-of-sample optimal portfolio returns ($| \lambda^*)$} &  &  & \\
          & \multicolumn{7}{c}{Fama-French Regressions} &  &  & \\
\cline{2-8}
Parameter & Mean & Std. dev. & 2.5\%ile & 25\%ile & median & 75\%ile & 97.5\%ile & $(\overline{\theta})$ GMM & Mean & Std dev\\
          &      &           &          &         &        &         &           & $t-$stat & \% Var  & \% Var\\
$\alpha$ &    1.66 &    0.17 &    1.33 &    1.55 &    1.66 &    1.77 &    1.99 & 2.77 & 60.59 & 2.26\\
$\beta_{R_p}$ & 0.71 &    0.06 &    0.60 &    0.67 &    0.71 &    0.75 &    0.83 & 4.34 & 11.66 & 1.81\\
$\beta_{SMB}$ &  -0.29 &    0.09 &   -0.48 &   -0.35 &   -0.29 &   -0.23 &   -0.11 & -0.80 & 0.97 & 0.57\\
$\beta_{HML}$ & 1.55 &    0.13 &    1.30 &    1.47 &    1.55 &    1.64 &    1.80 &  4.38 & 24.96  & 3.94\\
$\beta_{MOM}$ & 1.44 &    0.09 &    1.25 &    1.37 &    1.44 &    1.50 &    1.62 &  6.84 & 32.58 & 3.99\\
$\beta_{RMW}$ & 0.33 &    0.27 &   -0.19 &    0.15 &    0.33 &    0.51 &    0.85 &  0.56 & 1.24 & 1.35\\
$\beta_{CMA}$ & -0.43 &    0.12 &   -0.67 &   -0.51 &   -0.43 &   -0.35 &   -0.19 & -0.88 & 0.94 & 0.50\\
\hline
\multicolumn{11}{c}{ }\\
\multicolumn{11}{l}{Out-of-sample period 2: 2001-2024}\\
$\alpha$ &     -0.24 &    0.15 &   -0.54 &   -0.34 &   -0.24 &   -0.14 &    0.06 &  -0.90 & 52.72 & 3.22\\
$\beta_{R_p}$ &  0.85 &    0.05 &    0.76 &    0.82 &    0.85 &    0.89 &    0.95 & 13.32 & 28.95 & 4.24\\
$\beta_{SMB}$ &  0.11 &    0.10 &   -0.09 &    0.04 &    0.11 &    0.18 &    0.31 & 0.69 & 0.34 & 0.40\\
$\beta_{HML}$ & 0.78 &    0.10 &    0.59 &    0.72 &    0.78 &    0.84 &    0.97 &  6.03 & 12.06 & 2.50\\
$\beta_{MOM}$ & 0.68 &    0.11 &    0.47 &    0.61 &    0.68 &    0.75 &    0.89 &   6.36 & 21.58 & 5.46\\
$\beta_{RMW}$ & 0.83 &    0.15 &    0.53 &    0.73 &    0.83 &    0.93 &    1.12 &   5.13 & 7.06 & 2.16\\
$\beta_{CMA}$ & -0.01 &    0.15 &   -0.30 &   -0.11 &   -0.01 &    0.09 &    0.28 &  -0.04 & 0.19 & 0.27\\
\hline

\end{tabular}

\end{center}

\newpage

\begin{center}
{\Large {\bf Table 4} (Cont'd.)}\\
{\bf Gibbs posteriors on Fama-French factor regression coefficients}\\
\end{center}

\small
\noindent
Properties of the Gibbs posteriors on optimal out-of-sample portfolio monthly \% returns over the period 1980--2024.
The posterior comprises 300,000 post-burn-in Metropolis draws; the posterior on the parameter vector $\theta$ and $\lambda^*$ are estimated using the most recent 240 months at the
beginning of each year in the out-of-sample period.
Monthly \% returns are projected on the 6 Fama-French factors: the market risk premium, $R_p$, SMB, HML, MOM, RMW, and CMA.
$\lambda$ governs the relative weight of the data and the prior.  The Gibbs posterior values corresponding to the optimal $\lambda$ values;
$\lambda^*$ is selected at the beginning of each year using a KNEEDLE algorithm.  Also reported: the GMM $t$-statistic for the coefficient estimates from projecting the single out-of-sample
path constructed with the mean $\theta$ vector (at $\lambda^*$) onto the Fama-French factors.  (The coefficient from that regression equals the posterior mean coefficient.)
Mean \% Var reports the posterior mean of  the percentage of the portfolio variance attributed to each factor (and for $\alpha$, independent of the factors).  These sum to 
more than 100\% as I do not report all of the covariance effects--many of which are negative.  Std dev \% Var is the posterior standard deviation of this percentage.


\baselineskip 12pt

\vskip .1in

\begin{center}

\begin{tabular}{crrrrrrrccc}

\multicolumn{11}{l}{\bf{Panel C: Power utility} \boldmath{$\gamma = 3$}}\\
\multicolumn{11}{c}{}\\

\multicolumn{11}{l}{Out-of-sample period 1: 1980-2000}\\
          & \multicolumn{7}{c}{Gibbs posterior on out-of-sample optimal portfolio returns ($| \lambda^*)$} &  &  & \\
          & \multicolumn{7}{c}{Fama-French Regressions} &  &  & \\
\cline{2-8}
Parameter & Mean & Std. dev. & 2.5\%ile & 25\%ile & median & 75\%ile & 97.5\%ile & $(\overline{\theta})$ GMM & Mean & Std dev\\
          &      &           &          &         &        &         &           & $t-$stat & \% Var  & \% Var\\
$\alpha$ & 1.28 &    0.16 &    0.97 &    1.18 &    1.29 &    1.39 &    1.60 & 2.86 & 56.97 & 2.71\\
$\beta_{R_p}$ & 0.70 &    0.05 &    0.59 &    0.66 &    0.70 &    0.73 &    0.80 & 5.75 & 17.92 & 2.58\\
$\beta_{SMB}$ &  -0.27 &    0.09 &   -0.45 &   -0.33 &   -0.27 &   -0.21 &   -0.10 & -1.06 & 1.39 & 0.82\\
$\beta_{HML}$ & 1.35 &    0.12 &    1.10 &    1.26 &    1.35 &    1.43 &    1.59 &  5.21 & 30.34 & 4.92\\
$\beta_{MOM}$ & 1.04 &    0.09 &    0.86 &    0.98 &    1.04 &    1.10 &    1.22 &  6.54 & 27.68 & 4.92\\
$\beta_{RMW}$ & 0.39 &    0.26 &   -0.13 &    0.21 &    0.39 &    0.57 &    0.90 & 0.92 & 2.42 & 2.33\\
$\beta_{CMA}$ &  -0.32 &    0.11 &   -0.54 &   -0.40 &   -0.32 &   -0.24 &   -0.09  & -0.88 & 0.89 & 0.57\\
\hline
\multicolumn{11}{c}{ }\\
\multicolumn{11}{l}{Out-of-sample period 2: 2001-2024}\\
$\alpha$ &    -0.14 &    0.14 &   -0.42 &   -0.24 &   -0.14 &   -0.05 &    0.13 & -0.65 & 50.16 & 3.47\\
$\beta_{R_p}$ & 0.85 &    0.05 &    0.76 &    0.82 &    0.85 &    0.88 &    0.94 & 15.82 & 38.89 & 5.20\\
$\beta_{SMB}$ &   0.12 &    0.09 &   -0.06 &    0.06 &    0.12 &    0.18 &    0.31 & 0.93 & 0.48 & 0.53\\
$\beta_{HML}$ & 0.58 &    0.09 &    0.40 &    0.52 &    0.58 &    0.64 &    0.76 &  5.21 & 9.09 & 2.42\\
$\beta_{MOM}$ & 0.58 &    0.10 &    0.39 &    0.51 &    0.58 &    0.65 &    0.77 & 6.95 & 21.32 & 5.92\\
$\beta_{RMW}$ & 0.70 &    0.14 &    0.43 &    0.61 &    0.70 &    0.79 &    0.97 &  5.08 & 6.81 & 2.24\\
$\beta_{CMA}$ & 0.03 &    0.14 &   -0.25 &   -0.07 &    0.03 &    0.12 &    0.30 &  0.12 & 0.22 & 0.31\\
\hline

\end{tabular}

\end{center}

\newpage

\begin{center}
{\Large {\bf Table 4} (Cont'd.)}\\
{\bf Gibbs posteriors on Fama-French factor regression coefficients}\\
\end{center}

\small
\noindent
Properties of the Gibbs posteriors on optimal out-of-sample portfolio monthly \% returns over the period 1980--2024.
The posterior comprises 300,000 post-burn-in Metropolis draws; the posterior on the parameter vector $\theta$ and $\lambda^*$ are estimated using the most recent 240 months at the
beginning of each year in the out-of-sample period.
Monthly \% returns are projected on the 6 Fama-French factors: the market risk premium, $R_p$, SMB, HML, MOM, RMW, and CMA.
$\lambda$ governs the relative weight of the data and the prior.  The Gibbs posterior values corresponding to the optimal $\lambda$ values;
$\lambda^*$ is selected at the beginning of each year using a KNEEDLE algorithm.  Also reported: the GMM $t$-statistic for the coefficient estimates from projecting the single out-of-sample
path constructed with the mean $\theta$ vector (at $\lambda^*$) onto the Fama-French factors.  (The coefficient from that regression equals the posterior mean coefficient.)
Mean \% Var reports the posterior mean of  the percentage of the portfolio variance attributed to each factor (and for $\alpha$, independent of the factors).  These sum to 
more than 100\% as I do not report all of the covariance effects--many of which are negative.  Std dev \% Var is the posterior standard deviation of this percentage.


\baselineskip 12pt

\vskip .1in

\begin{center}

\begin{tabular}{crrrrrrrrcc}

\multicolumn{11}{l}{\bf{Panel D: Power utility} \boldmath{$\gamma = 6$}}\\
\multicolumn{11}{c}{}\\

\multicolumn{11}{l}{Out-of-sample period 1: 1980-2000}\\
          & \multicolumn{7}{c}{Gibbs posterior on out-of-sample optimal portfolio returns ($| \lambda^*)$} &  &  & \\
          & \multicolumn{7}{c}{Fama-French Regressions} &  &  & \\
\cline{2-8}
Parameter & Mean & Std. dev. & 2.5\%ile & 25\%ile & median & 75\%ile & 97.5\%ile & $(\overline{\theta})$ GMM & Mean & Std dev\\
          &      &           &          &         &        &         &           & $t-$stat & \% Var  & \% Var\\
$\alpha$ &    0.73 &    0.14 &    0.46 &    0.64 &    0.73 &    0.82 &    1.00 & 2.57 & 46.61 & 3.66\\
$\beta_{R_p}$ & 0.72 &    0.05 &    0.63 &    0.69 &    0.72 &    0.75 &    0.82 & 9.43 & 34.74 & 4.34\\
$\beta_{SMB}$ &  -0.32 &    0.08 &   -0.48 &   -0.37 &   -0.32 &   -0.26 &   -0.16 & -2.03 & 3.23 & 1.49\\
$\beta_{HML}$ & 1.08 &    0.10 &    0.88 &    1.01 &    1.08 &    1.15 &    1.29 & 6.79 & 35.13 & 5.33\\
$\beta_{MOM}$ & 0.55 &    0.08 &    0.40 &    0.50 &    0.55 &    0.60 &    0.70 & 5.31 & 14.32 & 4.49\\
$\beta_{RMW}$ &  0.48 &    0.21 &    0.07 &    0.34 &    0.48 &    0.62 &    0.89 & 1.80 & 5.30 & 3.54\\
$\beta_{CMA}$ & -0.13 &    0.10 &   -0.33 &   -0.20 &   -0.13 &   -0.06 &    0.07 & -0.58 & 0.38 & 0.41\\
\hline
\multicolumn{11}{c}{ }\\
\multicolumn{11}{l}{Out-of-sample period 2: 2001-2024}\\
$\alpha$ &      -0.05 &    0.12 &   -0.29 &   -0.14 &   -0.06 &    0.03 &    0.18 &  -0.35 & 45.93 & 37.20\\
$\beta_{R_p}$ & 0.83 &    0.04 &    0.75 &    0.80 &    0.83 &    0.86 &    0.91 &  20.13 & 57.97 & 6.83\\
$\beta_{SMB}$ &   0.07 &    0.08 &   -0.09 &    0.01 &    0.07 &    0.12 &    0.22 & 0.69 & 0.35 & 0.45\\
$\beta_{HML}$ & 0.35 &    0.08 &    0.19 &    0.29 &    0.35 &    0.40 &    0.50 & 3.77 & 5.24 & 2.06\\
$\beta_{MOM}$ & 0.45 &    0.08 &    0.29 &    0.40 &    0.45 &    0.51 &    0.61 & 7.53 & 20.51 & 6.18\\
$\beta_{RMW}$ & 0.56 &    0.11 &    0.34 &    0.49 &    0.56 &    0.64 &    0.79 & 5.51 & 7.00 & 2.43\\
$\beta_{CMA}$ & 0.05 &    0.12 &   -0.18 &   -0.03 &    0.05 &    0.13 &    0.29 & 0.34 & 0.30 & 0.40\\
\hline

\end{tabular}
\end{center}
\end{landscape}

\normalsize

\baselineskip 19pt

\parindent 18pt

\pagestyle{plain}

\newpage

\pagestyle{empty}

\begin{center}
{\Large {\bf Table A-1}}\\
{\bf Multivariate Potential Scale Reduction Factor}
\end{center}
\baselineskip 12pt

\noindent
For a given utility function, 240-month sample period, and auxiliary parameter $\lambda$, I compare multiple independent Metropolis chains to assess convergence.

\vskip .1in

\hoffset=-0.3in

\begin{tabular}{ccccc}

\multicolumn{5}{l}{{\bf Panel A. Log utility}}\\
 &       &         &   &          \\
Period & $\lambda$ &  n (chain length) & m (number of chains) & MPSRF\\
\hline
1 & 7,500 & 100,000 & 3 & 1.0004703\\
9 & 7,500 & 100,000 & 3 & 1.0002440\\
18 & 7,500 & 100,000 & 4 & 1.0001230\\
22 & 6,250 & 100,000 & 3 & 1.0004106\\
24 & 6,250 & 100,000 & 5 & 1.0002022\\
24 & 6,250 & 300,000 & 3 & 1.0001144\\
34 & 5,000 & 100,000 & 3 & 1.0004019\\
44 & 5,000 & 100,000 & 3 & 1.0001230\\
\hline

 &       &         &   &          \\
\multicolumn{5}{l}{{\bf Panel B. Power utility \boldmath{$\gamma = 2$}}}\\
\hline
1 & 5,000 & 100,000 & 4 & 1.0001204\\
9 & 4,000 & 100,000 & 4 & 1.0000933\\
18 & 4,500 & 100,000 & 4 & 1.0002640\\
18 & 4,500 & 300,000 & 3 & 1.0000791\\
22 & 3,500 & 100,000 & 4 & 1.0002196\\
22 & 3,500 & 300,000 & 2 & 1.0000429\\
24 & 2,500 & 100,000 & 4 & 1.0000771\\
24 & 2,500 & 300,000 & 3 & 1.0000422\\
34 & 2,500 & 100,000 & 4 & 1.0004776\\
34 & 2,500 & 300,000 & 3 & 1.0000829\\
44 & 4,500 & 100,000 & 4 & 1.0001316\\
44 & 4,500 & 300,000 & 3 & 1.0000641\\
\hline

 &       &         &   &          \\
\multicolumn{5}{l}{{\bf Panel C. Power utility \boldmath{$\gamma = 3$}}}\\
\hline   
1 & 4,000 & 100,000 & 3 & 1.0005777\\
1 & 4,000 & 300,000 & 2 & 1.0000454\\
9 & 3,500 & 100,000 & 3 & 1.0001730\\
9 & 3,500 & 300,000 & 2 & 1.0001899\\
18 & 3,500 & 100,000 & 4 & 1.0002793\\
18 & 3,500 & 300,000 & 2 & 1.0000864\\
22 & 3,000 & 100,000 & 5 & 1.0000824\\
22 & 3,000 & 300,000 & 3 & 1.0000693\\
24 & 3,000 & 100,000 & 5 & 1.0002097\\
24 & 3,000 & 300,000 & 3 & 1.0000693\\
34 & 2,500 & 100,000 & 4 & 1.0002869\\
34 & 2,500 & 300,000 & 3 & 1.0000743\\
44 & 2,500 & 100,000 & 4 & 1.0000793\\
44 & 2,500 & 300,000 & 3 & 1.0000526\\ 
\hline

\end{tabular}

\newpage

\begin{center}     
{\Large {\bf Table A-1}  (Cont'd.)}\\
{\bf Multivariate Potential Scale Reduction Factor}
\end{center}
\baselineskip 12pt

\noindent
For a given utility function, 240-month sample period, and auxiliary parameter $\lambda$, I compare multiple independent Metropolis chains to assess convergence.

\vskip .1in

\hoffset=-0.3in

\begin{tabular}{ccccc}

\multicolumn{5}{l}{{\bf Panel D. Power utility}\boldmath{ $\gamma = 6$}}\\
 &       &         &   &          \\
Period & $\lambda$ &  n (chain length) & m (number of chains) & MPSRF\\
\hline
1 & 3,000 & 100,000 & 4 & 1.0001007\\
1 & 3,000 & 300,000 & 2 & 1.0001778\\
9 & 2,500 & 100,000 & 4 & 1.0001475\\
9 & 2,500 & 300,000 & 2 & 1.0000298\\
18 & 2,500 & 100,000 & 4 & 1.0004917\\
18 & 2,500 & 300,000 & 2 & 1.0001451\\
22 & 2,500 & 100,000 & 4 & 1.0004142\\
22 & 2,500 & 300,000 & 2 & 1.0001432\\
24 & 2,250 & 200,000 & 3 & 1.0003145\\
34 & 2,000 & 100,000 & 5 & 1.0002440\\
34 & 2,000 & 300,000 & 4 & 1.0001573\\
44 & 2,000 & 100,000 & 4 & 1.0000994\\
44 & 2,000 & 300,000 & 3 & 1.0001669\\
\hline

\end{tabular}


\newpage

\pagestyle{empty}

\begin{center}
{\Large {\bf Table A-2}}\\
{\bf Metropolis chain properties}
\end{center}
\baselineskip 12pt

\noindent
For each 20-year period and power utility function (indexed by $\gamma$), I report the acceptance probability and effective sample size from the $\lambda^*$ case used to
construct the Gibbs posteriors.  Each posterior is obtained with 300,000 draws following a burn-in of 200,000 draws.  The table reports these posterior properties for each of the 
six hyperparameters, $\theta_1, \dots, \theta_6$ and the in-sample certainty equivalent.

\vskip .1in

\hoffset=-0.3in

\begin{center}

\begin{tabular}{cccrrrrrc}
\multicolumn{9}{l}{{\bf Panel A. Log utility}}\\
 &       &         &   &          \\
Period   & Posterior     & $\theta_1$ & $\theta_2$ & $\theta_3$ & $\theta_4$ & $\theta_5$ & $\theta_6$ & In-sample\\
         & property      &            &            &            &            &            &            & Cert. Eq.\\
1 & Accept. Rate: & 44\%    &      50\%   &    45\%   &   40\%   &       49\%   &    42\%   &   97\%\\
         & Eff. S.Sz.   & 56,847 &   54,902  & 35,640 & 20,316 &     59,799 &   20,542 & 55,247\\
\hline
9 & Accept. Rate: & 42\%    &      47\%   &    44\%   &   38\%   &  49\%   &  40\%   &   97\%\\
   & Eff. S.Sz.   & 48,368 &   46,174  & 34,856 & 18,064 &  57,911 & 18,560 & 42,717\\
\hline
18 & Accept. Rate: & 47\%    &      48\%   &    49\%   &   44\%   &  52\%   &  46\%   &   98\%\\
   & Eff. S.Sz.   & 46,476 &   41,988  & 48,278 & 28,556 &  56,947 & 26,313 & 41,014\\
\hline
22 & Accept. Rate: & 43\%    &      43\%   &    48\%   &   35\%   &       51\%   &    38\%   &   97\%\\
   & Eff. S.Sz.   & 35,812 &   30,522  & 46,589 & 15,577 &     59,071 &   14,759 & 28,468\\
\hline
24 & Accept. Rate: & 40\%    &      41\%   &    47\%   &   39\%   & 49\%   & 41\% & 96\%\\
   & Eff. S.Sz.   & 37,603 &   30,668  & 42,246 & 10,998 &     55,637 &   10,969 & 29,189\\
\hline
34 & Accept. Rate: & 43\%    &      43\%   &    48\%   &   36\%   & 51\%   & 39\% & 97\%\\
   & Eff. S.Sz.   & 39,147 &   35,534  & 44,724 & 15,417 &     58,301 &   14,844 & 32,582\\
\hline
44 & Accept. Rate: & 49\%    &      44\%   &    48\%   &   34\%   & 58\%   & 37\% & 98\%\\
   & Eff. S.Sz.   & 49,326 &   43,834  & 47,750 & 34,223 &     57,645 &   36,559 & 40,262\\
\hline

 &       &         &   &          \\
\multicolumn{9}{l}{{\bf Panel B. Power utility, \boldmath{$\gamma = 2$}}}\\
 &       &         &   &          \\
Period   & Posterior     & $\theta_1$ & $\theta_2$ & $\theta_3$ & $\theta_4$ & $\theta_5$ & $\theta_6$ & In-sample\\
         & property     &            &            &            &            &            &            & Cert. Eq.\\
1 & Accept. Rate: & 47\%    &      56\%   &    49\%   &   49\%   &       49\%   &    49\%   &   98\%\\
         & Eff. S.Sz.   & 55,708 &   47,056  & 32,690 & 19,051 &     57,720 &   20,087 & 49,712\\
\hline 
9 & Accept. Rate: & 48\%    &      54\%   &    51\%   &   50\%   &  49\%   &  52\%   &   99\%\\
   & Eff. S.Sz.   & 51,871 &   43,156  & 36,718 & 21,430 &  58,062 & 21,422 & 44,717\\
\hline 
18 & Accept. Rate: & 46\%    &      46\%   &    47\%   &   45\%   &  47\%   &  49\%   &   99\%\\
   & Eff. S.Sz.   & 47,224 &   46,375  & 47,514 & 26,879 &  60,299 & 24,644 & 39,604\\
\hline
22 & Accept. Rate: & 45\%    &      48\%   &    50\%   &   48\%   &       49\%   &    50\%   &   98\%\\
   & Eff. S.Sz.   & 34,538 &   29,235  & 42,535 & 14,732 &     60,011 &   13,782 & 27,179\\
\hline
24 & Accept. Rate: & 50\%    &      51\%   &    53\%   &   49\%   & 53\%   & 49\% & 99\%\\
   & Eff. S.Sz.   & 38,816 &   33,923  & 44,633 & 14,531 &     53,273 &   14,045 & 35,770\\
\hline
34 & Accept. Rate: & 49\%    &      50\%   &    52\%   &   50\%   & 53\%   & 50\% & 99\%\\
   & Eff. S.Sz.   & 36,669 &   34,636  & 43,262 & 16,091 &     55,071 &   15,295 & 33,671\\
\hline
44 & Accept. Rate: & 48\%    &      49\%   &    49\%   &   45\%   & 49\%   & 50\% & 98\%\\
   & Eff. S.Sz.   & 40,465 &   33,220  & 41,635 & 26,037 &     60,610&   27,396 & 34,338\\
\hline

\end{tabular}

\end{center}

\newpage

\begin{center}
{\Large {\bf Table A-2} (Cont'd.)}\\
{\bf Metropolis chain properties}   
\end{center}
\baselineskip 12pt

\noindent 
For each 20-year period and power utility function (indexed by $\gamma$), I report the acceptance probability and effective sample size from the $\lambda^*$ case used to
construct the Gibbs posteriors.  Each posterior is obtained with 300,000 draws following a burn-in of 200,000 draws.  The posterior for Period 24 in Panel D
uses 600,000 draws.  The table reports these posterior properties for each of the
six hyperparameters, $\theta_1, \dots, \theta_6$ and the in-sample certainty equivalent.
 
\vskip .1in        

\hoffset=-0.3in    

\begin{center}

\begin{tabular}{cccrrrrrc}
\multicolumn{9}{l}{{\bf Panel C. Power utility, \boldmath{$\gamma = 3$}}}\\
 &       &         &   &          \\
Period   & Variable     & $\theta_1$ & $\theta_2$ & $\theta_3$ & $\theta_4$ & $\theta_5$ & $\theta_6$ & In-sample\\
         &              &            &            &            &            &            &            & Cert. Eq.\\
1 & Accept. Rate: & 43\%    &      52\%   &    50\%   &   47\%   &       47\%   &    50\%   &   98\%\\
         & Eff. S.Sz.   & 55,943 &   46,442  & 31,152 & 17,296 &     62,213 &   18,235 & 49,712\\
\hline 
1 & Accept. Rate: & 43\%    &      50\%   &  51\%   &   47\%   &       46\%   &    51\%   &   98\%\\
         & Eff. S.Sz.   & 49,582 &   41,348  & 33,671 & 18,123 &     60,118 &   18,291 & 42,198\\
\hline
18 & Accept. Rate: & 45\%    &      46\%   &    47\%   &   41\%   &  50\%   &  43\%   &   99\%\\
   & Eff. S.Sz.   & 47,026 &   43,806  & 46,798 & 24,587 &  57,605 & 23,321 & 39,604\\
\hline
22 & Accept. Rate: & 38\%    &      42\%   &    52\%   &   39\%   &       46\%   &    41\%   &   96\%\\
   & Eff. S.Sz.   & 32,014 &   26,881  & 39,388 & 12,198 &     59,793 &   11,746 & 24,713\\
\hline
24 & Accept. Rate: & 44\%    & 49\%   &  59\%   &   46\%   & 51\%   & 52\% & 98\%\\
   & Eff. S.Sz.   & 31,860 &   26,267  & 32,559 & 8,953 &     52,151 &   8,596 & 24,885\\
\hline
34 & Accept. Rate: & 45\%    &      47\%   &    51\%   &   49\%   & 48\%   & 47\% & 98\%\\
   & Eff. S.Sz.   & 30,504 &   28,864  & 38,097 & 11,591 &     57,058 &   11,247 & 26,452\\
\hline
44 & Accept. Rate: & 43\%    &      43\%   &    47\%   &   45\%   & 50\%   & 48\% & 98\%\\
   & Eff. S.Sz.   & 45,023 &   37,841  & 44,929 & 27,740 &     59,556 &   29,416 & 46,894\\
\hline

&       &         &   &          \\
\multicolumn{9}{l}{{\bf Panel D. Power utility, \boldmath{$\gamma = 6$}}}\\
 &       &         &   &          \\
Period   & Posterior     & $\theta_1$ & $\theta_2$ & $\theta_3$ & $\theta_4$ & $\theta_5$ & $\theta_6$ & In-sample\\
         & property     &            &            &            &            &            &            & Cert. Eq.\\
1 & Accept. Rate: & 48\%    &      48\%   &    54\%   &   47\%   &       49\%   &    48\%   &   98\%\\
         & Eff. S.Sz.   & 51,661 &   43,856  & 25,355 & 14,002 &     60,585 &   13,822 & 41,868\\ 
\hline 
9 & Accept. Rate: & 48\%    &      45\%   &    55\%   &   47\%   &  49\%   &  49\%   &   99\%\\
   & Eff. S.Sz.   & 46,058 &   35,561  & 26,821 & 14,075 &  58,939 & 14,331 & 37,067\\
\hline
18 & Accept. Rate: & 41\%    &      43\%   &    44\%   &   37\%   &  49\%   &  39\%   &   96\%\\
   & Eff. S.Sz.   & 40,519 &   38,749  & 40,759 & 18,527 &  57,274 & 18,418 & 35,989\\
\hline
22 & Accept. Rate: & 48\%    &      49\%   &    51\%   &   44\%   &       50\%   &    45\%   &   98\%\\
   & Eff. S.Sz.   & 24,923 &   19,107  & 30,000 & 7,394 &     56,592 &   7,154 & 17,042\\
\hline
24 & Accept. Rate: & 46\%    &      48\%   &    51\%   &   41\%   & 47\%   & 43\% & 98\%\\
   & Eff. S.Sz.   & 49,993 &   40,302  & 59,248 & 12,080 &    104,013 &   11,657 & 40,260\\
\hline
34 & Accept. Rate: & 39\%    &    43\%   &    52\%   &   44\%   & 45\%   & 48\% & 97\%\\
   & Eff. S.Sz.   & 32,224 &   30,625  & 42,691 & 9,307 &     74,651 &   9,121 & 25,010\\
\hline
44 & Accept. Rate: & 51\%    &      46\%   &    51\%   &   39\%   & 53\%   & 46\% & 98\%\\
   & Eff. S.Sz.   & 34,290 &   28,651  & 36,004 & 22,129 &     56,879 &   23,351 & 51,160\\
\hline

\end{tabular}

\end{center}

\newpage

\begin{center}
{\Large {\bf Table A-3}}\\
{\bf Posterior Correlations}
\end{center}
\baselineskip 12pt

\noindent
The table reports posterior correlations between select elements of $\theta$ conditional on $\lambda^*$.
   
\vskip .1in

\hoffset=-0.3in       

\begin{center}

{\bf Panel A. Correlations between $\bm{\beta-\theta}$ and residual volatility-$\bm{\theta}$ in \%}\\

\vskip .1in

\begin{tabular}{crrrr}
Period   & $\gamma = 1$     & $\gamma = 2$ & $\gamma = 3$ & $\gamma = 6$\\
1 & -61 & -61 & -63 & -69\\
9 & -63 & -59 & -63 & -67\\
18 & -51 & -53 & -55 & -60\\
22 & -63 & -63 & -66 & -72\\
24 & -75 & -71 & -78 & -82\\
34 & -72 & -72 & -78 & -83\\
44 & -45 & -54 & -52 & -59\\
\hline
   &    &         &             \\
   &    &         &             \\
\end{tabular}

{\bf Panel B. Correlations between momentum$\bm{-\theta}$ and book-to-market$\bm{-\theta}$ in \%}\\

\begin{tabular}{crrrr}

Period   & $\gamma = 1$     & $\gamma = 2$ & $\gamma = 3$ & $\gamma = 6$\\
1 & 10 & 14 & 14 & 16\\
9 & 25 & 27 & 29 & 33\\
18 & 22 & 24 & 25 & 28\\
22 & 29 & 29 & 32 & 38\\
24 & 43 & 39 & 48 & 55\\
34 & 45 & 46 & 54 & 61\\
44 & 30 & 38 & 35 & 42\\
\hline
   &    &         &             \\
   &    &         &             \\

\end{tabular}

{\bf Panel C. Correlations between log size$\bm{-\theta}$ and $\bm{\beta-\theta}$ in \%}\\

\begin{tabular}{crrrr}

Period   & $\gamma = 1$     & $\gamma = 2$ & $\gamma = 3$ & $\gamma = 6$\\
1 & 28 & 28 & 29 & 28\\
9 & 26 & 23 & 28 & 29\\
18 & 14 & 16 & 15 & 17\\
22 & 12 & 20 & 13 & 12\\
24 & 11 & 12 & 10 & 10\\
34 & 12 & 12 & 11 & 8\\
44 & 16 & 16 & 16 & 14\\
\hline
\end{tabular}
\end{center}

\end{document}